\documentclass[11pt,a4paper]{amsart}

\pdfoutput=1

\usepackage{natbib}

\usepackage{tikz}
\usepackage{mathrsfs}
\usepackage{amsfonts}

\usepackage{upgreek}

\DeclareFontFamily{OT1}{pzc}{}
\DeclareFontShape{OT1}{pzc}{m}{it}{<-> s * [1.10] pzcmi7t}{}
\DeclareMathAlphabet{\mathpzc}{OT1}{pzc}{m}{it}

\usepackage{geometry}                % See geometry.pdf to learn the layout options. There are lots.
\usepackage{graphicx}
\usepackage{amssymb}
\usepackage{epstopdf}
\usepackage{fullpage}
\usepackage{color}
\usepackage{float}
\usepackage{caption}
\usepackage{tablefootnote}

\usepackage{url}

\usepackage{float} 

\RequirePackage[OT1]{fontenc}
\RequirePackage{amsthm,amsmath}
\usepackage{amsmath, amssymb, amsfonts} % for math symbols like \Sigma
\usepackage{algorithm}
\usepackage{algpseudocode}
\usepackage{booktabs}  % for tables
\usepackage{graphicx, subfigure} % for figures
\usepackage{multirow} % for tables

\usepackage{mathabx}

\usepackage{longtable}

\usepackage{accents}

\newcommand{\bfzero} {\boldsymbol{0}}
\newcommand{\bfalpha} {\boldsymbol{\alpha}}

\newcommand{\bfdelta} {\boldsymbol{\delta}}

\newcommand{\bfmu} {\boldsymbol{\mu}}

\newcommand{\bflambda} {\boldsymbol{\lambda}}

\newcommand{\bfSigma} {\boldsymbol{\Sigma}}
\newcommand{\bfLambda} {\boldsymbol{\Lambda}}
\newcommand{\bfDelta} {\boldsymbol{\Delta}}
\newcommand{\bfOmega} {\boldsymbol{\Omega}}

\newcommand{\bff} {\mathbf{f}}
\newcommand{\bfP} {\mathbf{P}}
\newcommand{\bfepsilon} {\mathbf{\epsilon}}

\newcommand{\bfI} {\mathbf{I}}
\newcommand{\bfZ} {\mathbf{Z}}

\newcommand{\bfX} {\mathbf{X}}
\newcommand{\bfV} {\mathbf{V}}
\newcommand{\bfE} {\mathbf{E}}
\newcommand{\bfW} {\mathbf{W}}
\newcommand{\bfD} {\mathbf{D}}

\newcommand{\bfK} {\mathbf{K}}

\renewcommand{\Pr}{\mathsf{P}}
\newcommand{\p}{\mathsf{P}}

\newcommand{\Cov}{\mathsf{Cov}}

\DeclareMathOperator{\diag}{diag}

\newcommand{\normal}{\mathsf{N}}

\newcommand{\IGam}{\mathsf{IG}}

\newcommand{\Wis}{\mathsf{Wis}}

\newcommand{\GWis}{\mathsf{Wis}}

\newcommand{\bfS} {\mathbf{S}}

\renewcommand{\Pr}{\mathsf{P}}

\newcommand{\indep}{\ensuremath{\perp \!\!\! \perp}}

\begin{document}

\title[Single-factor Graphical Models]{Bayesian Inference for Single-factor Graphical Models}\thanks{CONTACT D. Marcano. Email: dmarcano@uw.edu}

\author{David Marcano\textsuperscript{$\ast$} and Adrian Dobra\textsuperscript{$\ast$}}
\address{\textsuperscript{$\ast$} Department of Statistics, University of Washington, Seattle, WA, USA}

\begin{abstract}
We introduce efficient MCMC algorithms for Bayesian inference for single-factor models with correlated residuals where the residuals' distribution is a Gaussian graphical model. We call this family of models single-factor graphical models. We extend single-factor graphical models to datasets that also involve binary and ordinal categorical variables and to the modeling of multiple datasets that are spatially or temporally related. Our models are able to capture multivariate associations through latent factors across time and space, as well as residual conditional dependence structures at each spatial location or time point through Gaussian graphical models. We illustrate the application of single-factor graphical models in simulated and real-world examples.\\
KEYWORDS: Bayesian modeling; graphical models; factor models; MCMC; structural learning;
\end{abstract}

\maketitle

\date{\today} 

\tableofcontents

\section{Introduction}

In this paper, we bridge the gap between two very popular statistical models: factor models and Gaussian graphical models. Both classes of models capture structured multivariate associations, but the manner in which this is done is quite different. Classical factor models \citep{anderson-rubin} assume that multivariate associations come from one, two, or several latent variables, and that, conditional on these latent variables, the observed variables are independent. In Gaussian graphical models \citep{dempster_1972,lauritzen1996graphical}, two components of a multivariate normal random vector are conditionally independent given the rest if the corresponding element of its precision matrix is zero. The starting point for the modeling framework we develop has been originally laid out by \citet{stanghellini-1997,vicard-2000} and assumes that the same conditional independence relationship takes place also by conditioning on a single latent variable. This leads to the class of single-factor models with correlated residuals in which the correlation structure is defined by a Gaussian graphical model.

The statistical foundations of factor analysis date back to \cite{anderson-rubin} with comprehensive treatments provided by \cite{lawley-maxwell} and \cite{harman}. The non-Gaussian treatment of factor analysis can be traced back to \cite{bartholomew} and an extension under a Bayesian framework to \cite{arminger}. The seminal work of \cite{stanghellini-1997} and \cite{vicard-2000} provides the necessary and sufficient constraints for the identifiability of a single-factor model with correlated residuals. Since then, the literature has gone away from single-factor models to infinite factor models \citep{bhattacharya} and sparse models \citep{fruhwirth-schnatter-lopes}. More recent contributions focus on fast estimation using modern statistical frameworks \citep{wang-li} or extensions to spatial or more complex data \citep{krupskii-2017}. Notable advances include \cite{wang_wall_2003,hogan_tchernis_2004} who develop single-factor models to represent the spatial latent structure among random variables observed at various locations. The models of \cite{lopes-et-2008} represent the temporal dependence using one, two, or more factors, and the spatial dependence through their corresponding factor loadings. All of these approaches make the assumption that the resulting residuals are independent. In this sense, they follow the path opened by numerous other factor analysis approaches for multivariate data \--- see, for example, \cite{carvalho-et-2008,ghosh-dunson-2009,yoshida-west-2010} and the references therein. However, the assumption of independence of residuals is quite strong. In particular, it is unclear why a single-factor actually captures all the associations among the observed variables. One can capture the remaining associations by allowing more than one factor to enter the models, but in that case the non-trivial key question of how many factors should be allowed in the model emerges. In the framework we plan to develop, instead of allowing the inclusion of two or more factors, we will create a structured representation of the associations of the residuals in the single-factor model through graphical models.

\begin{figure}
  \begin{center}
\includegraphics[width=0.5\textwidth]{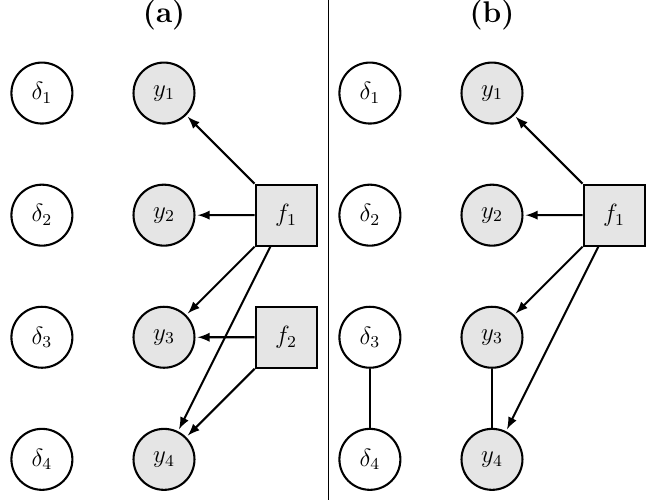}
  \end{center}
\caption{\label{fig:chaingraph1}Dependency structure in factor models for four observed variables $y_1,y_2,y_3,y_4$. Residuals, observed variables and latent variables are designated by white circles, gray circles and gray rectangles, respectively. Arrows indicate non-zero factor loadings. Panel (a): A factor model with two factors $f_1$ and $f_2$ and uncorrelated residuals $\delta_1,\delta_2,\delta_3,\delta_4$. Panel (b): Factor model obtained by marginalizing the two-factor model in Panel (a) over $f_2$. Marginalization creates dependence between $y_3$ and $y_4$ which is not accounted for by the first factor $f_1$. This dependence induces an edge between $\delta_3$ and $\delta_4$.}
\end{figure}

Figure \ref{fig:chaingraph1} shows the correspondence between a two-factor model with independent residuals and a one-factor model with correlated residuals. Our framework makes use of graphical models to represent the heterogeneity of the association structure of the residuals as well as the multivariate associations in the latent space. The graphs involved can be either known (e.g., if modeling spatial dependence) or considered unknown. The associations among a vector of observed variables are decomposed into associations in the latent space and associations in the residual space. This decomposition is advantageous for two reasons: (i) interpretability; (ii) additional associations can be modeled in the latent space in a coherent manner. In the data applications we present, we show that by loosening the constraint of the independence of the residuals, we obtain temporally and spatially varying structures in both the latent factor and the residual dependence graphs. This allows for a flexible modeling of complex spatio-temporal dependencies that are known in some dimensions but unknown in other dimensions.

The structure of this paper is as follows. In Section \ref{factor-models} we introduce the traditional factor model, discuss identifiability conditions, and extensions that allow correlated residuals. In Section \ref{sec:graphical-models} we give a brief overview of Gaussian graphical models and of the Bayesian approaches for inference based on the G-Wishart distribution that are relevant for our developments. In Section \ref{sec:gfm-onedataset}  we introduce our Bayesian framework in single-factor graphical models, and develop our main MCMC posterior sampling method which we call the Identifiable Direct Conditional Bayes Factor (IDCBF) algorithm. In the same section, we also show how to extend the single-factor graphical model to binary and ordinal categorical data via a copula framework. In Section \ref{sec:gfm-multiway}, we demonstrate how to use single-factor graphical models to analyze multiple datasets collected at different spatial locations or across varying time points. In Section \ref{sec:simulationstudy}, we demonstrate the numerical performance of our proposed sampling methods through a simulation study. In Section \ref{sec:rochdale}, we analyze a $8$-way binary contingency table. In Section \ref{sec:lbw}, we illustrate the use of the single-factor graphical model for multiple datasets in the analysis of $100$ contingency tables related to counties in North Carolina. In Section \ref{sec:hivanalysis}, we present our second real-world application, which focuses on modeling the temporal dynamics of risk factors associated with the acquisition of HIV. In Section \ref{sec:discussion}, we discuss our proposed framework along with several possible extensions.

\section{Factor Models}
\label{factor-models}
A factor model is a particular type of latent variable model in which the latent variables and the observable variables are continuous and the relationship between them is assumed to be linear \citep{alma99321088120801451}. Factor models are widely used across scientific fields with great success. For example, factor models have been extensively used in finance for asset allocation, risk management, and even to help interpret the stock market \citep{LHABITANT2017157}. In clinical psychology, factor models have been used to examine unobservable quantities such as neuroticism, depression, and learning and development \citep{wright_2020}.

The classical normal linear $k$-factor model can be written as:
\begin{align*}
    \bfX &= \bfLambda \bff + \bfdelta
\end{align*}
where $\bfX = (X_1,\ldots,X_p)$ is the $p$-dimensional vector of observed variables, $\bff$ is the $k$-dimensional ($k\le p$) vector of latent variables that is assumed to be component-wise independent, $\bfLambda$ is a $p\times k$ matrix of factor loadings, and $\bfdelta$ is a $p$-dimensional  vector of residuals with positive definite covariance matrix $\bfSigma$. The latent variables $\bff$ and the residuals $\bfdelta$ are assumed to be independent. In most applications, residuals are assumed to be independent, that is, $\bfSigma = \diag\{\sigma_1^2,\ldots,\sigma_p^2\}$. Therefore the resulting model is:
\begin{eqnarray} \label{factor-model1}
    \bfX\mid \bff,\bfLambda,\bfSigma &\sim \normal_p(\bfLambda \bff, \bfSigma)\\ \nonumber
    \bff &\sim \normal_k(\bf0, \bfV).
\end{eqnarray}
where $\bfV = \diag(v_1,\dots,v_k)$ is the covariance of the latent factors. Thus, the marginal covariance matrix of $\bfX$ is $\bfOmega = \bfLambda \bfV \bfLambda^{T} + \bfSigma$.  In its most general form, the factor model (\ref{factor-model1}) has two key issues: its identifiability and the choice of the number of factors $k$.

\subsection{Identifiability of the Classical Factor Model}
Making the classical factor model \eqref{factor-model1} identifiable is not straightforward for the following reasons. First, this model is invariant under orthogonal transformations. Let $\bfP$ be any orthogonal $k\times k$ matrix. Then $\bfLambda^* = \bfLambda P^{T}$ and $\bff^* = \bfP\bff$ imply $\bfX = \bfLambda^*\bff^* + \bfepsilon = \bfLambda \bff + \epsilon$. This indeterminacy can be eliminated by imposing $\Cov(f) = 
\bfI_k$. 

Another area of indeterminacy comes from the identifiability of $\bfLambda$. In a frequentist framework the identifiability of the factor loadings matrix is not an issue since the likelihood equations are satisfied regardless. It also does not have any bearing on the predictive inference or estimation of the covariance matrix $\bfOmega$. In a Bayesian framework, working with a model in which $\bfLambda$ is not identified leads to a multimodal posterior distribution $\p(\bfLambda\mid \cdot)$ that could be difficult to sample from. Additionally, this makes interpretation of the factors challenging, as both $\bfLambda$ and $f$ are not marginally identifiable. The factor loading matrix $\bfLambda$ can be made identifiable by constraining it to be lower triangular \citep{geweke-zhou, lopes-west, carvalho-et-2008}. Moreover,  the diagonal elements of $\bfLambda$ need to be constrained to be strictly positive in order to address the sign-switching problem: $\bfLambda \bff = (-\bfLambda)(-\bff)$. If some of the elements of $\bfLambda$ below the main diagonal are restricted to zero, the identifiability of $\bfLambda$ can be lost once again ------ see \cite{anderson-rubin}.

\subsection{Determining the number of latent factors}

Choosing the appropriate number of factors is still an open problem despite the numerous methods for addressing this question that have been proposed in the literature \--- see \cite{10.1214/15-STS539} for an in-depth review. In a frequentist framework, classical methods for finding the number of factors include the scree test \citep{cattell, cattell-vogelmann}, tests based on the likelihood ratio \citep{bartlett}, and information criteria methods such as minimum description length \citep{fishler-grosmann}.  In a Bayesian framework, \cite{lopes-west} propose an RJMCMC based on normal distributions and Student's $t$-distributions with preliminary draws within each iteration. However, there is a high computational cost associated with posterior moves between models of $(k, \bfLambda_k, \bfSigma) \to (k^{'}, \bfLambda_{k^{'}}, \bfSigma)$, which limits the applicability of this approach. Other Bayesian approaches that allow infinitely many columns in the loading matrix \citep{griffith} learn the number of factors intrinsically. These approaches require careful hyperparameter tuning.

Another approach is that of variable selection in a finite-dimensional overfitted factor model \citep{fruhwirth-schnatter-lopes} where they show that the number of factors may be recovered in such a model. This introduces several identifiability problems, in particular, a ranking deficient factor loading matrix $\bfLambda$ \citep{geweke-singleton}. To fix this, they introduce ordering constraints on the matrix $\bfLambda$ which allow the estimation of the rank of $\bfLambda$ to obtain an estimate of the number of factors in an over-fitted model.  \citet{10.1214/24-BA1423} provide the latest contribution to efficient estimation of the number of common factors in a sparse latent factor model with spike-and-slab priors in $\bfLambda$.

\subsection{Beyond the Classical Factor Model}
All the above methods for determining the number of factors require either additional constraints on model parameters or significantly higher computational costs. The need to properly estimate the number of factors comes from the strong assumption of independent residuals. What happens if this assumption is relaxed and allows residuals to be correlated (i.e. allow for a nondiagonal $\bfSigma$)? Figure \ref{fig:chaingraph1} shows the correspondence between a two-factor model with independent residuals and a one-factor model with correlated residuals. When we marginalize a multiple-factor model to one with fewer factors, we potentially induce dependence between the residuals. Furthermore, the two-factor model in panel (a) is not identifiable due to the result of \cite{anderson-rubin} (Lemma 5.3). 

The example in Figure \ref{fig:chaingraph1} implies that a model with a smaller number of factors can be considered as a replacement for a model with a larger number of factors by allowing the residuals to be correlated. In what follows, we develop the most extreme case of a model with a single-factor and correlated residuals. Compared to a model with $k$ factors that involves $pk+p$ parameters,  this model will involve $p+p(p+1)2$ parameters. In order to build models that are as parsimonious as possible, we will identify which elements of the precision matrix $\bfK = \bfSigma^{-1}$ can be set to zero while still providing a good fit for the data. To this end, we need to develop a method for selecting Gaussian graphical models for the distribution of the residuals.

\section{Gaussian Graphical Models}
\label{sec:graphical-models}

Let $\bfX  = (X_1,\ldots,X_p)$ be a vector of observed random variables that follows a multivariate normal distribution $\normal_p\left(\bf0, \bfK^{-1}\right)$. A Gaussian graphical model \citep{dempster_1972} with graph $G=(V,E)$  ($V=\{1,\ldots,p\}$, $E\subset V\times V$) is determined by requiring that the precision matrix $\bfK$ have nonzero elements only for pairs $(v,v^{\prime})\in V\times V$ such that $(v,v^{\prime})\notin E$. This implies that the conditional independence relationship $X_v \indep X_{v^{\prime}} \mid \bfX_{V\setminus\{v,v^{\prime}\}}$ is valid for every pair $(v,v^{\prime})\notin E$. We denote by $M^{+}(G)$ the cone of positive symmetric definite matrices that are consistent with the graph $G$, that is, have zero elements for all row and column pairs that do not correspond to an edge of $G$. Bayesian inference proceeds by specifying a prior $\Pr(\bfK\mid G)$ on $M^{+}(G)$ for the precision matrix $\bfK$, and a prior $\Pr(G)$ for the graph $G$ on the space $\mathcal{G}_p$ of $2^{p(p-1)/2}$ graphs with $p$ vertices. This leads to the joint posterior distribution
\begin{eqnarray} \label{eq:jointposteriorKG}
    \Pr(\bfK,G\mid \mathcal{D}) & \propto & \Pr(\mathcal{D}\mid \bfK) \Pr(\bfK\mid G)\Pr(G),
\end{eqnarray}
\noindent where $\mathcal{D}$ is the $n\times p$ data matrix with independent rows $\bfX^{(j)}\sim \normal_p\left(\bf0, \bfK^{-1}\right)$, $j=1,\ldots,n$.

One of the most popular priors on $M^{+}(G)$ is the $G$-Wishart prior $\GWis_G(\delta,\bfD)$  whose density with respect to the Lebesgue measure is 
\citep{roverato_2002,atay-kayis_massam_2005,letac_massam_2007}:
\begin{eqnarray} \label{eq:wishart}
\p\left(\bfK\mid G,\delta,\bfD\right) & = &\frac{1}{I_G(\delta,\bfD)}(\mbox{det}\; \bfK)^{(\delta-2)/2}\exp\left\{-\frac{1}{2}\langle \bfK,\bfD\rangle\right\},
\end{eqnarray}
where $\delta>2$ represents the number of degrees of freedom, $\bfD$ is a positive-definite rate matrix, and $\langle \bfK,\bfD\rangle = \mbox{tr}(\bfK^{T} \bfD)$ denotes the trace inner product. Under these conditions, the normalization constant $I_G(\delta,\bfD)$ is finite. If $G$ is the complete graph ($E=V\times V$), $\GWis_G(\delta,\bfD)$ is the Wishart distribution $\Wis_{p}(\delta,\bfD)$ \citep{muirhead_2005}. The $G$-Wishart prior is conjugate to the normal likelihood and produces a conditional posterior distribution on $M^{+}(G)$: $\p\left(\bfK \mid  G, \mathcal{D}\right) = \Wis_G(\delta + n, \bfD + \bfS)$ where $\bfS = \mathcal{D}^T\mathcal{D}$.

\cite{mohammadi-wit-2015} show that Bayesian inference with the G-Wishart prior has key advantages over frequentist methods for Gaussian graphical model determination such as the graphical lasso. Posterior estimation is usually done using MCMC sampling schemes \citep{wang-li, hinne-et-2014} or sequential Monte Carlo (SMC) \citep{tan-2017, boom-2022}. The high computational cost of some of these sampling methods has been alleviated by restricting the graph search to smaller subspaces of the graph space, such as decomposable graphs \citep{carvalho-et-2008, carvalho_scott_2009, wang_carvalho} since the corresponding marginal posterior distributions distributions $\Pr(G\mid \mathcal{D})$ over the graph space can be determined up to a normalizing constant using formulas. However, decomposability is quite hard to justify in applied settings. In particular, as the dimensionality of the data increases, constraining the graphs to be decomposable results in spurious edges ------ edges that are not present in the true Gaussian graphical model that generated the data \--- to receive high posterior inclusion probabilities \citep{fitch-2014, niu-2021}. As such, it is desirable to perform an unrestricted search over graphs that are both decomposable and non-decomposable.

In earlier literature, sampling from the G-Wishart distribution has been performed with the acceptance-rejection algorithm of \citet{wang_carvalho}, the random walk Metropolis-Hastings algorithm of \cite{dobra-lenkoski-rodriguez-2011} or the block Gibbs sampler algorithm of \cite{wang-li}. However, these sampling methods are relatively expensive from a computational point of view. One way to alleviate this problem is to replace the G-Wishart prior with shrinkage priors on the precision matrix \citep{wang-2015, li-2019, sagar-2021}. Efficient sampling schemes from the joint posterior $\Pr(\bfK,G\mid \mathcal{D})$  became possible after the direct sampler algorithm of \cite{lenkoski-2013} was developed. Because this sampler is a fast projection of the space of positive symmetric definite matrices onto $M^{+}(G)$, it opens up the possibility for numerous applications of the G-Wishart distribution in Bayesian modeling that do not rely on approximate samples \citep{wang_carvalho,dobra-lenkoski-rodriguez-2011,wang-li}.

Although several such algorithms for sampling from the joint posterior distribution \eqref{eq:jointposteriorKG} have been proposed in the literature (see, for example, \citet{dobra-lenkoski-rodriguez-2011} and the references therein), these approaches rely on approximate sampling schemes from the $G$-Wishart distribution and also on numerical methods for estimating the corresponding ratios of normalizing constants of the $G$-Wishart distributions involved in Metropolis-Hastings ratios. These algorithms are computationally demanding. The first algorithm for sampling from the joint posterior \eqref{eq:jointposteriorKG} that was truly efficient is the exchange algorithm of \cite{cheng-lenkoski-2012} that uses the block Gibbs sampler of \cite{wang-li} to sample precision from G-Wishart distributions. \cite{hinne-et-2014} suggest that the exchange algorithm of \cite{cheng-lenkoski-2012} can be made more efficient by using the direct sampler of \cite{lenkoski-2013}. They subsequently developed the double conditional Bayes factor algorithm (DCBF), which is a key part of the MCMC sampling algorithms that we propose in this paper. Additional improvements of the DCBF algorithm can be achieved in terms of mixing, convergence speed and computing time with the recently developed G-Wishart weighted proposal algorithm (WWA) \citet{boom-2022}, which is based on an informed proposal on graph space and a delayed acceptance scheme.  

\section{Single-factor graphical Models}
\label{sec:gfm-onedataset}

We consider the single-factor model \--- take $k=1$ in \eqref{factor-model1}:
\begin{eqnarray}\label{eq:singlefactor}
 \bfX & = &\bflambda f + \bfdelta,\\ \nonumber
 f & \sim & \normal(0,1),\\ \nonumber
 \bfdelta & \sim & \normal_p(\bf0,\bfSigma),
\end{eqnarray}
\noindent where $\bflambda = (\lambda_1,\ldots,\lambda_p)^T$ is the vector of factor loadings and $f$ is the common factor.

The single-factor graphical model (SFGM, heneforth) is obtained by further assuming that, in the single-factor model \eqref{eq:singlefactor}, the distribution of the residuals follows a Gaussian graphical model with conditional independence graph $G\in \mathcal{G}_p$, i.e.
\begin{eqnarray*}
    \bfK = \bfSigma^{-1}\in M^+_{G}.
\end{eqnarray*}
Throughout this paper, we assume that the graph $G$ is unknown and must be inferred from the data. After integrating the single-factor $f$ in \eqref{eq:singlefactor}, the marginal distribution of $\bfX$ is $\normal_p(\bf0,\bfOmega)$ with $\bfOmega=\bflambda\bflambda^T + \bfK^{-1}$. The single-factor graphical model \eqref{eq:singlefactor} is identified if the elements of $\bfK$ can be uniquely determined from the elements of $\bfOmega$.

The graph $G$ that defines the Gaussian graphical model of the residuals needs to have a particular structure to guarantee that the single-factor graphical model is identifiable. \cite{vicard-2000} builds on the results of \citet{stanghellini-1997} and shows that a necessary and sufficient identifiability condition is that every connected component of the complementary graph $\bar{G}$ of $G$ (that is, all edges of $\bar{G}$ are not present in $G$ and viceversa) contains at least one odd cycle. This constraint can be checked quite efficiently by first decomposing by $\bar{G}$ in its connected components, and determining whether each connected component is bipartite by performing a two-way coloring breadth-first search. In the sequel, we call a graph identifiable if it corresponds to a single-factor graphical model that is identifiable. Related identifiability results, although in a more general setting, are given in \citet{stanghellini-wermuth-2005}. In what follows, we denote by $\mathcal{G}^{id}_p$ the graphs with $p$ vertices that make the single-factor graphical model \eqref{eq:singlefactor} identifiable.

\citet{tarantola-vicard-2002} provide algorithms based on graph tree-spanning representations to determine which edges can be added to or deleted from a graph $G$ so that the idenfiability of the resulting single-factor graph model is preserved. Their approach can therefore increase the computational efficiency of MCMC algorithms that traverse the space of identifiable graphs by avoiding the need for separate identifiability checks for all $p(p-1)/2$ edges. \citet{giudici-stanghellini-2001} proposed a reversible jump MCMC algorithm for factor models with one, two, or more factors in which the distribution of the residuals follows a Gaussian graphical model. However, their methodology is constrained to decomposable graphs and needs to visit models that are not identifiable. In our proposed methodology, we allow graphs to be arbitrary (not necessarily decomposable) and we never have to visit models that are not identifiable.

\subsection{Prior specification and posterior inference}
\label{sec:gibbs_sampler}

Let $\mathcal{D}$ be the $n\times p$ data matrix with independent rows $\bfX^{(1)},\ldots,\bfX^{(n)}$ sampled from the single-factor graphical model \eqref{eq:singlefactor}:
\begin{equation*}
    \left(\bfX^{(j)}\mid  \bfalpha, \bflambda, f_j, \bfK, G\right) \sim \normal_p\left(\bfalpha + \bflambda f_j,\bfK^{-1}\right), \mbox{ for } j=1,\ldots,n.
\end{equation*}
We choose the following prior distributions for the model parameters $\bfalpha$, $\bflambda$, $\bfK$ and $G$:
\begin{eqnarray} \label{eq:onefactorprioor}
\bfalpha & \sim & \normal_p\left(\mathbf{0}, n_0^{-1}\bfI_p\right),\\ \nonumber
(\bflambda \mid \bfK, \Delta) & \sim & \normal_p\left( \mathbf{0}, \Delta \bfK^{-1}\right),\\ \nonumber
  \Delta & \sim &  \IGam(c,cd/2),\\ \nonumber
    \left( \bfK \mid G \right) & \sim & \Wis_{G}(\delta,\bfD),\quad \bfK\in M^{+}(G),\\ \nonumber
  \Pr(G) & \propto & \Pr(\mbox{size}(G))\Pr(G\mid \mbox{size}(G)) =  \frac{1}{m+1}\binom{m}{size(G)}^{-1},\quad G\in \mathcal{G}^{id}_p,
\end{eqnarray}
\noindent where $\bfalpha$ is a vector of mean parameters, $\IGam$ is the inverse gamma distribution, $\bfI_p$ is the $p$-dimensional identity matrix, and $n_0$, $\Delta$, $c$ and $d$ are positive real numbers. 

The conditional prior for $\bflambda$ that involves the scale hyperparameter $\Delta$ was selected for computational convenience because it leads to a normal conditional posterior that is easy to sample from. We use a $G$-Wishart prior for the precision matrix of the residual distribution, and the size-based prior for $G\in\mathcal{G}_p$ where $\mbox{size}(G)$ represents the number of edges of the graph $G$ and $m=\binom{p}{2}$ \citep{armstrong-et-2009}. Other priors on the graph space (e.g., uniform and sparsity--inducing priors \--- see, for example, Section 3 of \citet{dobra-lenkoski-rodriguez-2011} can easily be accommodated. In the simulated and real data examples that follow, we use a standard specification for the hyperparameters in \eqref{eq:onefactorprioor}: $\delta = 3$, $\bf D = \bfI_p$, $n_0=0.1$, $c=2$, and $d=1$. 

We employ a Gibbs algorithm to draw from the joint posterior distribution of all the parameters of the single-factor graphical model \citep{gelfand-smith-1990}:
\begin{align} \label{eq:singlefactorposterior}
    \Pr(\bfalpha,\bflambda,\bff,\bfK,G,\Delta\mid \mathcal{D}),
\end{align}
\noindent where $\bff=(f_1,\ldots,f_n)$ are the single-factors associated with the observed samples.

In order to sample from the joint posterior conditional distribution of $\bfK\in M^+(G)$ and $G\in \mathcal{G}^{id}_p$ given $\bff$, $\bfalpha$, $\bflambda$ and $\Delta$, we use a modified version of the DCBF algorithm of \citet{hinne-et-2014} \--- see Algorithm \ref{alg:idcbf}. We call it the identified double conditional Bayes factor algorithm (IDCBF) because it only visits the graphs in $\mathcal{G}^{id}_p$. The results of \citet{tarantola-vicard-2002} guarantee that the graphs in $\mathcal{G}^{id}_p$ can be visited by sequentially changing one edge at a time. The identifiability requirement in the IDCBF algorithm can be enforced by changing an edge (adding it if the edge is absent or deleting it if the edge is present) and applying the criterion of \citet{vicard-2000} to determine whether the new graph is indeed in $\mathcal{G}^{id}_p$. Alternatively, one can follow the methods of \citet{tarantola-vicard-2002} to determine which edges can be changed in a given graph in such a way $\mathcal{G}^{id}_p$ that the resulting graph is still in $\mathcal{G}^{id}_p$. The complete Gibbs sampling algorithm is presented in Algorithm \ref{alg:idcbfmcmc}.

\begin{algorithm}[H]
\caption{A single IDCBF MCMC Step.}\label{alg:idcbf}
\begin{algorithmic}
\item[\bf Input:] $G$,$\bfK$, $\bfD$, $\mathcal{D}$, $\bfS=\mathcal{D}^T\mathcal{D}$.
\item[\bf Output:] MCMC update for $G$ and $\bfK$ that preserves the posterior $\Pr(\bfK,G\mid \mathcal{D})$ over $\bfK\in M^{+}(G)$ and $G\in \mathcal{G}^{id}_p$.
\item[\bf FOR]{each pair $(v, v^{\prime}) \in V\times V$, $v<v^{\prime}$} {\bf DO}:
\hspace*{\algorithmicindent} \State {\bf 1.} Create a permutation of the variables so that $(v, v^{\prime}) \to (p - 1, p)$. Permute $G,\bfK,\bfD$ and $\bfS$ accordingly.
\hspace*{\algorithmicindent} \State {\bf 2.} Let $\Tilde{G}$ be the graph that is obtained from $G$ by flipping the edge $(p - 1, p)$ (adding it if absent, and deleting it if present).
\hspace*{\algorithmicindent} \State {\bf 3.} If $\Tilde{G}\notin \mathcal{G}^{id}_p$, continue to the next edge.
\hspace*{\algorithmicindent} \State {\bf 4.} Draw an auxiliary random matrix $\Tilde{K}_0 \sim W_{\Tilde{G}}(\delta,\bfD)$ using the direct sampling algorithm of \citet{lenkoski-2013}.
\hspace*{\algorithmicindent} \State {\bf 5.} Accept the move from $G$ to $\Tilde{G}$ with probabilities that depend on $\Tilde{K}_0$ from \citet{cheng-lenkoski-2012}.
\hspace*{\algorithmicindent} \State {\bf 6.} Restore the original ordering of $G$, $\bfD$ and $\bfS$.
\hspace*{\algorithmicindent} \State {\bf 7.} Draw $\bfK \sim W_{G}(\delta + n, \bfD + \bfS)$ using the direct sampling algorithm of \citet{lenkoski-2013}.
\end{algorithmic}
\end{algorithm}

\begin{algorithm}[H]
\caption{A single iteration of the Gibbs sampling algorithm for the full posterior distribution \eqref{eq:singlefactorposterior}}\label{alg:idcbfmcmc}
\begin{algorithmic}
\item[\bf Input:] $\bfalpha$, $\bflambda$, $\bff$, $G$, $\bfK$, $\Delta$, $\bfD$, $\mathcal{D}$, $n_0$, $\delta$, $c$, $d$.
\item[\bf Output:] MCMC update for $\bfalpha$, $\bflambda$, $\bff$, $G$, $\bfK$, $\Delta$ that preserve the full posterior distribution \eqref{eq:singlefactorposterior}.
\item[\bf Perform the following steps:]
\State {\bf Step 1.} Update the precision matrix $\bfK$ and the graph $G$ given the data matrix $\Tilde{\mathcal{D}}$ with rows $\bfX^{(j)}-\bfalpha-\bflambda f_j$, j=$1,\ldots,n$ using Algorithm \ref{alg:idcbf}.
\State {\bf Step 2.}  Update the single-factors $\bff$ by sampling for each $j=1,\ldots,n$ from the full conditional of $f_j$ given $\bff_{-j}$ and the rest of the parameters:
\begin{equation*}
 (f_j\mid \--) \sim \normal\left( \frac{\bflambda^{T} \left(\bfX^{(j)} -\bfalpha\right)}{\bflambda^{T}\bflambda+ \frac{\bflambda^T\bfK^{-1}\bflambda}{\bflambda^{T}\bflambda}},\frac{\frac{\bflambda^T\bfK^{-1}\bflambda}{\bflambda^{T}\bflambda}}{\bflambda^{T}\bflambda+ \frac{\bflambda^T\bfK^{-1}\bflambda}{\bflambda^{T}\bflambda}}\right),
\end{equation*}
\State {\bf Step 3.} Update the vector of means $\bfalpha$ by sampling from the full conditional: 
\begin{equation*}
 (\bfalpha \mid \--)\sim \normal_p\left( \left[ \left( n\bfK + n_0\bfI_p\right)\right]^{-1}\bfK \sum\limits_{j=1}^n\left( \bfX^{(j)}-\bflambda f_j\right), \left[ \left( n\bfK + n_0\bfI_p\right)\right]^{-1}\right).
\end{equation*}
\State {\bf Step 4.} Update the factor loadings $\bflambda$ by sampling from the full conditional:
\begin{equation*}
    (\bflambda \mid \-- ) \sim \normal_p \left(\bf\hat{\mu} , \bf\hat{K}^{-1}\right).
\end{equation*}
\noindent where
\begin{align*}
    \bf\hat{K} = \left(\bfDelta^{-1} + \sum\limits_{j=1}^n f_j^2\right) \bfK, \quad \mbox{and}\quad \bf\hat{\mu} = \bf\hat{K}^{-1}\left( \sum\limits_{j=1}^n f_j ({\bfX}^{(j)}-\bfalpha)\right).
\end{align*}
\State {\bf Step 5.} Update $\Delta$ by sampling from the full conditional:
\begin{equation*}
 (\Delta \mid \-- ) \sim \IGam\left( c+ \frac{p}{2},\frac{1}{2}\left( cd + \hat{\bflambda}^{T}\hat{\bflambda}\right)\right),
\end{equation*}
\noindent where $\hat{\bflambda} = \bfK^{1/2}\bflambda \sim \normal_p \left( \mathbf{0}, \Delta \bfI_p\right)$ with $\bfK = \left(\bfK^{1/2}\right)^T\bfK^{1/2}$.
\end{algorithmic}
\end{algorithm}

It would be of interest to infer what observed variables the single factor loads on. To this end, for each $v=1,\ldots,p$, we can perform a Bayesian test of the interval hypothesis $H_{0}: |\lambda_v|\le\epsilon$ against hypothesis $H_{1}: |\lambda_v|>\epsilon$, for some $\epsilon>0$. Given equal apriori probabilities of the two hypotheses, the Bayes factor
\begin{eqnarray}
B_{1,0} = \frac{\Pr\left(H_{1}\mid \mathcal{D}\right)}{\Pr\left(H_{0}\mid \mathcal{D}\right)}
    \label{eq:bayesfactor}
\end{eqnarray}
\noindent can be estimated as the number of posterior samples of $\lambda_v$ whose absolute value is larger than $\epsilon$ divided by the number of posterior samples whose value is less than $\epsilon$.

A simulation study that illustrates the performance of Algorithms \ref{alg:idcbf} and \ref{alg:idcbfmcmc} in simulated data with $p=5$ variables is presented in Section \ref{sec:simulationstudy}.

\subsection{Extensions to binary and ordinal categorical data} \label{sec:copulas}

Categorical datasets can be modeled in this framework by introducing latent continuous random variables that are in a one-to-one correspondence with the observed variables that are binary or ordinal categorical. single-factor graphical models are employed as the distribution of these latent variables.

We start by assuming that the observed variables are binary. The probability distribution of the $j$-th sample $\bfX^{(j)}$ is represented through a multivariate probit model constructed based on latent continuous variables $\bfZ^{(j)}=(Z^{(j)}_{1}\ldots,Z^{(j)}_{p})^{T}$ sampled from a single-factor graphical model \eqref{eq:singlefactor}:
\begin{eqnarray} \label{eq:binarylatent}
 X^{(j)}_{v} \mid \Tilde{Z}^{(j)}_{v} & \equiv & \bfI\{\Tilde{Z}^{(j)}_{v}>0\},\\ \nonumber
  \Tilde{Z}^{(j)}_{v} & = & Z^{(j)}_{v}/\left(\bfK^{-1}\right)^{1/2}_{vv},\\ \nonumber
  \left(\bfZ^{(j)}\mid  \bfalpha, \bflambda, f_j, \bfK, G\right) & \sim & \normal_p\left(\bfalpha + \bflambda f_j,\bfK^{-1}\right),\\ \nonumber
  \bfK & \in & M^+(G), \quad G\in \mathcal{G}^{id}_p,
\end{eqnarray}
\noindent for $j=1,\ldots,n$ and $v\in V$. Here, $\bfI(A)$ represents the indicator function of the event $A$. The prior specification for the multivariate probit model \eqref{eq:binarylatent} is completed as described in Section \ref{sec:gibbs_sampler}. Draws from the joint posterior distribution of latent data and model parameters
\begin{align*}
\Pr\left(\{\bfZ^{(1)},\ldots,\bfZ^{(n)}\},\bfalpha,\bflambda,\bff,\bfK,G,\Delta\mid \mathcal{D}\right),
\end{align*}
\noindent proceeds by resampling from the full conditional of model parameters given the current latent data $\Pr\left(\bfalpha,\bflambda,\bff,\bfK,G,\Delta\mid \{\bfZ^{(1)},\ldots,\bfZ^{(n)}\}\right)$ using Algorithm \eqref{alg:idcbfmcmc}, then by resampling the latent data $\{\bfZ^{(1)},\ldots,\bfZ^{(n)}\}$ by making random draws from the full conditionals 
$$\left(Z^{(j)}_{v}\mid \left\{ Z^{(j)}_{v^{\prime}}: v^{\prime}\ne v\right\},\bfalpha,\bflambda,\bff,\bfK,G,\Delta\right)$$
\noindent for each $j=1,\ldots,n$ and $v\in V$:
\begin{equation}\label{eq:truncnormal}
  \normal\left(\alpha_i+\lambda_v f_j -\sum\limits_{v^{\prime}\in bd_G(v)} \frac{K_{vv^{\prime}}}{K_{vv}}\left(Z^{(j)}_{v^{\prime}}-\alpha_{v^{\prime}}-\lambda_{v^{\prime}}f_j\right),\frac{1}{K_{vv}}\right),
 \end{equation}
\noindent truncated below at $0$ if $X^{(j)}_{v}=1$ and above at $0$ if $X^{(j)}_{v}=0$. Here $bd_G(v)$ comprises all $v^{\prime}$ such that $v$ and $v^{\prime}$ are linked by an edge in $G$. Binary data that are missing at random can easily be accommodated. If $X^{(j)}_{v}$ is missing, $Z^{(j)}_{v}$ is simulated from the normal distribution \eqref{eq:truncnormal} without truncation.

Next we consider the more general case where the observed variables can be binary, ordinal categorical, or continuous. We model the joint distribution of $\bfX^{(j)}$, $j=1,\ldots,n$ as:
\begin{eqnarray}\label{eq:copulasinglefactor}
    X^{(j)}_v & = & F^{-1}_v\left(\Phi\left(\Tilde{Z}^{(j)}_v\right)\right),\quad \mbox{ for }v\in V,\\ \nonumber
    \Tilde{Z}^{(j)}_{v} & = & Z^{(j)}_{v}/\left(\bfK^{-1}\right)^{1/2}_{vv},\\ \nonumber
  \left(\bfZ^{(j)}\mid  f_j, \bfK, G\right) & \sim & \normal_p\left(\bflambda f_j,\bfK^{-1}\right),\\ \nonumber
  \bfK & \in & M^+(G), \quad G\in \mathcal{G}^{id}_p.
\end{eqnarray}
In \eqref{eq:copulasinglefactor}, $F_v$ is the CDF of $X_v$, and $\Phi(\cdot)$ is the CDF of the standard normal distribution. This model says that the joint distribution of $\bfX$ is given by the composition of the Gaussian copula with the correlation matrix obtained by scaling $\bfK$ and the univariate distributions $\{F_v:v\in V\}$. We call \eqref{eq:copulasinglefactor} the copula single-factor graphical model (CSFGM) and contrast it with the copula Gaussian graphical model (CGGM) of \citet{dobra-lenkoski-2011}:
\begin{eqnarray}\label{eq:copulaggm}
    X^{(j)}_v & = & F^{-1}_v\left(\Phi\left(\Tilde{Z}^{(j)}_v\right)\right),\quad \mbox{ for }v\in V,\\ \nonumber
    \Tilde{Z}^{(j)}_{v} & = & Z^{(j)}_{v}/\left(\bfK^{-1}\right)^{1/2}_{vv},\\ \nonumber
  \left(\bfZ^{(j)}\mid  \bfK, G\right) & \sim & \normal_p\left(\bf0,\bfK^{-1}\right),\\ \nonumber
  \bfK & \in & M^+(G), \quad G\in \mathcal{G}_p.
\end{eqnarray}
The CGGM is based on the copula multivariate normal model of \citet{hoff-2007}, which is a particular case of \eqref{eq:copulaggm} obtained by keeping the graph $G$ fixed throughout the graph. Leaving arbitrary graphs in $\mathcal{G}_p$, the CGGMs are very likely to be sparser than the full graph model of \citet{hoff-2007}, which is the key reason for their development. The CSFGMs defined in \eqref{eq:copulasinglefactor} differ from the CGGMs through the introduction of the single-factor $f$ which in turn requires constraining the set of permissible graphs to $\mathcal{G}^{id}_p$ and adding factor loadings $\bflambda$ to the set of model parameters. Since the additional parameters together with a reduced set of permissible graphs might not lead to more parsimonious high posterior probability CSFGMs with respect to the high posterior probability CGGMs, the utility of CSFGMs when compared to CGGMs can be seen as comparable but not necessarily superior. However, as we will discuss in Section \ref{sec:gfm-multiway}, the relevance of single-factor graphical models as well as of CSFGMs becomes apparent in the analysis of multiple data sets that involve the same observed random variables.

The CSFGMs can be fitted by treating the univariate distributions $\{F_v:v\in V\}$ as nuisance parameters. The extended rank likelihood of \citet{hoff-2007} replaces $X^{(j)}_v = F^{-1}_v\left(\Phi\left(\Tilde{Z}^{(j)}_v\right)\right)$ for $v\in V$ and $j=1,\ldots,n$ with a relationship that preserves the ranks of observed and latent samples:
\begin{align*}
    X_v^{(j_1)} < X_v^{(j_2)} {\implies} \Tilde{Z}_v^{(j_1)} < \Tilde{Z}_v^{(j_2)},\quad \Tilde{Z}_v^{(j_1)} < \Tilde{Z}_v^{(j_2)} {\implies} X_v^{(j_1)} \le X_v^{(j_2)},
\end{align*}
\noindent for all $1\le j_1\ne j_2\le n$ and $v\in V$. Given the observed data $\mathcal{D}$, the latent samples $\mathcal{ZD}=\{\tilde{\bfZ}^{(1)},\ldots,\tilde{\bfZ}^{(n)}\}$ are constrained by the following bounds for each $v\in V$:
\begin{eqnarray} \label{eq:bounds}
  L^{j}_{v}(\mathcal{ZD}) & < & \Tilde{Z}^{(j)}_{v}<U^{j}_{v}(\mathcal{ZD}),\\ \nonumber
  L^{j}_{v}(\mathcal{ZD}) & = & \max\left\{ \Tilde{Z}^{(k)}_{v}:X^{(k)}_{v}<X^{(j)}_v\right\},\\ \nonumber
  U^{j}_{v}(\mathcal{ZD}) & = & \min\left\{ \Tilde{Z}^{(k)}_{v}:X^{(j)}_{v}<X^{(k)}_v\right\}.
\end{eqnarray}
\noindent We define the lower and upper bounds for which the defining sets are empty to be $-\infty$ and $\infty$, respectively. The choice of prior distributions and the sampling procedures from the corresponding joint posterior distribution of latent data and model parameters of CSFGMs mirrors those of the multivariate probit model \eqref{eq:binarylatent} with two key differences. First, the vector of means $\bfalpha$ is set to zero. Second, resampling the latent data $\mathcal{ZD}$ is performed by sampling from the full conditionals $\left(Z^{(j)}_{v}\mid \left\{ Z^{(j)}_{v^{\prime}}: v^{\prime}\ne v\right\},\bflambda,\bff,\bfK,G,\Delta\right)$ for each $j=1,\ldots,n$ and $v\in V$:
\begin{equation*}
  \normal\left(\lambda_v f_j -\sum\limits_{v^{\prime}\in bd_G(v)} \frac{K_{vv^{\prime}}}{K_{vv}}\left(Z^{(j)}_{v^{\prime}}-\lambda_{v^{\prime}}f_j\right),\frac{1}{K_{vv}}\right),
 \end{equation*}
\noindent truncated to the interval $[L^{j}_{v}(\mathcal{ZD}),U^{j}_{v}(\mathcal{ZD})]$ given in \eqref{eq:bounds}.

Related Bayesian copula factor models have also been developed by \citet{murray-et-2013}. However, their work involves models with several factors and uncorrelated residuals. We illustrate the application of the multivariate probit model \eqref{eq:binarylatent} and the CSFGM in the analysis of a binary contingency table with $p=8$ observed variables in Section \ref{sec:rochdale}.

\section{Single-factor graphical Models for Multiple Datasets}
\label{sec:gfm-multiway}

We present a generalized version of single-factor graphical models for data in which several random variables are measured in disjoint groups of observed samples. Groups can be associated with multiple spatial locations or several points in time. The samples belonging to each group can be viewed as a dataset. Our key goal is to develop a statistical model that captures the associations among variables within groups, while also taking into account group-level associations. In other words, we are interested in jointly modeling multiple datasets to represent the associations both within and across these datasets. We argue that single-factor graphical models are ideal for this setting because their individual group-level factors can be modeled through their joint distribution, allowing for the representation of correlations across groups. The Gaussian graphical models for the group-level residuals are interpretable models that illustrate the within-group correlations through their conditional independence graphs. 

The idea of using factors to model group-level associations is certainly not new. \citet{wang_wall_2003} propose multiple single-factor models with uncorrelated residuals to represent multivariate associations among random variables recorded at several spatial locations. In their framework, the spatially-correlated single-factors are able to also capture correlations among the observed variables at every given location. \citet{hogan_tchernis_2004} build on this approach and develop factor-analytic models for areal data in which the correlated factors are area-specific latent variables. \citet{lopes-et-2008} put forward a non-separable, non-stationary spatiotemporal model that involves multiple factors associated with each time point that determine associations across spatial locations. These factors are assumed to be uncorrelated at any given time point but follow a dynamic linear model across time points. Spatial dependencies are represented as structured correlations in their factor loadings matrix. The existent literature contains many other contributions along the core modeling solutions we discussed. The difference between these approaches and ours is the flexibility of modeling within-group associations through graphs that are assumed to be unknown and are subsequently inferred from the data. Edges that are common among the group level graphs are interpreted as dependencies that remain the same across groups, while edges that differ across groups show how the associations among observed variables change from group to group.

We assume that the observed data $\mathcal{D}$ involve $L$ disjoint groups of $n_1,\ldots,n_L$ samples, and that the same variables are measured in each group, i.e.  $\mathcal{D} = \{\mathcal{D}^1,\ldots,\mathcal{D}^L\}$. Here $\mathcal{D}^l$ is the $n_l\times p$ matrix with independent rows $\bfX^{(l1)},\ldots,\bfX^{(ln_l)}$ where $\bfX^{(lj)}=\left(X^{(lj)}_1,\ldots,X^{(lj)}_p\right)^T$. We assume that the $p$ observed variables are continuous. However, if $\mathcal{D}$ involve variables that are binary or ordinal categorical, straightforward extensions of the models below are available through the multivariable probit and the CSFGMs from Section \ref{sec:copulas}.

The data in each group $l=1,\ldots,L$ are modeled with the single-factor graphical model \eqref{eq:singlefactor}:
\begin{eqnarray} \label{eq:latentfactorgroups}
 \left(\bfX^{(lj)}\mid  \bfalpha_l, \bflambda_l, f_{lj}, \bfK_{l}, G_l\right) & \sim & \normal_p\left(\bfalpha_l + \bflambda_l f_{lj},\bfK_l^{-1}\right), \mbox{ for } j=1,\ldots,n_l.
\end{eqnarray}
The sets of group-specific parameters $\{\bfalpha_l, \bflambda_l, \bfK_{l}, G_l\}$, $l=1,\ldots,L$ are assumed to be independent of each other. They are assigned independent prior distributions \eqref{eq:onefactorprioor} with hyperparameters $\{\Delta_l:l=1,\ldots,L\}$ that are also assumed to be independent of each other. The common factors $\left\{ f_{lj}:l=1,\ldots,L,j=1,\ldots,n_l\right\}$ are assumed to be independent within groups, but dependent across groups:
\begin{eqnarray} \nonumber
 f_{lj} &\sim &\normal\left( \mu_l, 1\right), \mbox{ for } l=1,\ldots,L,\quad j=1,\ldots,n_l,\\ \label{eq:groupsgraphical}
 \bfmu = \left( \mu_1,\ldots,\mu_L\right)^T & \sim & \normal_L\left(\bf0,\bfK_{\mu}^{-1}\right),\\ \nonumber
 \bfK_{\mu} & \in & M^+(G_{\mu}), \quad G_{\mu}\in \mathcal{G}_L,\\ \nonumber
 \bfK_{\mu} & \sim & \Wis_{G_{\mu}}\left(\delta_{\mu},\bfD_{\mu}\right).
\end{eqnarray}
\noindent The mean vector $\bfmu$ is sampled from a Gaussian graphical model with conditional independence graph $G_{\mu}$. Although group-specific graphs $\{G_l:l=1,\ldots,L\}$ are assumed to be unknown and will be inferred from the data, graph $G_{\mu}$ can be regarded as either known or unknown. If information about associations between groups is available, it is appropriate to choose the edges of $G_{\mu}$ accordingly to incorporate these associations into the Gaussian graphical model \eqref{eq:groupsgraphical}. We note that by fixing the graph $G_{\mu}$, we only constrain the structure of the associations between groups. The strength of these associations is conveyed through the elements of the precision matrix $\bfK_{\mu}$, which are estimated from the data. 

Let us discuss the case where each group $l=1,\ldots,L$ comprises data $\mathcal{D}_l$ collected in a spatial area $l$. The spatial dependencies of areas are defined by constructing a symmetric proximity matrix $\bfW=(w_{l_{1}l_{2}})$ where $w_{l_1l_2}$ is equal to the distance between the centroids of the areas $l_1$ or $l_2$. Alternatively, $\bfW$ can be chosen as an adjacency matrix in which $w_{l_1l_2}=1$ if the areas $l_1$ and $l_2$ are neighbors (that is, share a border) and $w_{l_1l_2}=0$ otherwise \--- see \citep{cressie-1973}. Following \citep{dobra-lenkoski-rodriguez-2011}, an adequate $G$-Wishart prior for $\bfK_{\mu}$ is obtained by setting $\bfD_{\mu} = (\delta_{\mu}-2)(\bfE_{\bfW}-\rho \bfW)^{-1}$ with $\bfE_{\bfW} = \diag\{ w_{1+},\ldots,w_{L+}\}$ and $w_{l+} = \sum_{l^{\prime}}w_{ll^{\prime}}$. With this choice, the matrix $\bfD_{\mu}^{-1}$ is the mode of the $G$-Wishart prior $\Wis_{G_{\mu}}\left(\delta_{\mu},\bfD_{\mu}\right)$ which is consistent with a proper CAR model specification with proximity matrix $\bfW$ \citep{cressie-1973}. Here $\rho$ is a spatial autocorrelation parameter that must be restricted to belong to the interval defined by the reciprocals of the minimum and maximum eigenvalues of $\bfW$. The higher values of $\rho$ give, a priori, a stronger degree of positive spatial association \citep{besag-1974,gelfand-vounatsou-2003}. With these choices, the matrix $\bfD_{\mu}$ is positive definite. The graph $G_{\mu}$ is defined to be the neighborhood graph associated with $\bfW$ with vertices $\{1,2,\ldots,L\}$ and edges that link vertices that correspond to areas that are neighbors.

Next, we consider the case where each group $l=1,\ldots,L$ comprises data $\mathcal{D}_l$ collected at time points or over time periods $t_1<t_2<\ldots<t_L$. Suitable choices for $G_{\mu}$ involve graphs $G^1_{\mu}$, $G^2_{\mu}$, $G^3_{\mu}$ and $G^4_{\mu}$ with vertices $\{1,2,\ldots,L\}$ and edges $E^1$, $E^2$, $E^3$ and $E^4$, where
\begin{eqnarray*}
    E^1 & = & \{(l-1,l):2\le l\le L\},\\
    E^2 & = & E^1 \cup \{ (l-2,l):3\le l\le L\},\\
    E^3 & = & E^2 \cup \{ (l-3,l):4\le l\le L\},\\
    E^4 & = & E^3 \cup \{ (l-4,l):5\le l\le L\}.
\end{eqnarray*}
\noindent These four graphs define autoregressive AR(1), AR(2), AR(3) and AR(4) models. The hyperparameters of the $G$-Wishart prior to $\bfK_{\mu}$ are set to $\delta_{\mu}=3$ and $\bfD_{\mu}=\bfI_L$.

Sampling from the joint posterior distribution of the single-factor graphical model for multiple datasets and the known structure of associations between groups specified by a fixed graph $G_{\mu}$ proceeds as follows. The group-specific parameters $\{\bfalpha_l, \bflambda_l, \bfK_{l}, G_l, \Delta_l\}$, $l=1,\ldots,L$ are resampled as in Algorithm \ref{alg:idcbfmcmc}. The common factors are updated by sampling from the full conditionals:
\begin{eqnarray*}
     (  f_{lj} \mid \-- ) & \sim & \normal\left( \frac{\bflambda_l^{T} \left(\bfX^{(lj)} -\bfalpha_l\right)+\frac{\bflambda_l^T\bfK_l^{-1}\bflambda_l}{\bflambda_l^{T}\bflambda_l}\mu_l}{\bflambda_l^{T}\bflambda_l+ \frac{\bflambda_l^T\bfK_l^{-1}\bflambda_l}{\bflambda_l^{T}\bflambda_l}},\frac{\frac{\bflambda_l^T\bfK_l^{-1}\bflambda_l}{\bflambda_l^{T}\bflambda_l}}{\bflambda_l^{T}\bflambda_l+ \frac{\bflambda_l^T\bfK_l^{-1}\bflambda_l}{\bflambda_l^{T}\bflambda_l}}\right).
\end{eqnarray*}
The mean vector $\bfmu$ is updated by sequentially sampling from the full conditional of each of its elements given the rest and current model parameters:
\begin{eqnarray*}
   \left( \mu_l \mid \left\{ \mu_{l^{\prime}}:l^{\prime}\ne l\right\},\-- \right) & \sim & \normal\left( -\sum\limits_{l^{\prime}\ne l} \frac{\left(\bfK_{\mu}\right)_{ll^{\prime}}}{\left(\bfK_{\mu}\right)_{ll}}\mu_{l^{\prime}}+\sum\limits_{j=1}^{n_l} f_{lj},\frac{1}{n_l+\left(\bfK_{\mu}\right)_{ll}}\right), 
\end{eqnarray*}
\noindent for $l=1,\ldots,L$. Finally, the precision matrix $\bfK_{\mu}$ is updated by sampling its $G$-Wishart full conditional:
\begin{eqnarray*}
     \left( \bfK_{\mu} \mid \-- \right) \sim \Wis_{G_{\mu}}\left(\delta_{\mu}+1,D_{\mu}+\bfmu\bfmu^T\right).
\end{eqnarray*}

\section{Simulation study} \label{sec:simulationstudy}

In this section we present an example of using the Gibbs algorithm and the IDCBF algorithm for making draws from the posterior distribution of graphical single-factor models. We consider the following model with $p=5$ observed variables:
\begin{eqnarray} \label{eq:factorsimstudy}
 \left( \begin{array}{c} X_1\\ X_2\\ X_3\\ X_4\\ X_5\end{array}\right) = \left( \begin{array}{c} \alpha_1\\ \alpha_2\\ \alpha_3\\ \alpha_4\\  \alpha_5\end{array}\right) + \left( \begin{array}{c} \bflambda_1\\ \bflambda_2\\ \bflambda_3\\ \bflambda_4\\  \bflambda_5\end{array}\right) f + \left( \begin{array}{c} \delta_1\\ \delta_2\\ \delta_3\\ \delta_4\\ \delta_5\end{array}\right).
\end{eqnarray}
Set $\alpha_1=\alpha_2=\alpha_3=0.1$, $\alpha_4=\alpha_5=-0.1$, $\lambda_1=0.8$, $\lambda_2=1.0$, $\lambda_3 = \lambda_4 = 1.2$, $\lambda_5=0.8$.  Consider two graphs $G_1$ and $G_2$ with five vertices such that the complementary of $G_1$ has edges $E_1 = \{ (1,4),(1,5),(2,4), (2,5), (3,4), (3,5), (4,5)\}$, and the complementary of $G_2$ has edges $E_2=\{ (1,3), (1,4), (2,4), (2,5), (3,5)\}$. Since the complementary of $G_1$ has a cycle of an odd length $3$ and the complementary of $G_2$ has a cycle of an odd length $5$, the single-factor models $M_1$ and $M_2$ obtained by assuming that the distribution of the residuals in \eqref{eq:factorsimstudy} are both identifiable: 
\begin{equation}
 \bfdelta = (\delta_1,\ldots,\delta_5)^T \sim \normal_5\left(\bf0,{\bfK}_{i}^{-1}\right),\quad i=1,2,
 \label{eq:simstudydistrib}
\end{equation}
such that $\bfK_1\in M^+_{G_1}$ for $M_1$ and $\bfK_2\in M^+_{G_2}$ for $M_2$. We take
\begin{eqnarray*}
 \bfK_1 = \left[ \begin{array}{ccccc} 1 & 0.5 & 0.4 & 0 & 0\\ 0.5 & 1 & 0.5 & 0 & 0\\ 0.4 & 0.5 & 1 & 0 & 0\\ 0 & 0 & 0 & 1 & 0\\ 0 & 0 & 0 & 0 &1\end{array}\right], \quad \bfK_2 = \left[ \begin{array}{ccccc} 1 & 0.5 & 0 & 0 & 0.4\\ 0.5 & 1 & 0.5 & 0 & 0\\ 0 & 0.5 & 1 & 0.5 & 0\\ 0 & 0 & 0.5 & 1 & 0.5\\ 0.4 & 0 & 0 & 0.5 &1\end{array}\right]
\end{eqnarray*}

We simulate $n=20,50,100$ samples from $M_1$ and $M_2$ by sampling $\bfdelta\sim \normal_5\left(\bf0,\bfK_i^{-1}\right)$, $i=1,2$, $f\sim \normal(0,1)$. 
The table below gives the performance of posterior estimation obtained by averaging samples from the Gibbs sampling algorithm over 10 simulations of 10,000 iterations each, with a burn-in time of 5,000 iterations.

We summarize the results in Table \ref{tab:single-idcbf}. We measure performance as the distance between the posterior estimates and the true model parameters. For $\bfalpha$ and $\bflambda$ we use the $\ell_2$ norm, for $\bfK$ the KL divergence, and the Frobenius norm for $G$. As the simulated sample size increases, the estimation performance also increases for each model parameter, including the residuals' conditional independence graphs. We see that we are able to successfully recover the true values of $\bfalpha$ and $\bflambda$, together with the true graphs and the precision matrices that define the distribution of the residuals \eqref{eq:simstudydistrib}.  

\begin{table}[h!]
\setlength\tabcolsep{0pt}
\small
\begin{tabular*}{\textwidth}{c@{\extracolsep{\fill}}*{6}{c}}
\toprule 
    $M_1$ & $||\hat{\bfalpha}-\bfalpha||_2$ & $||\hat{\bflambda}-\bflambda||_2$ & KL & $||\hat{G}_1-G_1||_F$ \\
    \midrule
    \multirow{1}{*}{n = 20} 
     & 0.2320 (1e-01)    & 0.1700 (5e-02)    &   0.1655 (2e-02)  & 1.755 (3e-01) \\
    \midrule
    \multirow{1}{*}{n = 50} 
    & 0.1216 (4e-02)    & 0.1666 (8e-02)    &   0.0962 (5e-03)  & 0.7137 (1e-01) \\
    \midrule
    \multirow{1}{*}{n = 100} 
    & 0.0787 (4e-03)    & 0.0933 (2e-02)    &   0.0027 (1e-03)  & 0.1931 (3e-02) \\
     \midrule
    \multirow{1}{*}{n = 1000} 
    & 0.0144 (5e-04)    & 0.0127 (4e-03)    &   0.0007 (1e-05)  & 0.1151 (2e-02) \\
    \bottomrule
\end{tabular*}
\newline
\vspace{0.5cm}
\newline
\begin{tabular*}{\textwidth}{c@{\extracolsep{\fill}}*{6}{c}}
\toprule 
    $M_2$ & $||\hat{\bfalpha}-\bfalpha||_2$ & $||\hat{\bflambda}-\bflambda||_2$ & KL & $||\hat{G}_1-G_1||_F$ \\
    \midrule
    \multirow{1}{*}{n = 20} 
     & 0.2987 (1e-01)    & 0.2739 (1e-01)    &   0.2258 (3e-02)  & 1.774 (3e-01) \\
    \midrule
    \multirow{1}{*}{n = 50} 
    & 0.1819 (7e-02)    & 0.2029 (9e-02)    &   0.0506 (3e-03)  & 0.9021 (1e-01) \\
    \midrule
    \multirow{1}{*}{n = 100} 
    & 0.0987 (8e-03)    & 0.1028 (7e-02)    &   0.0169 (2e-03)  & 0.4039 (8e-02) \\
     \midrule
    \multirow{1}{*}{n = 1000} 
    & 0.0255 (1e-03)    & 0.0327 (5e-03)    &   0.0027 (1e-04)  & 0.1871 (3e-02) \\
    \bottomrule
\end{tabular*}
\caption{\label{tab:single-idcbf}Results for the IDCBF sampler for different number of observations, averaged over 10 dataset simulations of  10 Markov chains for 10,000 iterations each for graphs $G_i, i=\{1,2\}$ for the corresponding models $M_i, i=\{1,2\}$, with a burnin of 5,000. Standard errors are shown in parentheses. The table contains the $\ell_2$ norms of the difference between the $\bfalpha$ and $\bflambda$ estimates and their true quantities, the Kullback-Leibler divergence of the estimated $K_i$ matrix with respected to $K_i$, and the Frobenius norm of the estimated graph with respect to $G_i$.}
\label{table:2}
\end{table}

\section{Analysis of a binary categorical dataset}\label{sec:rochdale}

We present an analysis of the Rochdale data \citep{whittaker1990} that comprises eight observed binary variables \--- see Table \ref{tab:rochdaledata}. These eight variables are: $a$, wife economically active (no,yes); $b$, age of wife $>38$ (no,yes); $c$, husband unemployed (no,yes); $d$, child $\le 4$ (no,yes); $e$, wife's education, high-school+ (no,yes); $f$, husband's education, high-school+ (no,yes); $g$, Asian origin (no,yes); $h$, other household member working (no,yes).

{\renewcommand{\baselinestretch}{1.0}
\begin{table}[h]
\begin{center}
\caption{\label{tab:rochdaledata}Rochdale data from \citet{whittaker1990}. The cells counts appear row by row in lexicographical order with variable $h$ varying fastest and variable $a$ varying slowest. The grand total of this table is $665$.}
\setlength{\tabcolsep}{1mm}
\begin{tabular}{cccccccccccccccc}\hline
 $5$  & $0$ & $2$  &  $1$  & $5$  & $1$ & $0$ & $0$ & $4$ & $1$ & $0$ & $0$ & $6$ & $0$ & $2$ & $0$\\
 $8$  & $0$ & $11$ & $0$ & $13$  & $0$ & $1$ & $0$ & $3$ & $0$ & $1$ & $0$ & $26$ & $0$ & $1$ & $0$\\
 $5$  & $0$ & $2$ & $0$ & $0$ & $0$ & $0$ & $0$ & $0$ & $0$ & $0$ & $0$ & $0$ & $0$ & $1$ & $0$\\
 $4$  & $0$ & $8$ & $2$ & $6$ & $0$ & $1$ & $0$ & $1$ & $0$ & $1$ & $0$ & $0$ & $0$ & $1$ & $0$\\
 $17$ & $10$ & $1$ & $1$ & $16$ & $7$ & $0$ & $0$ & $0$ & $2$ & $0$ & $0$ & $10$ & $6$ & $0$ & $0$\\
 $1$ & $0$ & $2$ & $0$ & $0$ & $0$ & $0$ & $0$ & $1$ & $0$ & $0$ & $0$ & $0$ & $0$ & $0$ & $0$\\
 $4$ & $7$ & $3$ & $1$ & $1$ & $1$ & $2$ & $0$ & $1$ & $0$ & $0$ & $0$ & $1$  & $0$ & $0$ & $0$\\
 $0$ & $0$ & $3$ & $0$ & $0$ & $0$ & $0$ & $0$ & $0$ & $0$ & $0$ & $0$ & $0$ & $0$ & $0$ & $0$\\
 $18$ & $3$ & $2$ & $0$ & $23$ & $4$ & $0$ & $0$ & $22$ & $2$ & $0$ & $0$ & $57$ & $3$ & $0$ & $0$\\
 $5$ & $1$ & $0$ & $0$ & $11$ & $0$ & $1$ & $0$ & $11$ & $0$ & $0$ & $0$ & $29$ & $2$ & $1$ & $1$\\
 $3$ & $0$ & $0$ & $0$ & $4$ & $0$ & $0$ & $0$ & $1$ & $0$ & $0$ & $0$ & $0$ & $0$ & $0$ & $0$\\
 $1$ & $1$ & $0$ & $0$ & $0$ & $0$ & $0$ & $0$ & $0$ & $0$ & $0$ & $0$ & $0$ & $0$ & $0$ & $0$\\
 $41$ & $25$ & $0$ & $1$ & $37$ & $26$ & $0$ & $0$ & $15$ & $10$ & $0$ & $0$ & $43$ & $22$ & $0$ & $0$\\
 $0$ & $0$ & $0$ & $0$ & $2$ & $0$ & $0$ & $0$ & $0$ & $0$ & $0$ & $0$ & $3$ & $0$ & $0$ & $0$\\
 $2$ & $4$ & $0$ & $0$ & $2$ & $1$ & $0$ & $0$ & $0$ & $1$ & $0$ & $0$ & $2$ & $1$ & $0$ & $0$\\
 $0$ & $0$ & $0$ & $0$ & $0$ & $0$ & $0$ & $0$ & $0$ & $0$ & $0$ & $0$ & $0$ & $0$ & $0$ & $0$\\
\hline
\end{tabular}
\end{center}
\end{table}

We employ the multivariate probit model and the CSFGM introduced in the main text of the paper. For each of these two models, we ran six different instances of the MCMC sampling algorithm that alternates between resampling the latent data and resampling the model parameters. Each instance of the Markov chain was run from a randomly generated starting value of model parameters for 30,000 iterations. In the calculation of posterior estimates,  the first 10,000 iterations were discarded as burn-in. For both models, convergence of the chains to their stationary posterior distribution occurs quite fast \--- see Figure \ref{fig:rochdale-numedges}.

\begin{figure}
    \centering
\includegraphics[width = 0.5\textwidth,angle=-90]{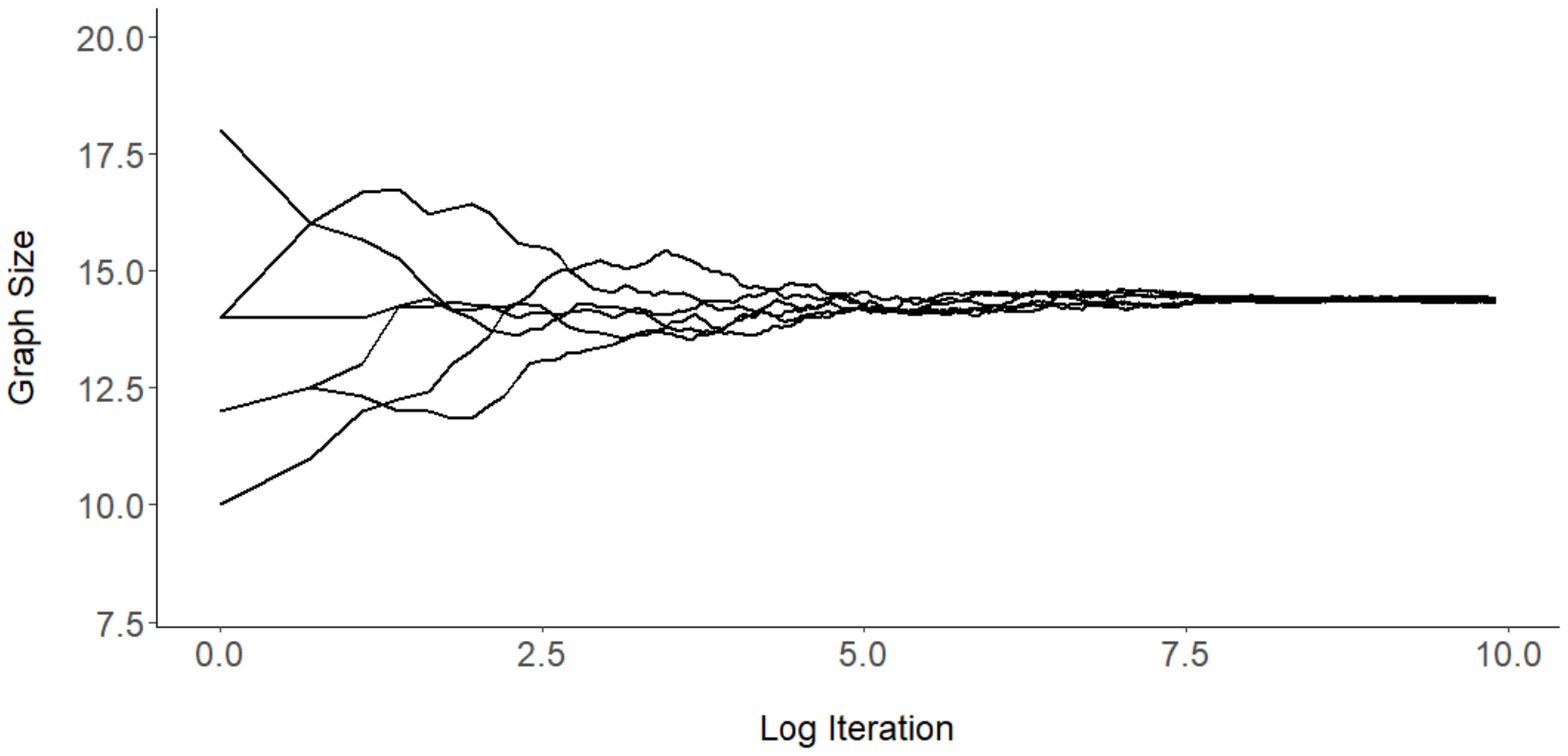}
\includegraphics[width = 0.5\textwidth,angle=-90]{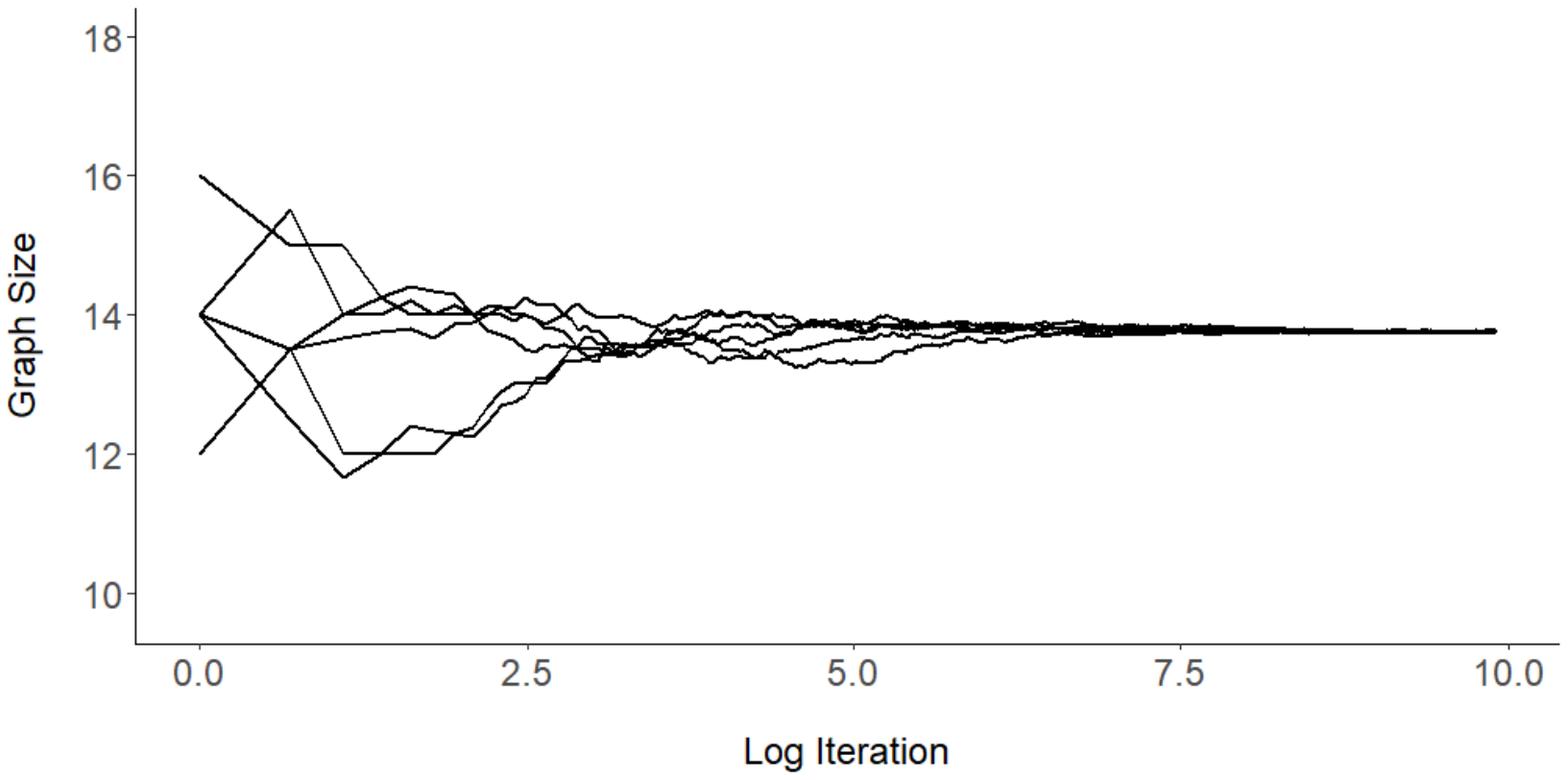}
    \caption{Cumulative estimates of the posterior expected number of edges for the CSFGM Gibbs sampler (top panel) and the multivariate probit model (bottom panel) in the Rochdale data.}
    \label{fig:rochdale-numedges}
\end{figure}

We calculate the expected cell counts as described in \citet{dobra-lenkoski-2011}. We show our results in Table \ref{tab:rochdaleresults} and compare them with the expected cell counts obtained from the CGGMs in Section 5.1 of \citet{dobra-lenkoski-2011}. We see that all three models can successfully capture the largest cell counts in these data.

\begin{table}[h!]
    \centering
    \begin{tabular}{c|c|c|c|c}
    \hline
        Cell & Observed &  Multivariate Probit & CSFGM  & CGGM \\
        \hline
  0  0  0  1  1  0  0  1  & 57 &     56.45 & 56.45 & 56.80\\
  1  0  0  1  1  0  0  1 & 43 &  49.87 & 49.87 & 47.55 \\
  1  0  0  0  0  0  0  1 & 41 &     35.37 & 35.37 & 36.12\\
  1  0  0  0  1  0  0  1  & 37 &    33.44 & 33.44 & 36.61\\
  0  0  1  1  1  0  0  1 & 29 &     35.50 & 32.43 & 32.40\\
  \hline
    \end{tabular}
    \caption{Expected cell counts for the top 5 lagest cell counts in the Rochdale data.}
    \label{tab:rochdaleresults}
\end{table}

\begin{figure}
    \centering
    \includegraphics[width=\textwidth]{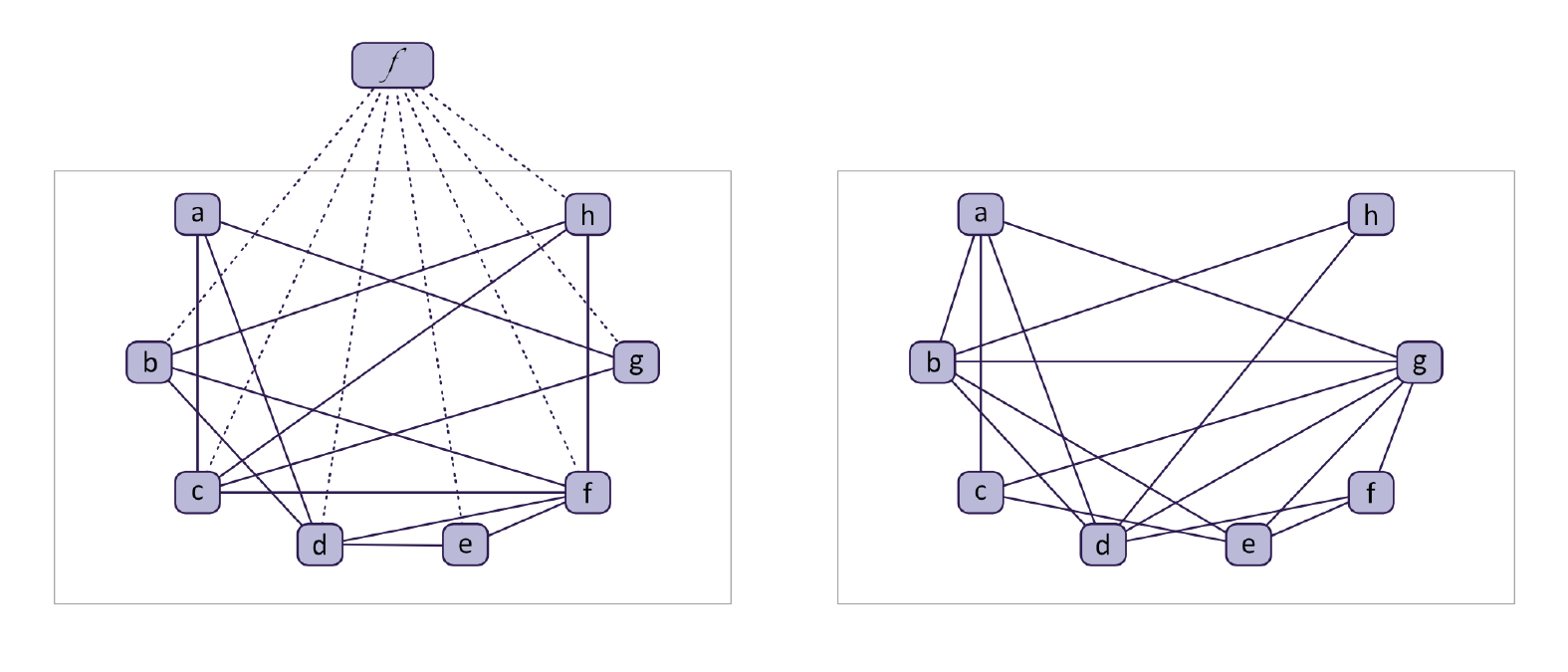}%
    \caption{Left panel: Estimated factor loadings and residual graph obtained from CSFGMs. Right panel: Estimated conditional independence graph obtained from CGGMs. The latent variable is denoted with $\mathsf{f}$ and appears at the top of the figure in the left panel. Solid lines represent edges in the residual graph with a posterior inclusion probability greater than $0.5$, and dotted lines represent factor loadings that have a Bayes factor $B_{1,0}$ above $3.2$.} 
    \label{fig:rochdale-estGraph}
\end{figure}

\begin{figure}
    \centering
    \includegraphics[width=0.7\textwidth]{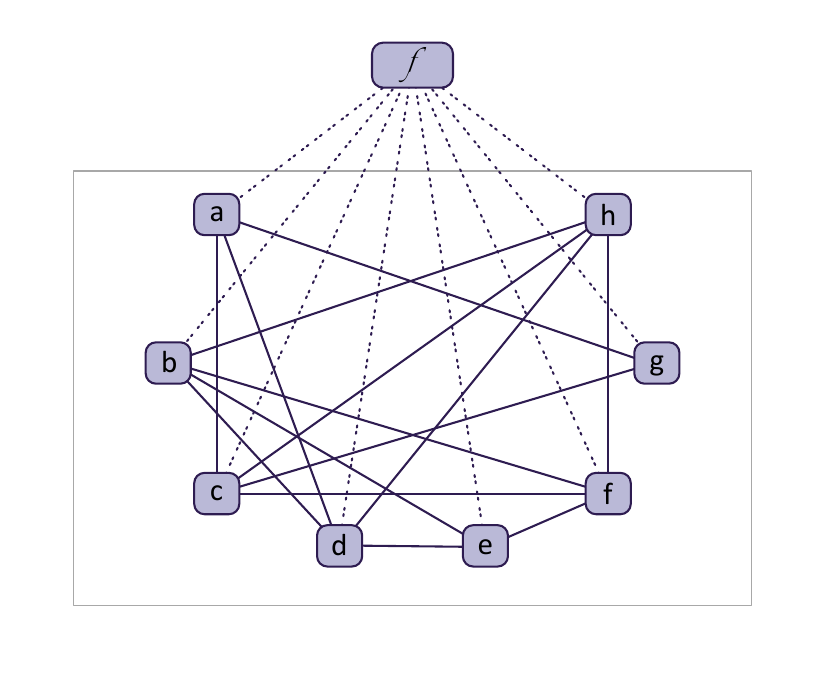}%
    \caption{Estimated factor loadings and residual graph obtained from the multivariate probit model. The latent variable is denoted with $\mathsf{f}$ and appears at the top of the figure in the left panel. Solid lines represent edges in the residual graph with a posterior inclusion probability greater than $0.5$, and dotted lines represent factor loadings that have a Bayes factor $B_{1,0}$ above $3.2$.} 
    \label{fig:rochdale-estGraph-probit}
\end{figure}

 In Figure \ref{fig:rochdale-estGraph} we show a comparison of the posterior estimates conditional independence graphs in the latent space obtained from the CSFGMs and the CGGMs. These are graphs whose edges receive an estimated posterior probability greater than $0.5$, also known as median graphs. We see that two conditional independence graphs involve similar edges with few differences. For example, variables $c$ and $h$ are linked by an edge in the CSFGM graph, but this edge is missing in the CGGM graph. Similarly, variables $g$ and $h$ are linked by an edge in the CGGM graph, but not linked in the CSFGM graph. Figure \ref{fig:rochdale-estGraph-probit} shows the posterior estimates conditional independence graphs in the latent space from the multivariate probit model. The key differences between Figure \ref{fig:rochdale-estGraph-probit} and the CSFGM graph in Figure \ref{fig:rochdale-estGraph} are the multivariate probit model residual graph includes an edge between variables $b$ and $e$, $d$ and $h$, but these edges are missing in the CSFGM graph. The multivariate probit model residual graph is also missing edges $d$ and $f$ while this edge is present in the CSFGM graph.
 
Tables \ref{tab:postsumprobitrochdale} and \ref{tab:postsumcopularochdale} give posterior means, 95\% credible intervals and Bayes factors $B_{1,0}$ for testing the  hypothesis $H_{0}: |\lambda_v|\le 0.01$ against the hypothesis $H_{1}: |\lambda_v|>0.01$ for each loading $\lambda_v$ assuming equal prior probabilities for these hypotheses. The interpretation of the Bayes factors is based on the guidelines provided in Section 3.2 of \citet{kass-raftery-1995}. We see that, with the exception of the loading associated with variable $a$ in the multivariate probit models, the data provides at least substantial evidence in favor of hypothesis $H_{1}: |\lambda_v|>0.01$. Thus, the single factor loads on all the variables with the exception of variable $a$ for the multivariate probit model. The posterior summaries for factor loadings are quite consistent between the multivariate probit model and the CSFGM.

\begin{table}[h!]
    \centering
    \begin{tabular}{c|c|c|c|c}
    \hline
        Variable & Posterior Mean &  95\% CI & $B_{1,0}$ & Evidence against $H_0$\\
        \hline
  $a$ & $0.040$ & $[0.002,0.115]$ & $4.685$ & Substantial\\
  $b$ & $0.057$ & $[0.002,0.169]$ & $7.636$ & Substantial\\
  $c$ & $0.050$ & $[0.002,0.139]$ & $7.503$ & Substantial\\
  $d$ & $0.056$ & $[0.002,0.158]$ & $7.873$ & Substantial\\
  $e$ & $0.067$ & $[0.003,0.183]$ & $9.661$ & Substantial\\
  $f$ & $0.055$ & $[0.002,0.157]$ & $7.795$ & Substantial\\
  $g$ & $0.070$ & $[0.003,0.190]$ & $10.655$ & Strong\\
  $h$ & $0.056$ & $[0.003,0.151]$ & $8.443$ & Substantial\\
  \hline
    \end{tabular}
    \caption{Posterior summaries of the absolute values of the factor loadings in the multivariate probit model for the Rochdale data: posterior means, 95\% credible intervals, Bayes factors and their interpretation.}
    \label{tab:postsumprobitrochdale}
\end{table}

\begin{table}[h!]
    \centering
    \begin{tabular}{c|c|c|c|c}
    \hline
        Variable & Posterior Mean &  95\% CI & $B_{1,0}$ & Evidence against $H_0$\\
        \hline
  $a$ & $0.024$ & $[0.001,0.073]$ & $2.472$ & Not worth more\\
      &        &               &        & than a bare mention\\
  $b$ & $0.043$ & $[0.002,0.122]$ & $5.494$ & Substantial\\
  $c$ & $0.030$ & $[0.001,0.085]$ & $3.890$ & Substantial\\
  $d$ & $0.040$ & $[0.001,0.116]$ & $5.161$ & Substantial\\
  $e$ & $0.049$ & $[0.002,0.139]$ & $6.599$ & Substantial\\
  $f$ & $0.036$ & $[0.001,0.101]$ & $4.695$ & Substantial\\
  $g$ & $0.045$ & $[0.002,0.124]$ & $6.057$ & Substantial\\
  $h$ & $0.037$ & $[0.001,0.103]$ & $4.949$ & Substantial\\
  \hline
    \end{tabular}
    \caption{Posterior summaries of the absolute values of the factor loadings in the CSFGM for the Rochdale data: posterior means, 95\% credible intervals, Bayes factors and their interpretation.}
    \label{tab:postsumcopularochdale}
\end{table}

\section{Analysis of Low Birth Weight in North Carolina Data}\label{sec:lbw}

In this section, we illustrate modeling spatially correlated contingency tables with the copula graphical single-factor models for multiple datasets. Low birth weight babies ($\le2,500$ grams) are at increased risk of developing serious health problems in their early years or serious disabilities/illnesses later in their lives \--- see, for example, \citet{tassone-et-2010}. We investigated three risk factors that could potentially lead to low birth weights defined as newborns whose weight at birth is less than 2500 grams (LBW, yes/no):  sex (S, male/female), mother's age at birth dichotomized as younger or older than $35$ (MAGE, yes/no) and race (R, white non-hispanic/other). We use data collected by the North Carolina State Center for Health Statistics that are available from the North Carolina Vital Statistics Dataverse (\url{http://arc.irss.unc.edu/dvn/dv/NCVITAL}). The births recorded in 2006, 2007 and 2008 are grouped by the $100$ counties in which the mothers reside. Thus, the data consist of $100$ four-way dichotomous contingency tables. Figure \ref{fig:lowbirths} shows the proportion of low birth weight infants in each county in North Carolina; the overlaid spaghetti plot connects the centroids of the counties that share a common border. The plot seems to suggest that counties that are neighbors of each other are more likely to have similar proportions of LBW than counties that do not share a border. 

\begin{figure}[h!]
\begin{center}
\includegraphics[width = \linewidth]{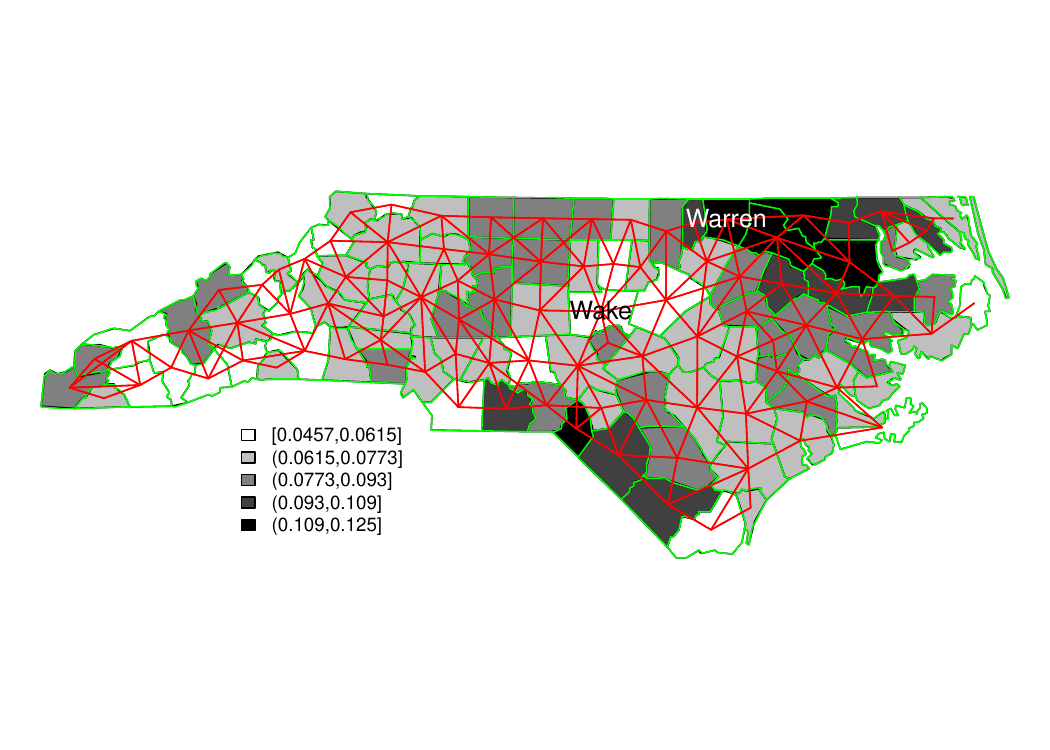}
\caption{Proportion of low birth weight infants in the $100$ counties of North Carolina. The red lines show the neighborhood structure of the counties thats share a border as an undirected graph.}\label{fig:lowbirths}
\end{center}
\end{figure}

\indent To gain a better understanding of these data, we used the MC3 log-linear model determination methodology with conjugate priors \citep{massam-liu-dobra-2009} to identify representative decomposable graphical models associated with the $100$ contingency tables. First, we analyze each table independently and assume that the patterns of interactions among LBW, S, MAGE, and R can freely vary across counties. Figure \ref{fi:lbwmaps} shows that the strength of the first-order interaction between LBW and sex (S), MAGE, and race (R) varies considerably from county to county, with race (R) having a strong effect on LBW for many counties, while mother's age at birth (MAGE) and sex (S) having a much weaker effect on LBW for most counties.  If we assume complete homogeneity of the interaction patterns across counties and put together all the $100$ contingency tables, the posterior probabilities of the edges linking LBW with S, MAGE and R are equal with $1$, $0.98$ and $1$.  These results suggest that merging all county-level contingency tables can hide important regional differences and leads to misleading results, and that any method of joint analysis for these contingency tables needs to account for the similarities that seem to be present among neighboring geographical regions.

\begin{figure}
\includegraphics[width=0.6\textwidth]{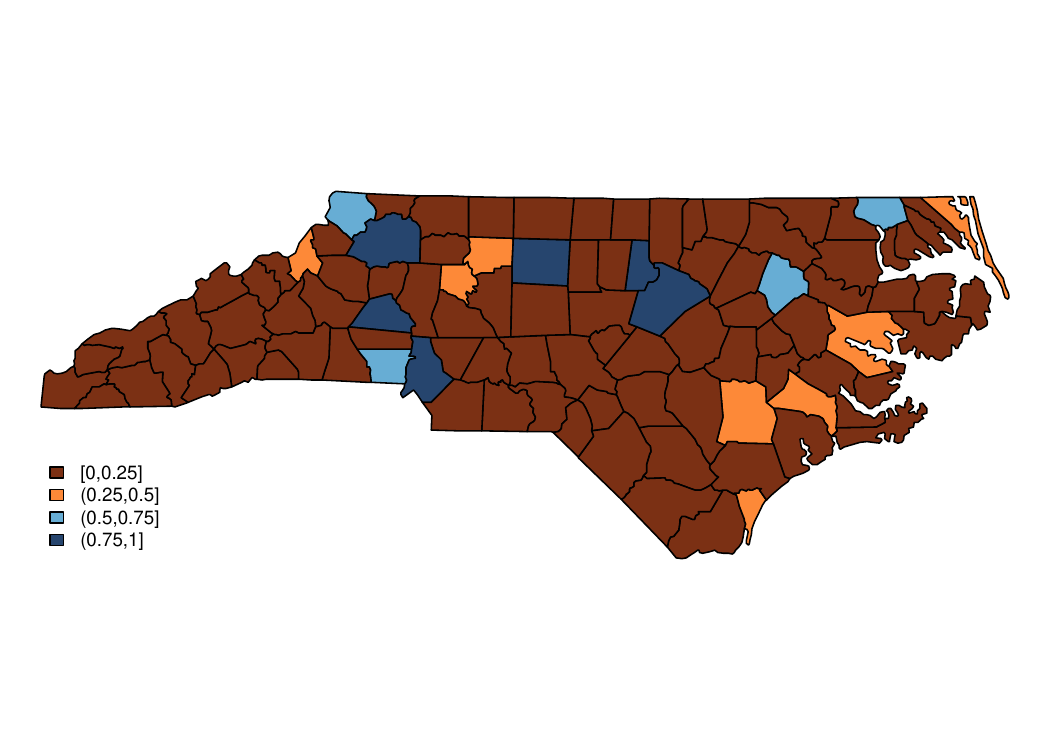}
\includegraphics[width=0.6\textwidth]{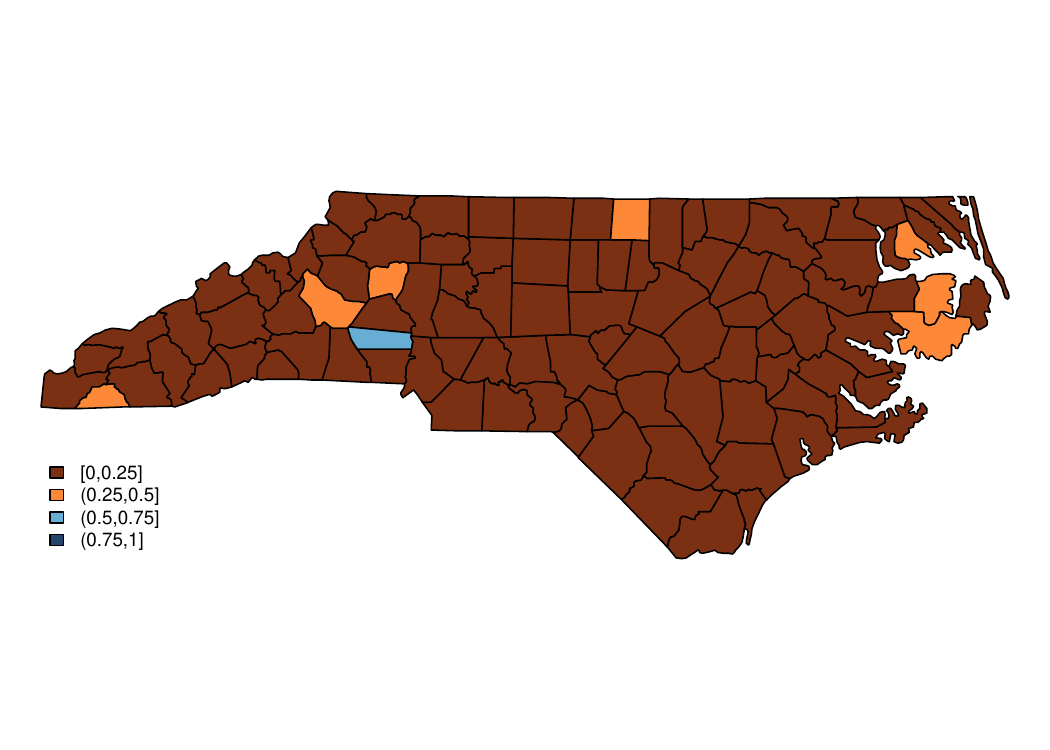}
\includegraphics[width=0.6\textwidth]{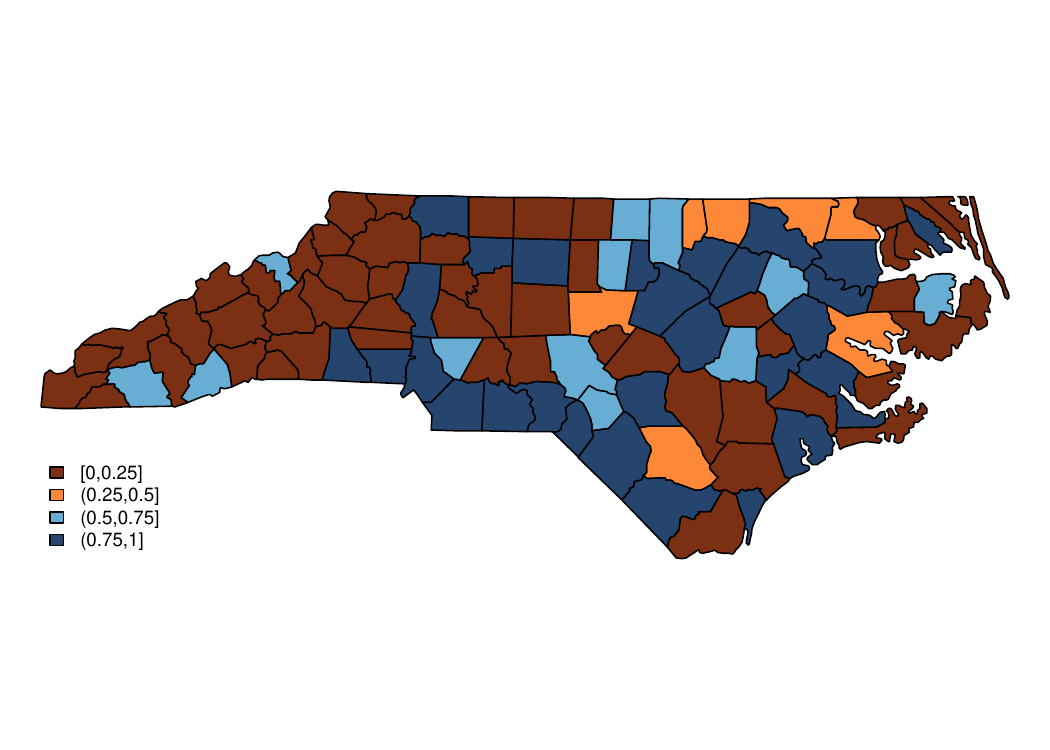}
        \caption{Choropleth maps for the posterior inclusion probabilities of edges that connect low birth weight (LBW) with sex (S) (top panel), low birth weight (LBW) with mother's age at birth (MAGE) (middle panel), and low birth weight (LBW) with race (R) (bottom panel).}
        \label{fi:lbwmaps}
\end{figure}

\indent To highlight the impact of spatial heterogeneity in the analysis, we take a closer look at two counties: Wake and Warren. Wake County is one of the most affluent and densely populated counties in North Carolina and has one of the lowest incidence rates of LBW ($5.74$\%). Warren County is at the opposite end of the spectrum and has one of the highest incidence rates of LBW ($12.46$\%). The most relevant decomposable graphical models associated with the Wake and Warren contingency tables are presented in Table \ref{tab:wakewarren}. The number of recorded births for Wake county is $37,935$, leading to a higher degree of certainty around the log-linear model that best represents the data: the log-linear model [LBW,R][MAGE,R][S,LBW] has a posterior probability of $0.983$. Here [A,B] denotes the interaction between variables A and B. Only $626$ births have been recorded for Warren County. This leads to much greater uncertainty about the most suitable log-linear model: the log-linear model of independence has a posterior probability of $0.458$ while the log-linear model that includes the interaction between LBW and race (R) is about half as probable. The data from Wake County seem to offer substantial evidence that LBW is related to both race and gender, while the data from Warren County offer only partial evidence that LBW is dependent on race.
\begin{table}[h!]
\begin{center}
{\footnotesize
\begin{tabular}{cc|cc} \hline
 \multicolumn{2}{c|}{Wake County} & \multicolumn{2}{c}{Warren County}\\
  Posterior prob.  & Model & Posterior prob.  & Model\\ \hline
 0.9825 &	[LBW,R][MAGE,R][S,LBW] & 0.4581 &	[R][LBW][MAGE][S]\\
 0.0107 &	[LBW,R][MAGE,R][S] & 0.2498 & 	[LBW,R][MAGE][S]\\
 0.0065 &	[MAGE,LBW,R][S,LBW] & 0.0773 &	[R][MAGE,LBW][S]\\
 0.0001 &	[LBW,R][MAGE,R][S,MAGE] & 0.0422 &	[LBW,R][MAGE,LBW][S]\\
 0.0001 &	[MAGE,LBW,R][S] & 0.0334 & [R][MAGE][S,LBW]\\ 
 \hline
\end{tabular}
}
\end{center}
\caption{Five most probable decomposable graphical log-linear models for Wake and Warren counties.\label{tab:wakewarren}}
\end{table}

We use the CSFGMs for multiple datasets introduced in Section 5 of the main article. The areal dependencies are defined through the neighborhood graph shown in Figure \ref{fig:lowbirths}. The associated proximity matrix $\bfW$ has elements equal to $1$ for counties that share a border and $0$ otherwise. We set the spatial autocorrelation parameter $\rho = 0.9$ and $\delta_{\mu}=3$. We sampled from the posterior distribution of model parameters using five Markov chains initialized at random starting points and ran for 30,000 iterations. The Gibbs sampling algorithm of Section 5 shows good mixing and convergence with respect to all model parameters. Figures \ref{fig:lbw:wakeNumedges} and \ref{fig:lbw:warrenNumedges} illustrate the convergence of the five Markov chains for the posterior expected number of edges in the residual graphs for Wake and Warren counties, respectively. The posterior parameter estimates we present are based on the 100,000 samples obtained from all 5 chains after discarding the first 10,000 iterations in each chain as burnin.   

\begin{figure}
    \centering
    \includegraphics[width = \textwidth]{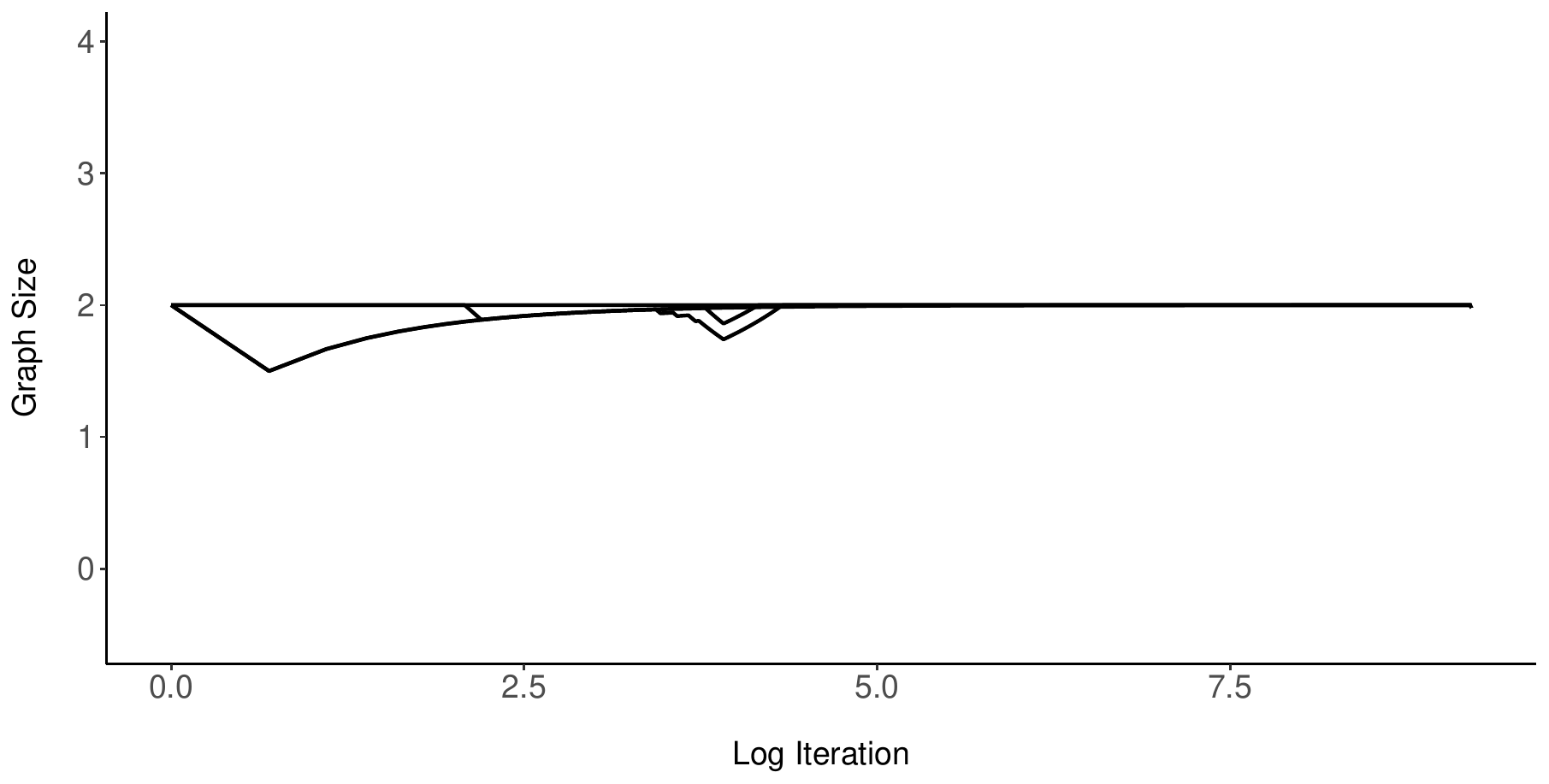}
    \caption{Convergence plot of the posterior expected number of edges in residual graph for Wake County. Each of the 5 lines corresponds with a single Markov chain.}
    \label{fig:lbw:wakeNumedges}
\end{figure}

\begin{figure}
    \centering
    \includegraphics[width = \textwidth]{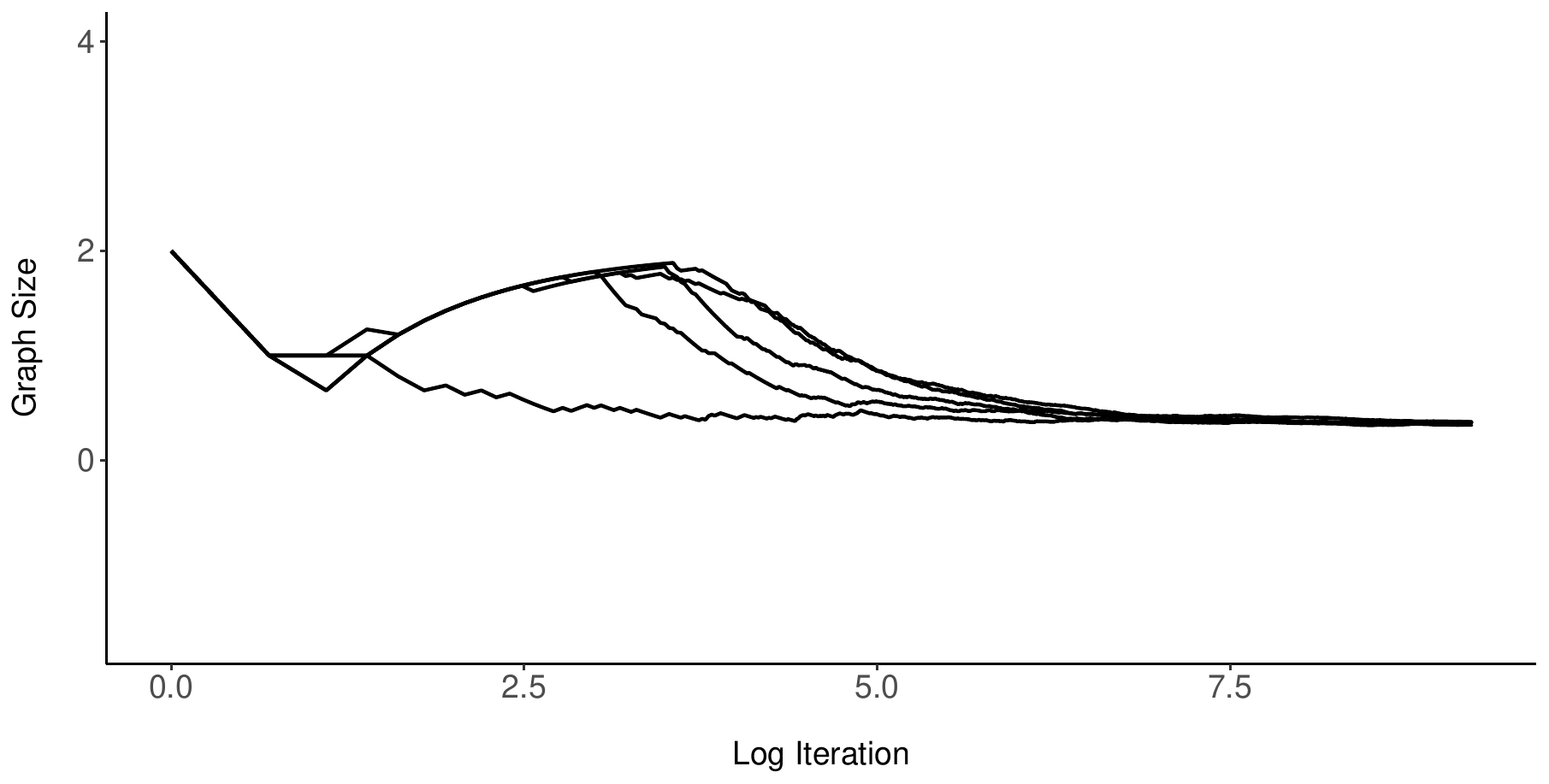}
    \caption{Convergence plot of the posterior expected number of edges in residual graph for Wake County. Each of the 5 lines corresponds with a single Markov chain.}
    \label{fig:lbw:warrenNumedges}
\end{figure}

\begin{table}[h!]
    \centering
    \begin{tabular}{c|c|c|c|c}
    \hline
        Variable & Posterior Mean &  95\% CI & $B_{1,0}$ & Evidence against $H_0$\\
        \hline
  LBW & $0.066$ & $[0.028,0.087]$ & $85.948$ & Strong\\
  S & $4.809$ & $[4.385,5.668]$ & $98.99$ & Strong\\
  MAGE & $0.020$ & $[0.001,0.036]$ & $2.471$ & Not worth more\\
  &&&& than a bare mention\\
  R & $0.014$ & $[0.001,0.030]$ & $1.626$ & Not worth more\\
    &&&& than a bare mention\\
  \hline
    \end{tabular}
    \caption{Posterior summaries of the absolute values of the factor loadings for Wake County: posterior means, 95\% credible intervals, Bayes factors and their interpretation.}
    \label{tab:postsumwakecounty}
\end{table}

\begin{table}[h!]
    \centering
    \begin{tabular}{c|c|c|c|c}
    \hline
        Variable & Posterior Mean &  95\% CI & $B_{1,0}$ & Evidence against $H_0$\\
        \hline
  LBW & $0.359$ & $[0.015,0.910]$ & $47.539$ & Strong\\
  S & $4.809$ & $[4.385,5.668]$ & $98.99$ & Strong\\
  MAGE & $0.684$ & $[0.006,4.872]$ & $29.117$ & Strong\\
  R & $0.014$ & $[0.001,0.030]$ & $1.626$ & Not worth more\\
    &&&& than a bare mention\\
  \hline
    \end{tabular}
    \caption{Posterior summaries of the absolute values of the factor loadings for Warren County: posterior means, 95\% credible intervals, Bayes factors and their interpretation.}
    \label{tab:postsumwarrencounty}
\end{table}

Tables \ref{tab:postsumwakecounty} and \ref{tab:postsumwarrencounty} give posterior means, 95\% credible intervals and Bayes factors $B_{1,0}$ to test hypothesis $H_{0}: |\lambda_v|\le 0.01$ against hypothesis $H_{1}: |\lambda_v|>0.01$ for each factor loading $\lambda_v$ associated with variables LBW, sex (S), MAGE and race (R) assuming equal prior probabilities for these hypotheses in Wake and Warren counties. The interpretation of the Bayes factors is based on the guidelines provided in Section 3.2 of \citet{kass-raftery-1995}. In Wake County, the data provide strong evidence in favor of hypothesis $H_{1}: |\lambda_v|>0.01$ for LBW and sex (S), but do not provide evidence for this hypothesis for MAGE and race (R). Thus, the common factor loads on variables LBW and sex (S), but does not load on MAGE and race (R). However, in Warren County, the common factor loads on LBW, sex (S) and MAGE, but does not load on race (R). Figures \ref{fig:wake} and \ref{fig:warren} show the estimated residual graphs and factor loadings for Wake and Warren Counties. The estimated residual graph for Wake County has two edges that link MAGE and race (R), and LBW and race (R). However, the estimated residual graph for Warren County has no edges.

\begin{figure}
    \centering
    \includegraphics[width = 0.5\textwidth]{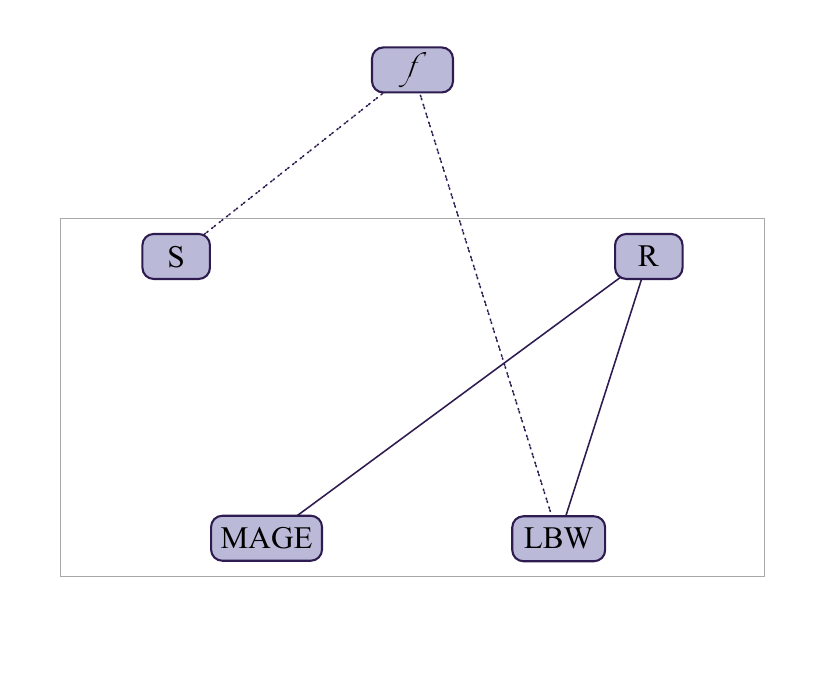}
    \caption{Estimated factor loadings and residual graph for Wake County. The latent variable is denoted with $\mathsf{f}$. Solid lines represent edges in the residual graph with a posterior inclusion probability greater than $0.95$, and dotted lines represent factor loadings that have a Bayes factor $B_{1,0}$ above $3.2$.}
    \label{fig:wake}
\end{figure}

\begin{figure}
    \centering
    \includegraphics[width = 0.5\textwidth]{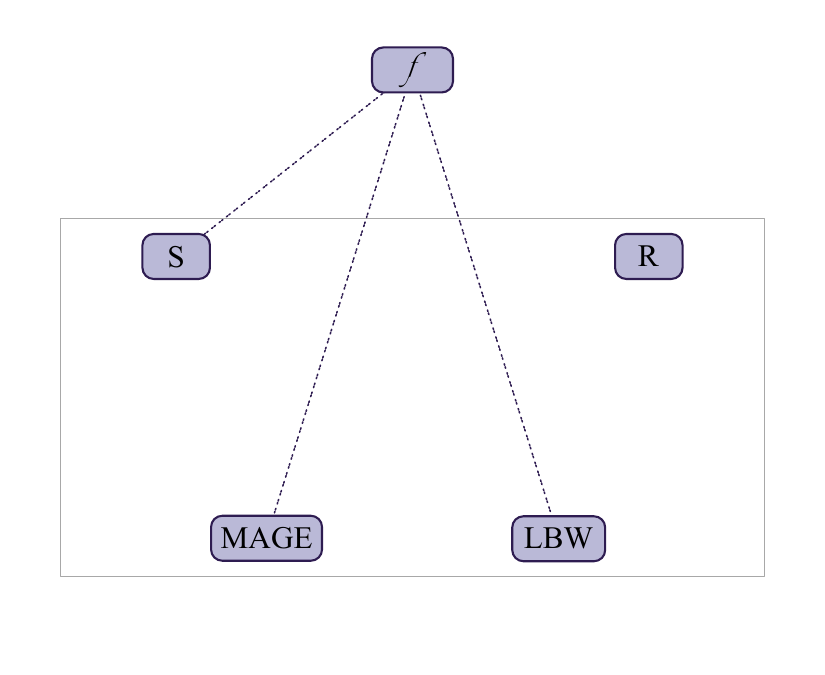}
    \caption{Estimated factor loadings and residual graph for Warren County. The latent variable is denoted with $\mathsf{f}$. Dotted lines represent factor loadings that have a Bayes factor $B_{1,0}$ above $3.2$.}
    \label{fig:warren}
\end{figure}

Figures \ref{fig:lbwindicesmap}, \ref{fig:raceindicesmap}, \ref{fig:sexindicesmap} and \ref{fig:mageindicesmap} present posterior estimates of factor loadings on each of the four variables observed in the North Carolina counties. The spatially correlated common factors load on the LBW in all counties. The common factors load on race (R) in all but 8 counties, on sex (S) in all but all but 10 counties, and on MAGE in all but 4 counties. This implies the existence of a relevant spatial dependence of the associations among LBW, race, sex and MAGE that is captured by the CSFGMs. Figure \ref{fi:lbwmapssinglefactor} displays the posterior inclusion probabilities of the edges connecting LBW with  sex (S), MAGE and race (R) in the county-level residual graphs. Almost all the edges that connect sex (S) and MAGE with LBW have very small posterior inclusion probabilities. Several edges between race (R) and LBW have larger posterior inclusion probabilities. We see that these findings are consistent with those in Figure \ref{fi:lbwmaps} with the key difference that the results from CSFGMs involve fewer edges in the residual graphs compared to the log-linear models fitted separately for each county-level contingency table. This makes sense since the spatial correlation of the common factors account for some of the associations in Figure \ref{fi:lbwmaps}.

\begin{figure}
\includegraphics[width=0.7\textwidth]{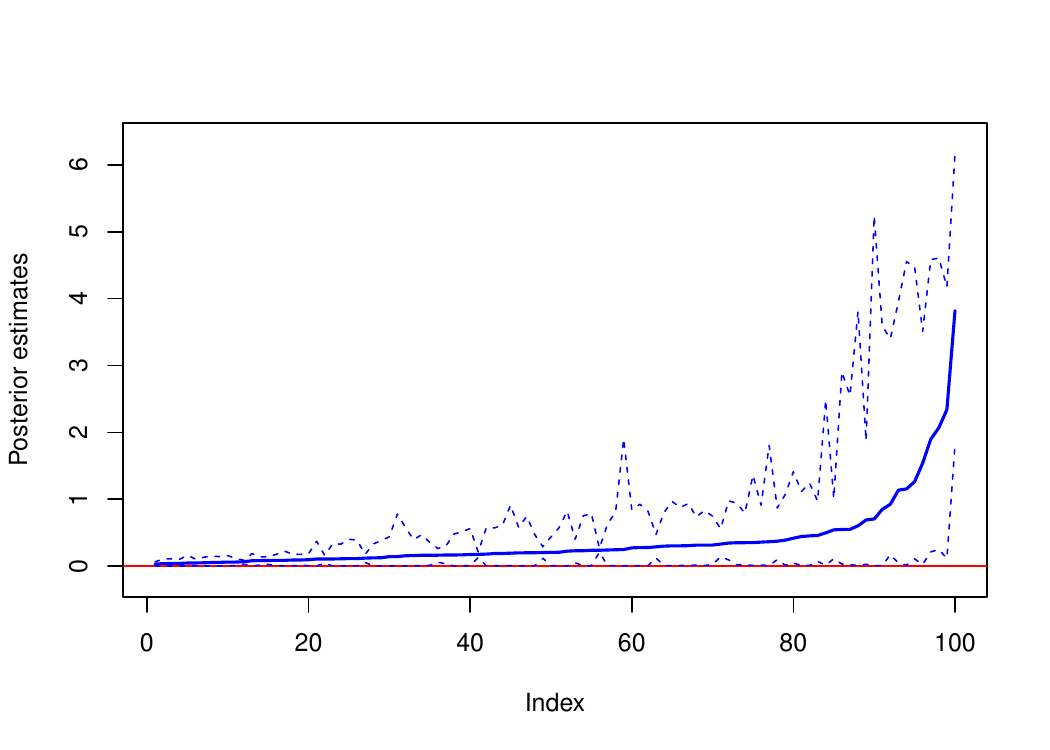}
\includegraphics[width=0.7\textwidth]{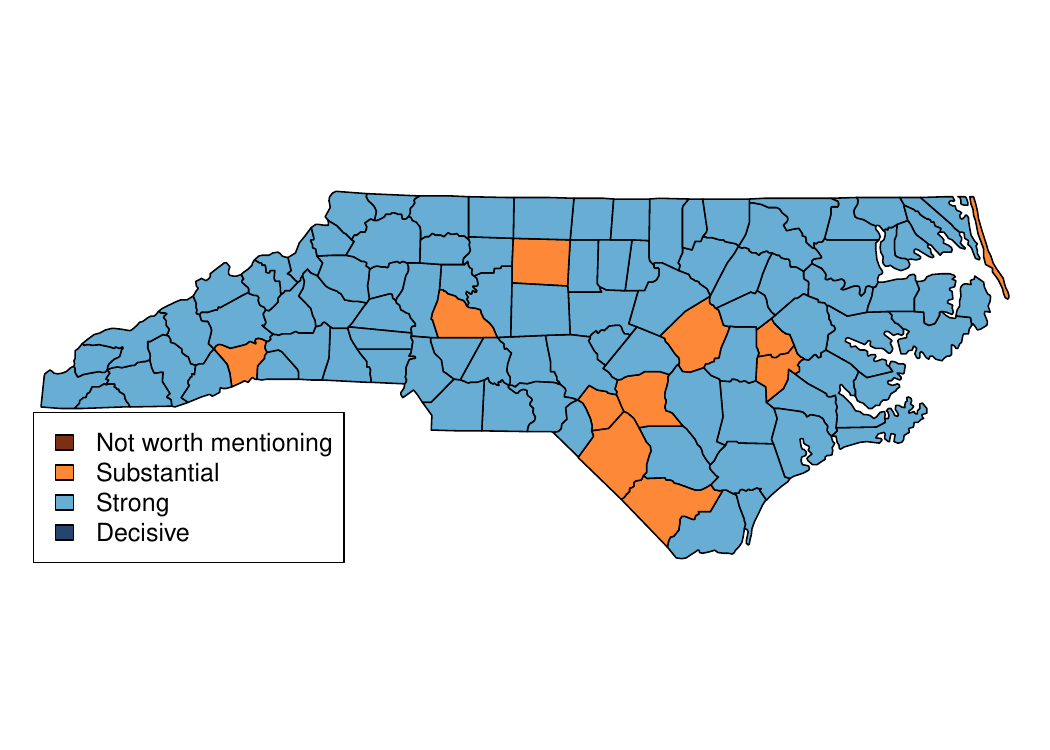}
    \caption{Top panel: Posterior means (solid line) and 95\% credible intervals (dotted lines) of the absolute values of the factor loadings on LBW arranged in increasing order with respect to their posterior means. Bottom panel: Choropleth map showing the Bayes factor $B_{1,0}$ associated with factor loadings on low birth weight (LBW).}
        \label{fig:lbwindicesmap}
\end{figure}

\begin{figure}
\includegraphics[width=0.7\textwidth]{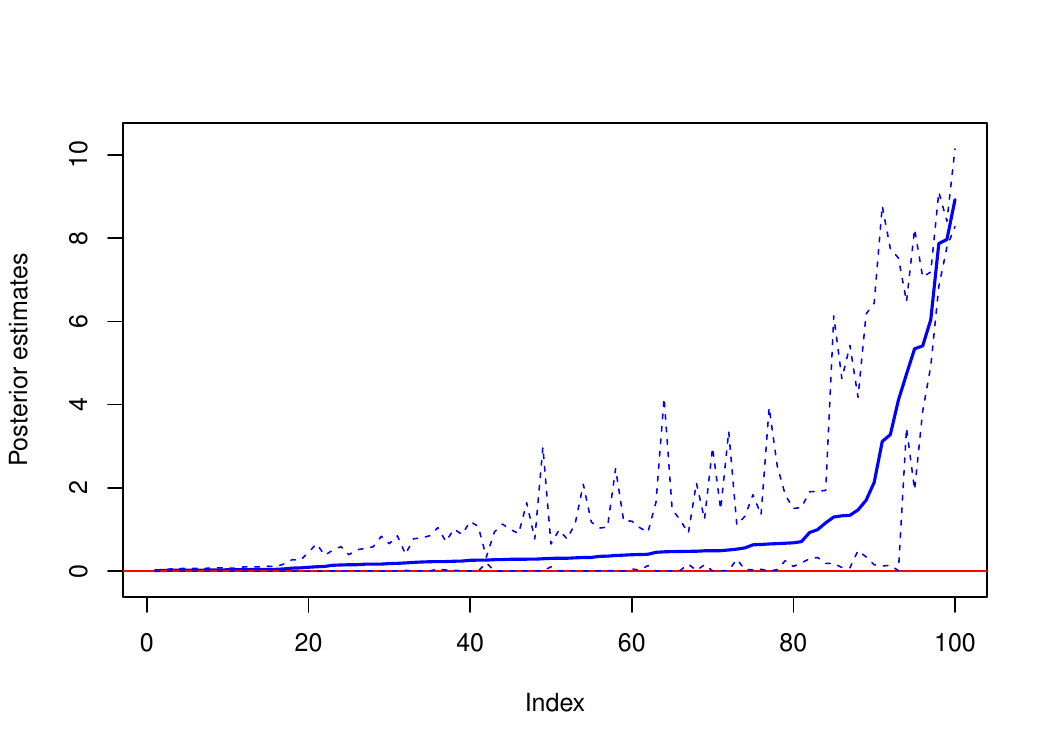}
\includegraphics[width=0.7\textwidth]{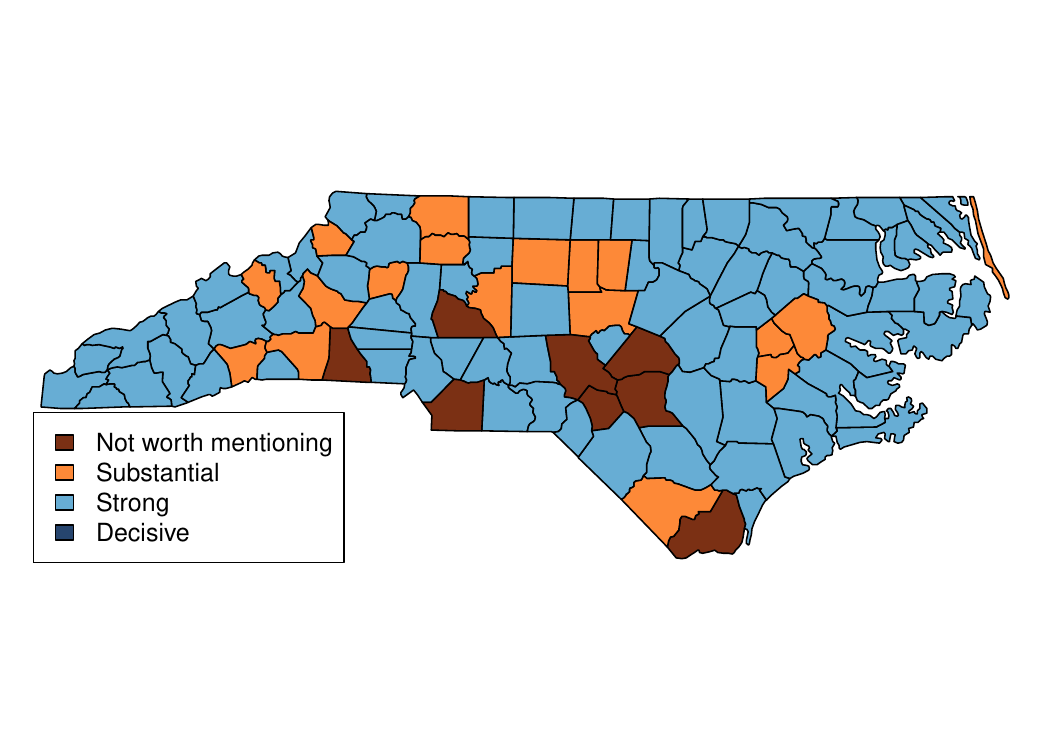}
    \caption{Top panel: Posterior means (solid line) and 95\% credible intervals (dotted lines) of the absolute values of the factor loadings on race (R) arranged in increasing order with respect to their posterior means. Bottom panel: Choropleth map showing the Bayes factor $B_{1,0}$ associated with factor loadings on race (R).}
        \label{fig:raceindicesmap}
\end{figure}

\begin{figure}
\includegraphics[width=0.7\textwidth]{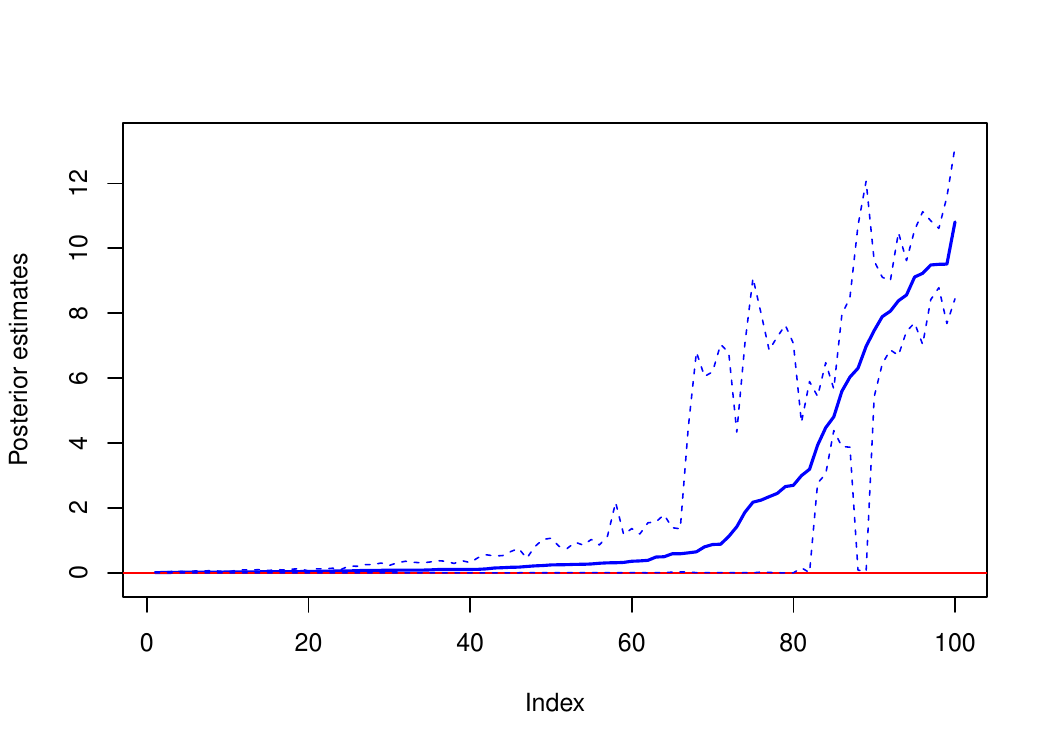}
\includegraphics[width=0.7\textwidth]{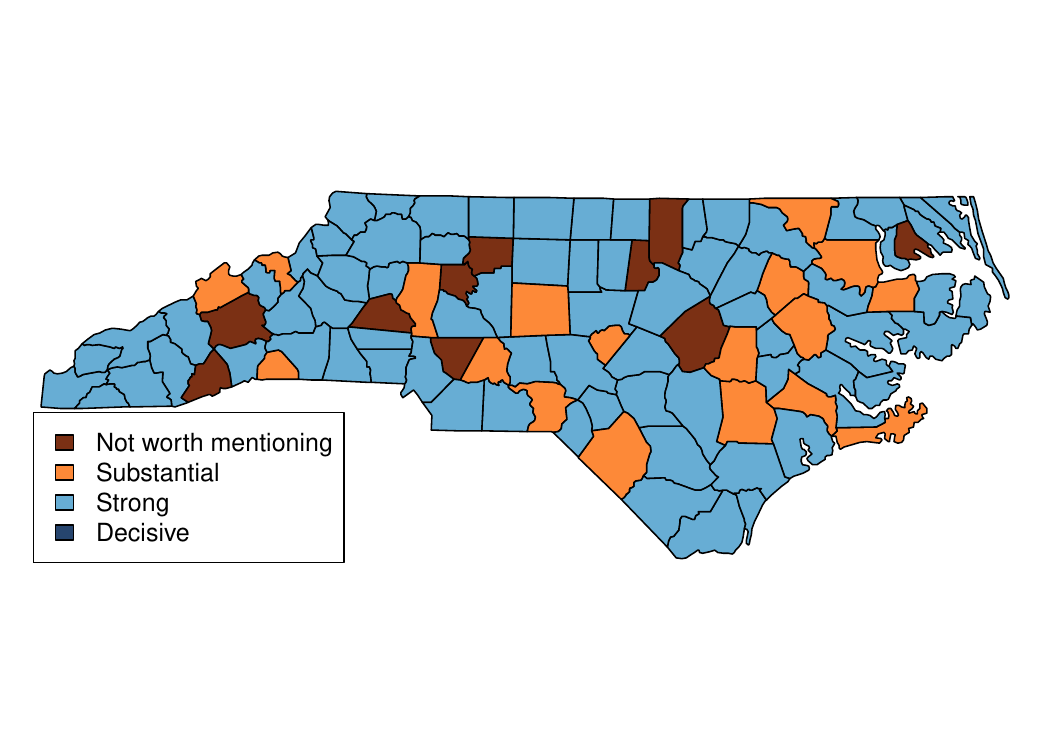}
    \caption{Top panel: Posterior means (solid line) and 95\% credible intervals (dotted lines) of the absolute values of the factor loadings on sex (S) arranged in increasing order with respect to their posterior means. Bottom panel: Choropleth map showing the Bayes factor $B_{1,0}$ associated with factor loadings on sex (S).}
        \label{fig:sexindicesmap}
\end{figure}

\begin{figure}
\includegraphics[width=0.7\textwidth]{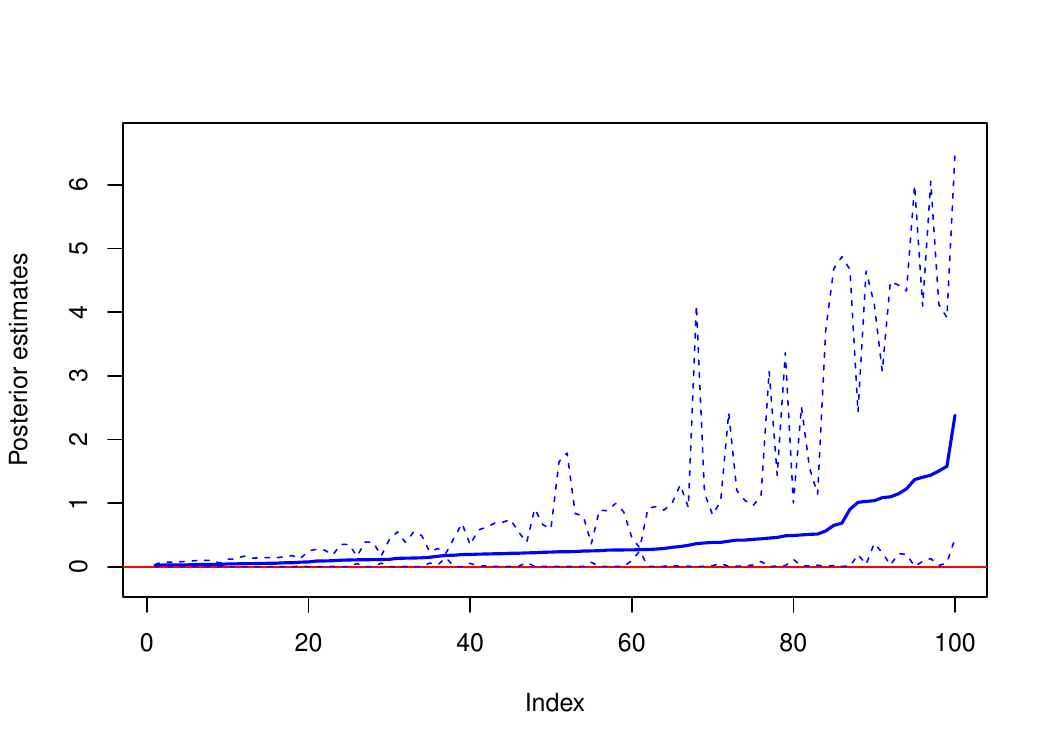}
\includegraphics[width=0.7\textwidth]{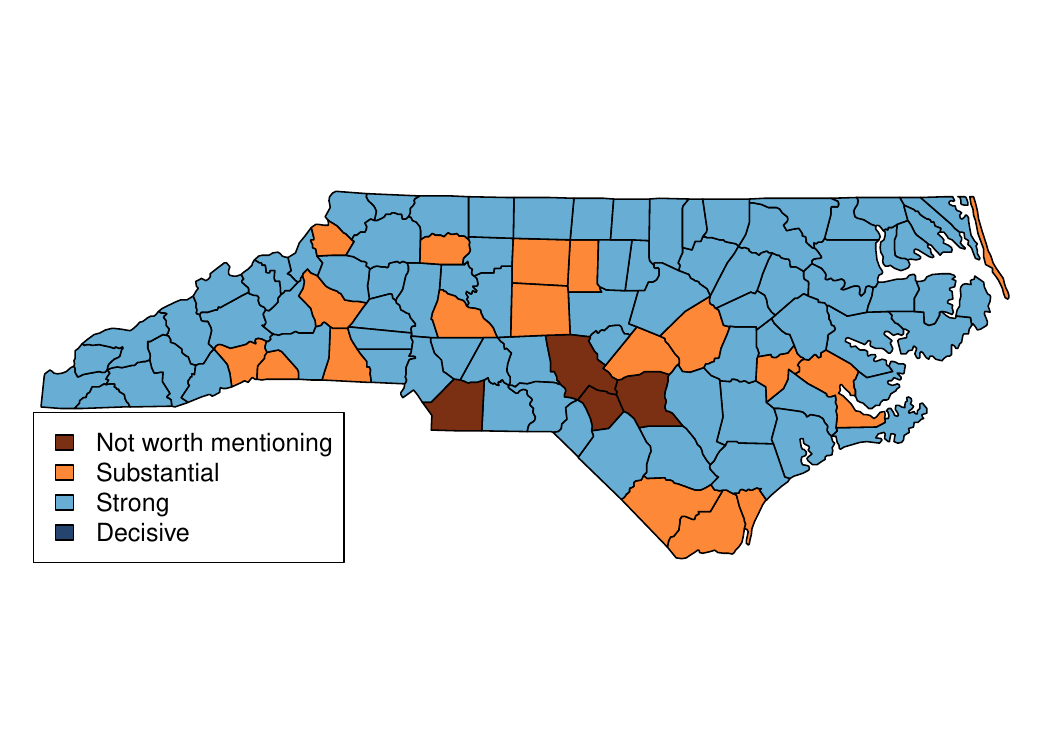}
    \caption{Top panel: Posterior means (solid line) and 95\% credible intervals (dotted lines) of the absolute values of the factor loadings on MAGE arranged in increasing order with respect to their posterior means. Bottom panel: Choropleth map showing the Bayes factor $B_{1,0}$ associated with factor loadings on mother's age at birth (MAGE).}
        \label{fig:mageindicesmap}
\end{figure}

\begin{figure}
\includegraphics[width=0.6\textwidth]{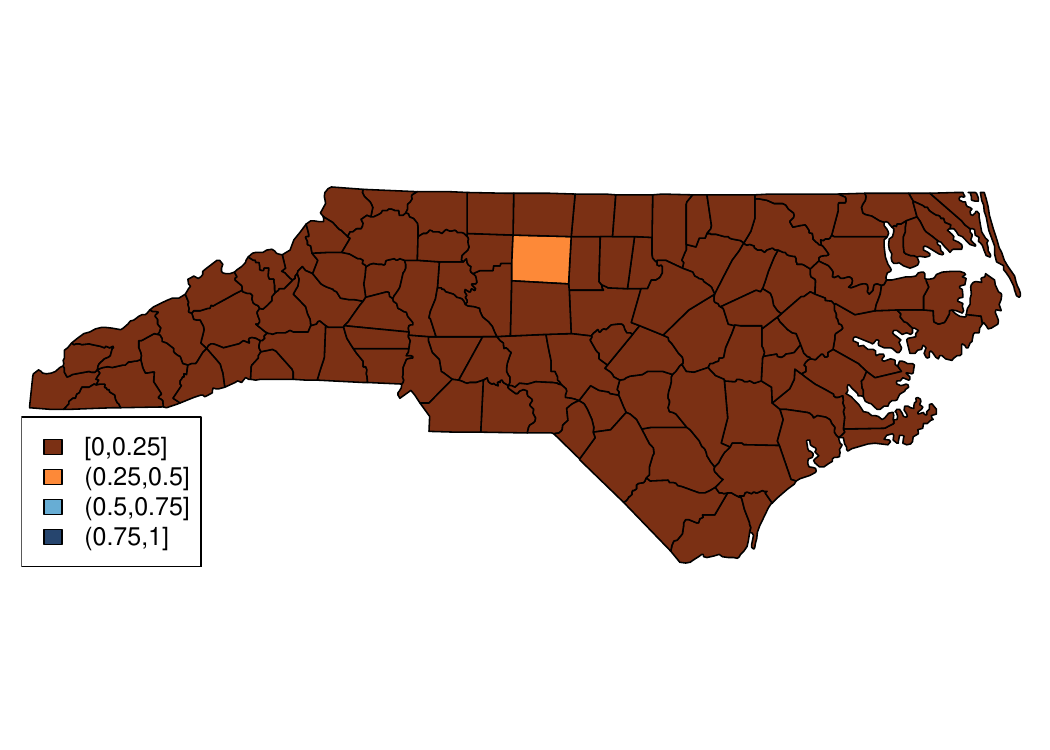}
\includegraphics[width=0.6\textwidth]{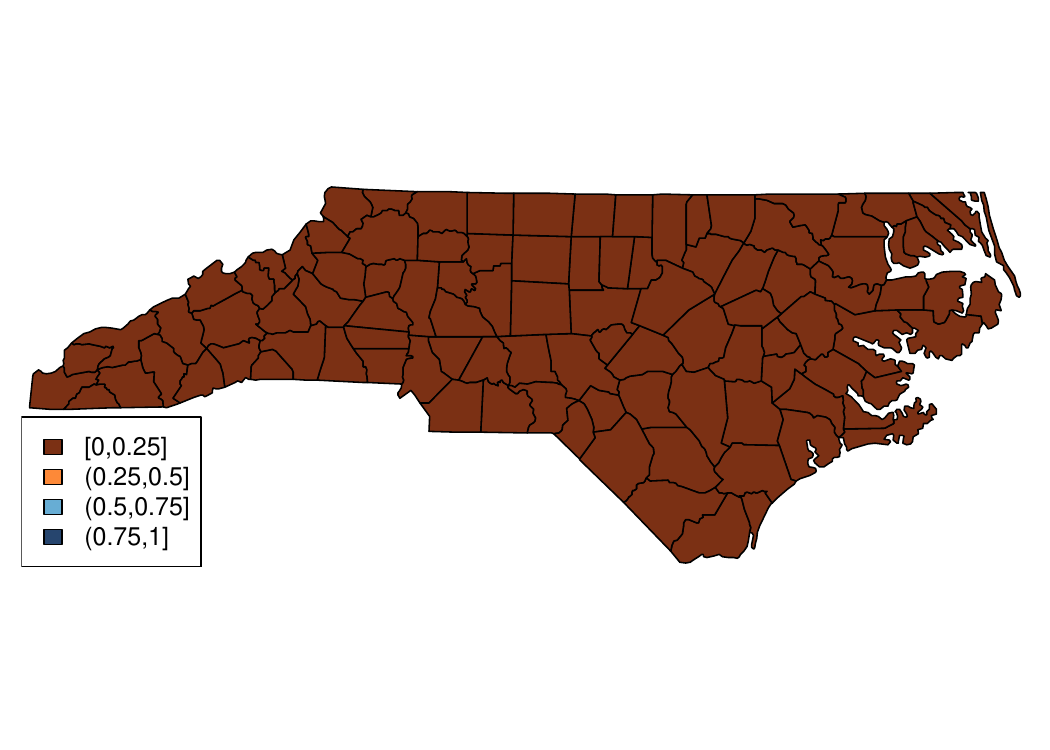}
\includegraphics[width=0.6\textwidth]{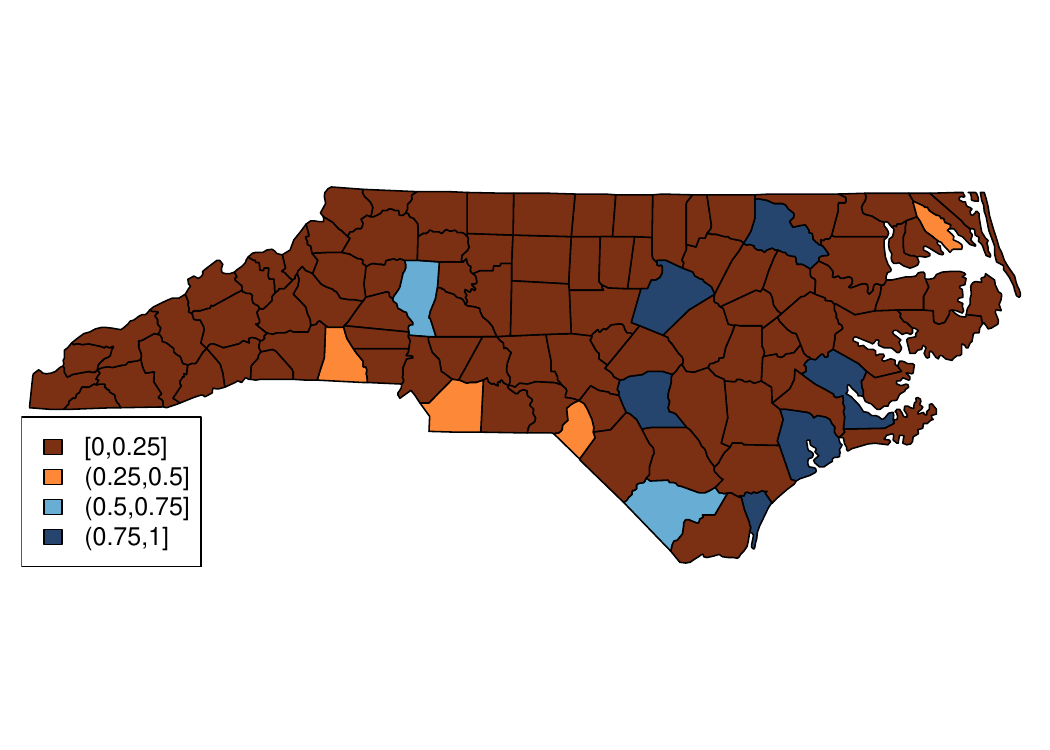}
        \caption{Choropleth maps for the posterior inclusion probabilities of edges that connect low birth weight (LBW) with sex (S) (top panel), low birth weight (LBW) with mother's age at birth (MAGE) (middle panel), and mother's age at birth (MAGE) with race (R) (bottom panel) based on the CSFGMs.}
        \label{fi:lbwmapssinglefactor}
\end{figure}

\section{Application to HIV Dataset} \label{sec:hivanalysis}

We analyze data collected from a demographic surveillance site in an African country. This rural study area is characterized by high adult HIV prevalence, high levels of poverty and unemployment, and frequent residential changes. A version of these data has been previously analyzed in \citet{dobra-et-2019}. From the entire population under surveillance between January 1, 2004 and December 31, 2016, we selected those individuals who consented to test at least twice for HIV after the age of 15, and whose first test was negative. A number of $9,827$ men and $15,073$ women satisfy these inclusion criteria. The date of HIV seroconversion was assumed to occur according to a uniformly random distribution between the date of the last negative and the first positive HIV test \citep{vandormael-et-2018}. Here, seroconversion refers to the transition from infection with the HIV virus to the detectable presence of HIV antibodies in the blood. 

The surveillance site collects longitudinal residential information about the exact periods of time  each study participant spent living inside or outside the study area. Repeat-testers can change their place of residence multiple times: they can move between two residencies located inside the rural study area, between two residencies located outside the rural study area, or between a residency inside the rural study area and another residency outside the rural study area. The relevance of looking at whether repeat-testers have resided outside the rural study area comes from the findings of \citet{dobra-et-2017}. Their results indicate that the risk of HIV acquisition is significantly increased for both men and women when they spend more time outside the rural study area. We note that the exposure period for a repeat-tester starts at the time of their first HIV test, and ends at their HIV seroconversion date for seroconverters, or at the time of their last HIV negative test for those that did not seroconvert. The residential locations occupied by seroconverters after they acquired HIV were discarded.

For each calendar year between 2004 and 2016, we recorded for each study participant the following binary variables: \textbf{Age} \--- the individual's age in years, \textbf{Seroconverted} (0=No,1=Yes) \--- whether the individual has seroconverted that year, \textbf{Sex} (0 = Men; 1 = Women), \textbf{Outside} (0 = spent less than 185 days living outside the study area; 1 = spent more than 185 days living outside the study area) \--- whether the individual was away from their home for most of the year, \textbf{Single} (0 = No; 1 = Yes) \--- whether the individual's marital status was single that year and \textbf{Education} (0 = never went to school; 1 = some formal education) \--- whether the individual had formal education up to that year. These variables have been included in the data because the an individual's risk of HIV acquisition is influenced by their age, by their amount of time spent away from home, by their marital status and by their level of education \citep{dobra-et-2017}.

We conducted separate analyzes for men and women based on 13 years of data that involve 5 variables: Age, Seroconverted, Outside, Single, and Education. For each sex, we used the CSFGMs for multiple datasets specified in Eq. \eqref{eq:groupsgraphical}. Temporal dependencies were modeled using the conditional independence graph $G_{\mu}$, which corresponds to an autoregressive model of order one (AR(1)). To expedite the resampling of the latent data, we transformed Age into an ordinal categorical variable using quantile binning. The age groups were defined based on the 0.1, 0.25, 0.5, 0.75, and 0.9 quantiles of the recorded ages of the study participants across all years of observation, which approximately correspond to ages 13, 18, 25, 46, and 63. We sampled from the posterior distribution of model parameters using six Markov chains, which were initialized at random starting points and ran for 30,000 iterations. The posterior estimates we present are based on the 120,000 samples obtained from all six chains after discarding the first 10,000 iterations in each chain as burnin. 

Figure \ref{fig:HIVfactorloadings} displays the posterior estimates of the absolute values of the factor loadings for each year of data. The largest absolute loadings correspond to the variable Single in both men and women. In 2013, the absolute loadings for Education spiked in both men and women, whereas these loadings remained low in other years. The variable Age has higher absolute loadings for most years in women. In men, the loadings for Age were slightly larger before 2009 and subsequently remained at lower levels.

The posterior inclusion probabilities of edges in the residual graphs that connect the variable Seroconverted to other variables are illustrated in Figure \ref{fig:HIVppiSeroconverted}. For men, the edge between Seroconverted and Age became stronger in 2009 and remained so until 2015, while other relationships involving the variable Seroconverted received only small posterior inclusion probabilities. For women, starting with 2009, all four edges that involve the variable Seroconverted exhibit considerable volatility in their strength. The posterior inclusion probabilities of all edges in the residual graphs for all years are included in the Appendix. These figures indicate that the temporal variability in the strength of most edges is considerably larger in women than in men, although it is quite similar for edges involving the variables Age or Education.

Figure \ref{fig:hiv2013} displays the estimated factor loadings and the edges of the median residual graphs for both men and women in 2013. In women, the single factor loads on all five variables. In men the single factor does not load on variable Outside. The residual graphs for men and women contain the triangle formed by edges between Age, Single and Outside. However, the residual graph for women contains two additional edges that connect Seroconverted with Age and Single. Median residual graphs and factor loadings for the other years can be found in the Appendix. 

This analysis shows that the CSFGMs for multiple datasets are highly effective in capturing the temporal dynamics of multivariate interactions among the variables Age, Seroconverted, Outside, Single, and Education, as well as the differences in these interactions between men and women. These results are highly relevant for studying the evolution of the HIV epidemic in relation to the risk factors for HIV acquisition in both men and women.

\begin{figure}
\includegraphics[scale=.4]{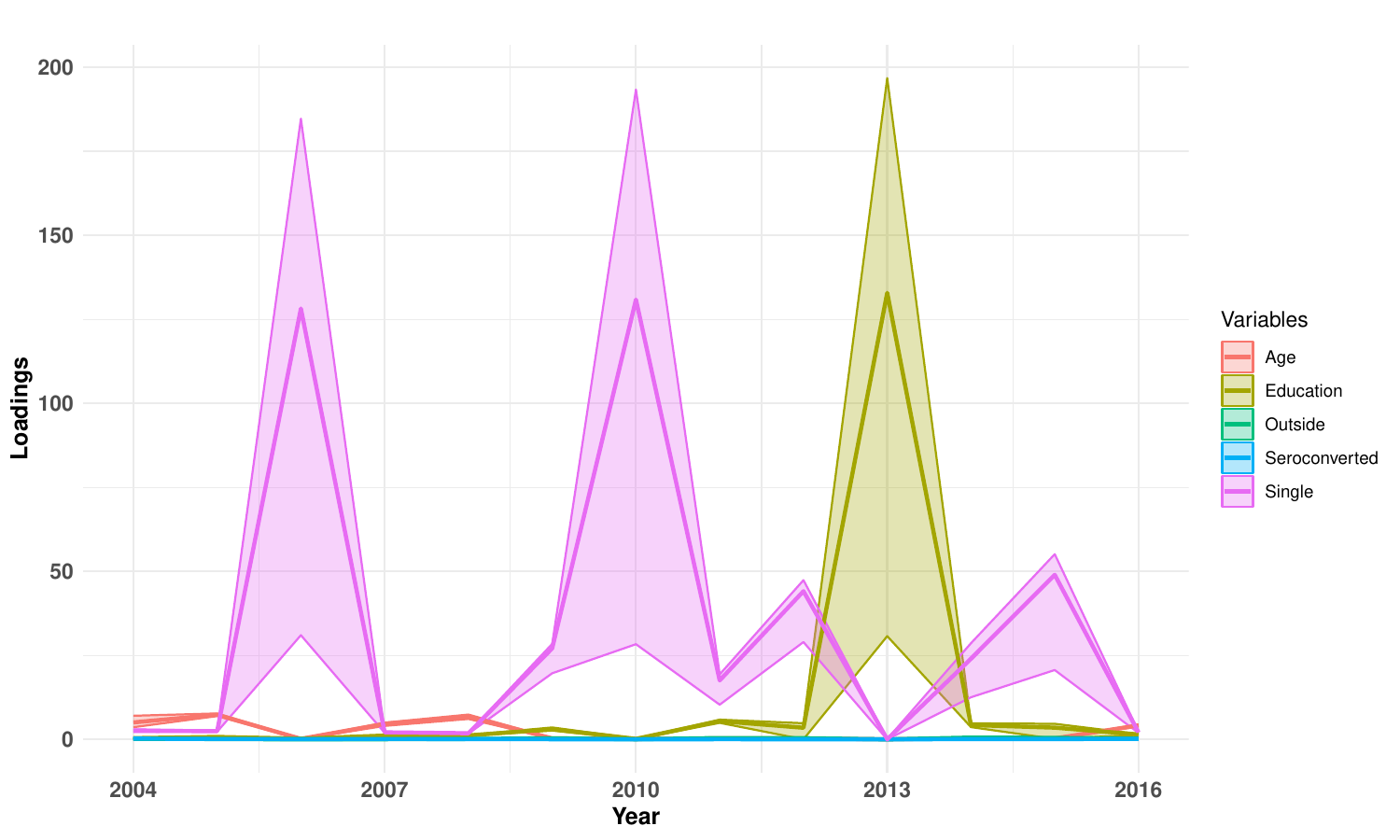} 
\includegraphics[scale=.4]{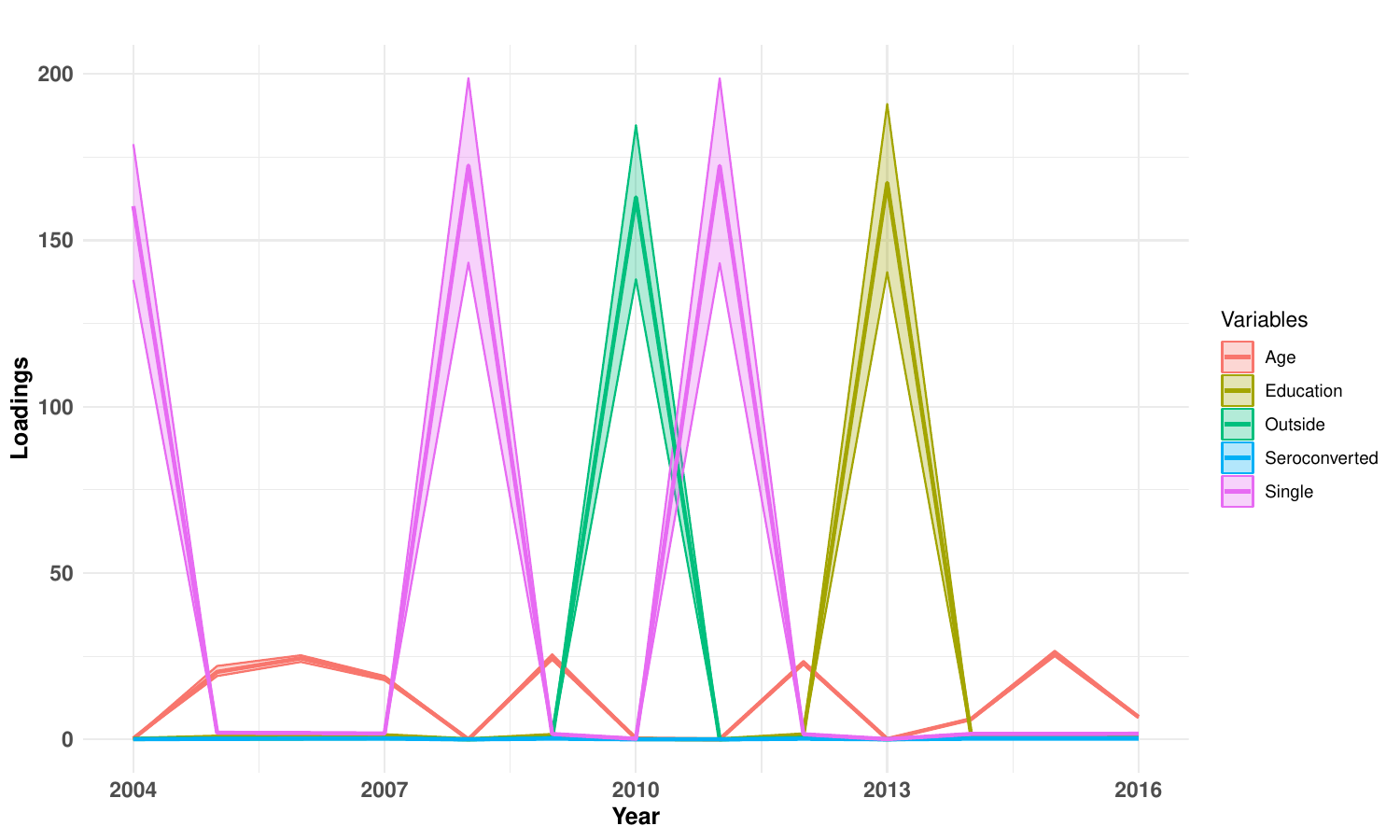} 
    \caption{Posterior inclusion probabilities for edges that involve variable Seroconverted in the residual graphs in the HIV data for men (top panel) and women (bottom panel).}
    \label{fig:HIVfactorloadings}
\end{figure}

\begin{figure}
\includegraphics[scale=.4]{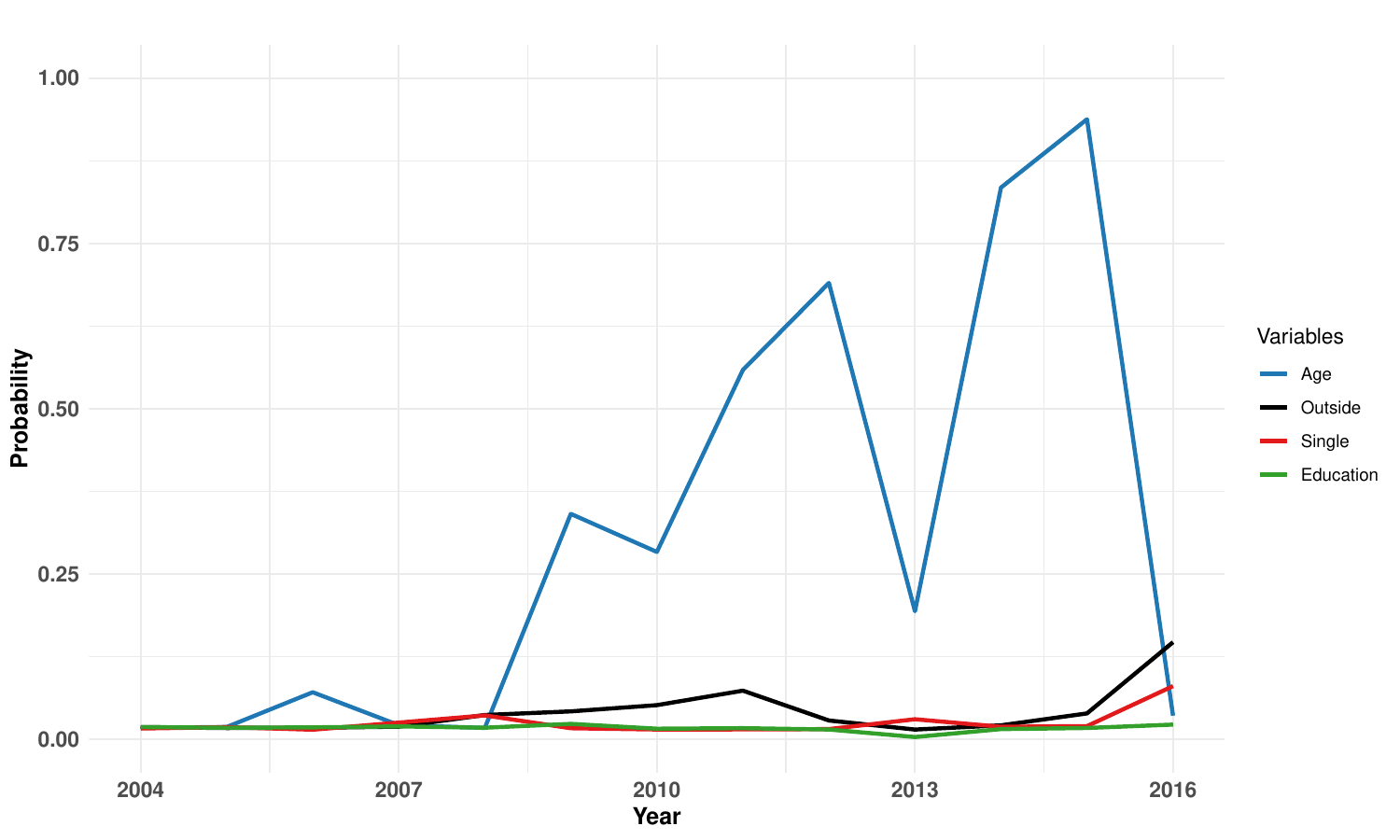} 
\includegraphics[scale=.4]{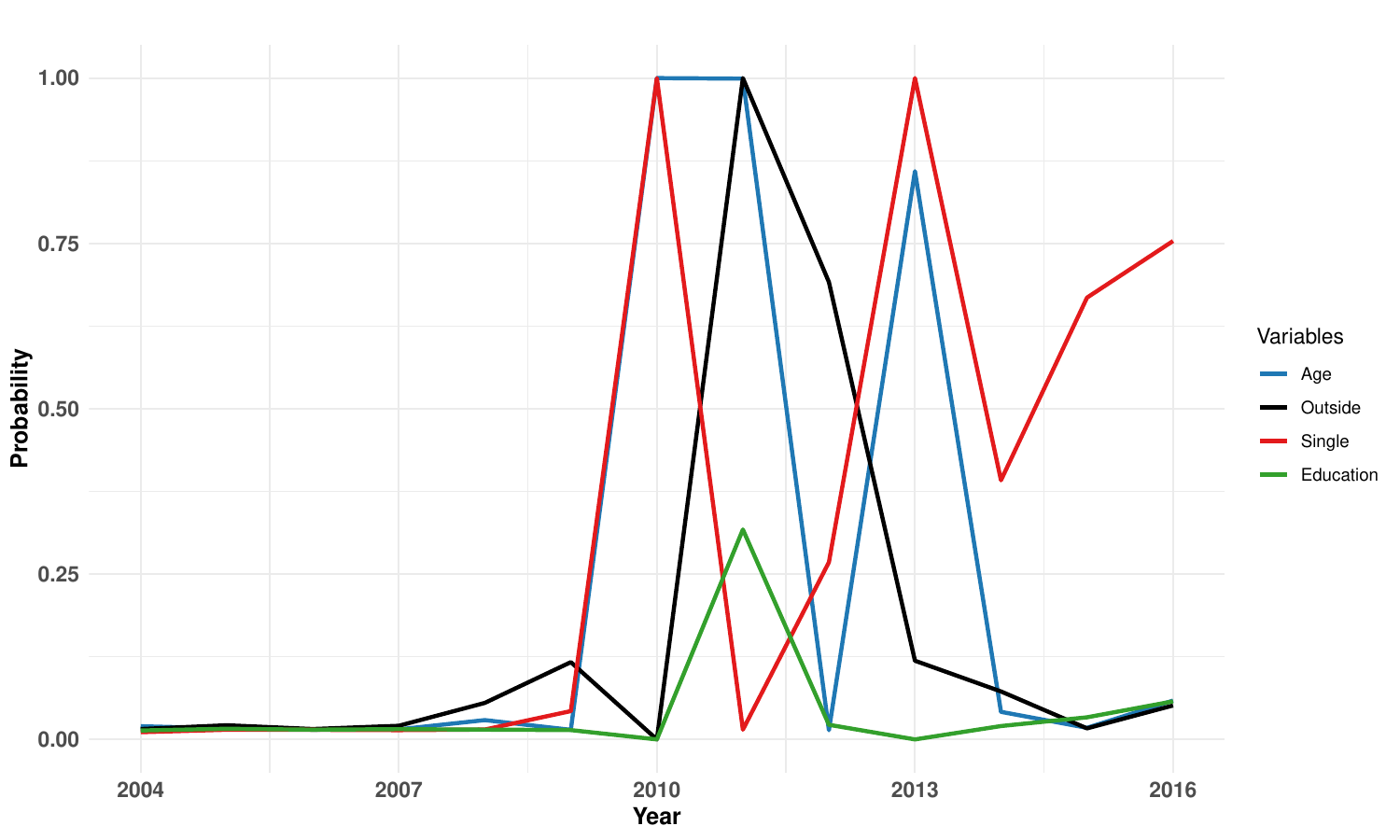} 
    \caption{Posterior inclusion probabilities for edges that involve variable Seroconverted in the residual graphs in the HIV data for men (top panel) and women (bottom panel).}
    \label{fig:HIVppiSeroconverted}
\end{figure}

\begin{figure}
\includegraphics[scale=.8]{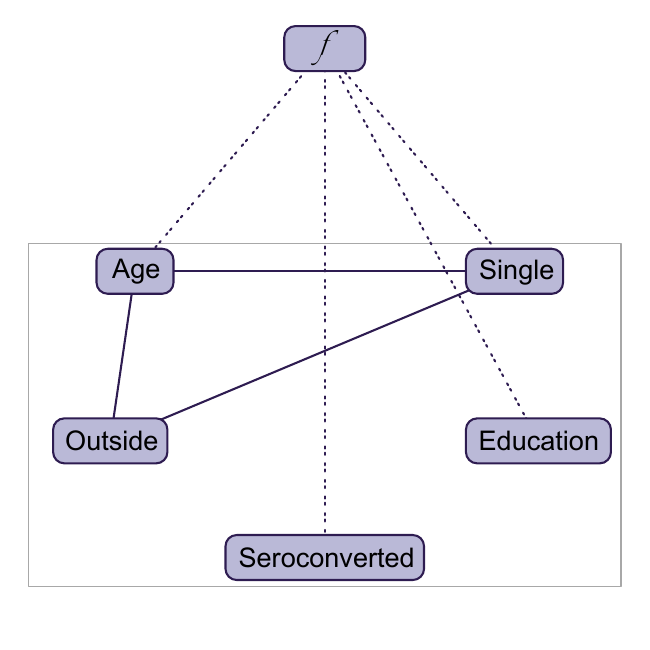} 
\includegraphics[scale=.8]{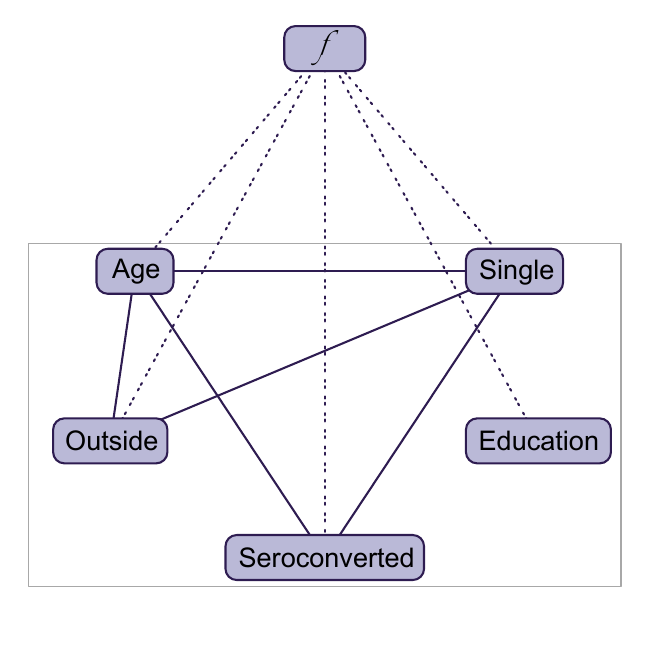} 
    \caption{Estimated factor loadings and residual graph in the HIV data for year 2013 for men (top panel)and women (bottom panel). The latent variable is denoted with $f$. Solid lines represent edges in the residual graph with a posterior inclusion probability greater than 0.5, and dotted lines represent factor loadings that have a Bayes factor $B_{1,0}$ above $3.2$.}
    \label{fig:hiv2013}
\end{figure}

\section{Discussion} \label{sec:discussion}

In this paper, we develop single-factor graphical models together with their extensions to datasets that involve binary and categorical ordinal variables. Their defining feature is their structure that comprises one latent variable and a graphical model for the joint distribution of the residuals. The key rationale for using single-factor graphical models becomes apparent in the analysis of multiple datasets that are spatially or temporally correlated. The single-factors allow for a coherent representation of associations across datasets, while residuals' graphs capture spatial or temporal dynamics of the associations between the observed variables after adjusting for associations across datasets. We note that the graph $G_{\mu}$ does not need to be fixed. If the structure of associations across datasets is unknown, $G_{\mu}$ can be updated by sampling from its full conditional distribution with the DCBF algorithm of \citet{hinne-et-2014}. The only limitation is computational as larger number of datasets $L$ might be difficult to handle. As we have seen in the examples we presented, $L=100$ datasets can be handled when $G_{\mu}$ remains fixed, but updating $G_{\mu}$ could turn out to be problematic.

The framework from Section \ref{sec:gfm-multiway} can be extended to represent associations across datasets that are simultaneously spatial and temporal. Let us assume that the data consist of groups of samples observed in $L_1$ spatial areas at $L_2$ time points.  A joint model for datasets $\{\mathcal{D}^{l_1,l_2}:1\le l_1\le L_1, 1\le l_2\le L_2$ involves single-factor graphical models \eqref{eq:latentfactorgroups} for each $\mathcal{D}^{l_1,l_2}$ and a $L_1\times L_2$ random matrix of means $\bfmu$ in \eqref{eq:groupsgraphical} that follows a matrix-variate distribution \citep{gupta-nagar-2000}:
\begin{eqnarray*}
 \mbox{vec}\left(\bfmu\right)\mid \bfK_{R},\bfK_{C} & \sim & \normal_{L_1L_2}\left(\bfzero,\left[\bfK_{C}\otimes \bfK_{R}\right]^{-1}\right).
\end{eqnarray*}
Here $\bfK_{R}$ is a $L_1\times L_1$ row precision matrix that represents spatial association and $\bfK_{C}$ is a $L_2\times L_2$ column precision matrix that represents temporal association. Furthermore, we assume that $\bfK_{R}\in M^+(G_R)$ and $\bfK_{C}\in M^+(G_C)$ where $G_R\in \mathcal{G}_{L_1}$ and $G_C\in \mathcal{G}_{L_2}$. The row graph $G_R$ and the column graph $G_C$ define graphical models for the rows and columns of the random matrix $\bfmu$ \citep{wang-west-2009}. These graphs encode spatial and temporal associations and can be fixed if the structure of these associations is known. Alternatively, $G_R$ and $G_C$ can be left unspecified and inferred from the data. \citet{dobra-lenkoski-rodriguez-2011} provide a $G$-Wishart prior specification for $\bfK_{R}$ and $\bfK_{C}$ and develop a RJMCMC sampler for the matrix-variate GGM of $\bfmu$. A version of this sampler, possibly improved with the DCBF algorithm of \citet{hinne-et-2014} can be integrated into the posterior sampling algorithms of Section \ref{sec:gfm-multiway}.

Our Bayesian framework for single-factor graphical models can be extended to graphical multi-factor models based on the identifiability conditions of \citet{grzebyk-et-2004} for factor models with zero constraints on the factor loadings and on the elements of the correlation matrix of the residuals. The sufficient conditions for the identifiability of \citet{giudici-stanghellini-2001} are also likely to be relevant. Another potentially useful extension could involve the development of priors that induce sparsity in the factor loadings and of their related posterior sampling methods. Sparsity is important since the latent variable $f$ could be associated with a subset of observed variables as illustrated in Figure \ref{fig:chaingraph1}. One option is the spike and slab priors on the factor loadings \citep{west-2003,carvalho-et-2008,hahn-carvalho-scott-2012}. A second option involves generalized double Pareto (GDP) independent priors \citep{armagan-et-2013}. As opposed to the spike and slab priors that set some factor loadings to zero, the GDP priors are shrinkage priors that place significant probability mass at or near zero. However, priors that induce sparsity in the factor loadings could lead to graphical factor models that are not identifiable if too many factor loadings are set to zero. For example, Theorem 5.5 from \citet{anderson-rubin} shows that for a single-factor model at least $3$ factor loadings must be non-zero for its identifiability to be preserved. Our numerical experiments with spike and slab priors and GDP priors have demonstrated a lack of convergence and mixing of MCMC samplers when many factor loadings were set to zero or were numerically close to zero.

\section*{Acknowledgments}
The authors would like to thank Abel Rodriguez (Department of Statistics, University of Washington) and Francesco Stingo (Department of Statistics, Computer Science, and Applications ``G. Parenti'', University of Florence) for their valuable discussions.

\section*{Funding}
The first author received support from a Graduate Research Fellowship provided by the National Science Foundation.

\section*{Appendix}

The analysis of the HIV data is presented in Section \ref{sec:hivanalysis}. Here, we present the estimated factor loadings and residual graphs for each year of the data, separately for men and women—see Figures \ref{fig:hiv2004} to \ref{fig:hiv2016}. Figures \ref{fig:HIVppiAge}, \ref{fig:HIVppiOutside}, \ref{fig:HIVppiSingle}, and \ref{fig:HIVppiEducation} illustrate the posterior inclusion probabilities for edges involving the variables Age, Outside, Single, and Education in the residual graphs.

\begin{figure}
\includegraphics[scale=.6]{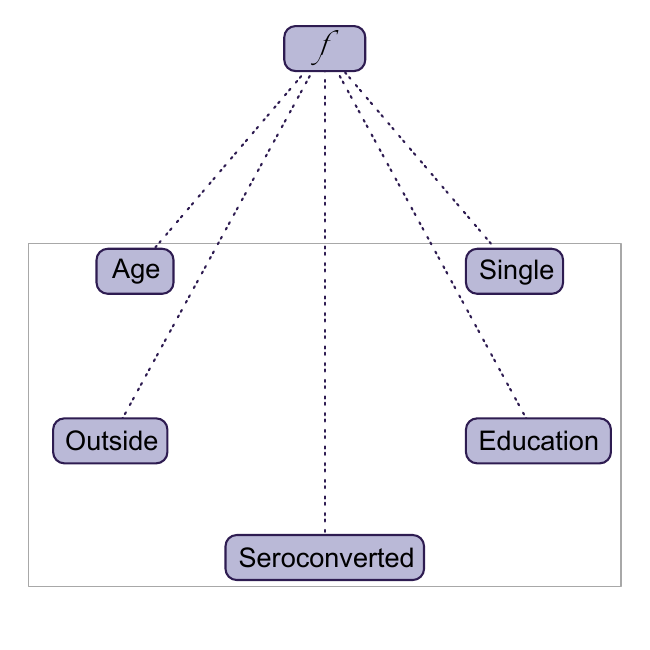} 
\includegraphics[scale=.6]{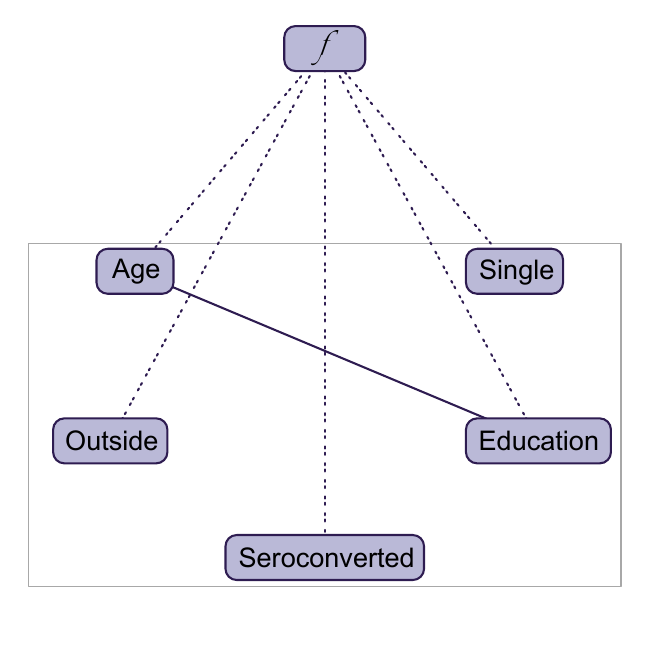} 
    \caption{Estimated factor loadings and residual graph in the HIV data for year 2004 for men (left panel)and women (right panel). The latent variable is denoted with $f$. Solid lines represent edges in the residual graph with a posterior inclusion probability greater than 0.5, and dotted lines represent factor loadings that have a Bayes factor $B_{1,0}$ above $3.2$.}
    \label{fig:hiv2004}
\end{figure}

\begin{figure}
\includegraphics[scale=.6]{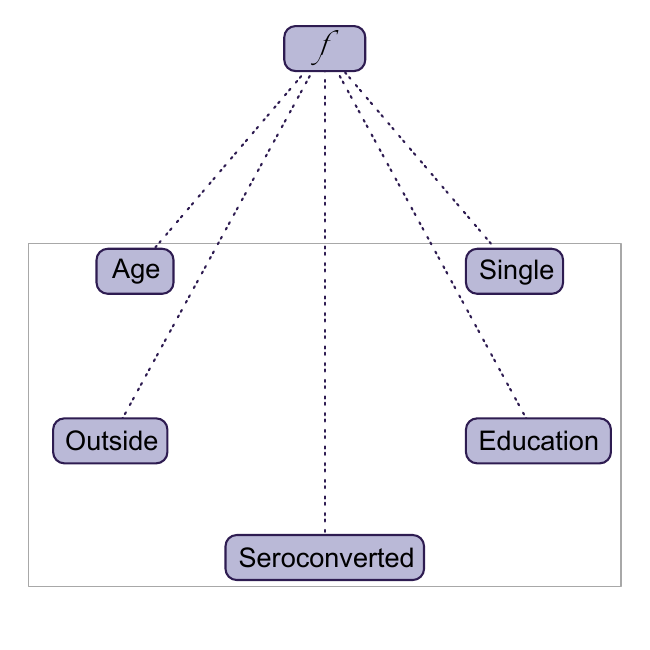} 
\includegraphics[scale=.6]{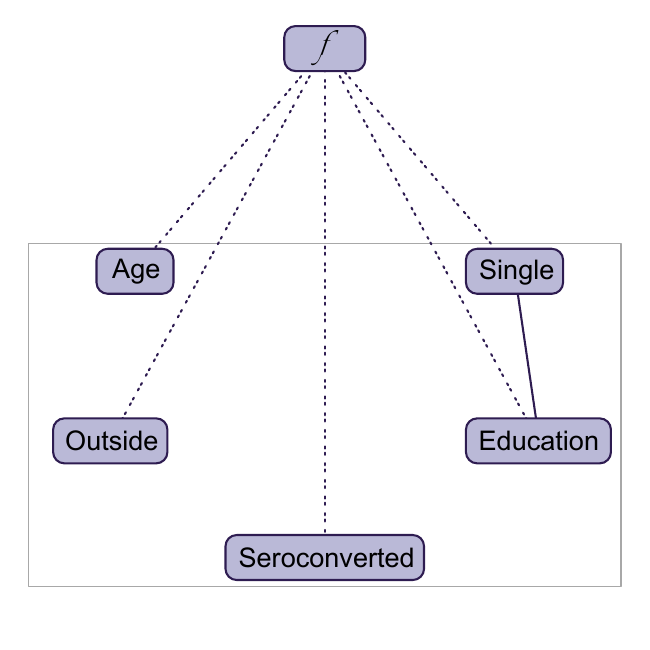} 
    \caption{Estimated factor loadings and residual graph in the HIV data for year 2005 for men (left panel)and women (right panel). The latent variable is denoted with $f$. Solid lines represent edges in the residual graph with a posterior inclusion probability greater than 0.5, and dotted lines represent factor loadings that have a Bayes factor $B_{1,0}$ above $3.2$.}
    \label{fig:hiv2005}
\end{figure}

\begin{figure}
\includegraphics[scale=.6]{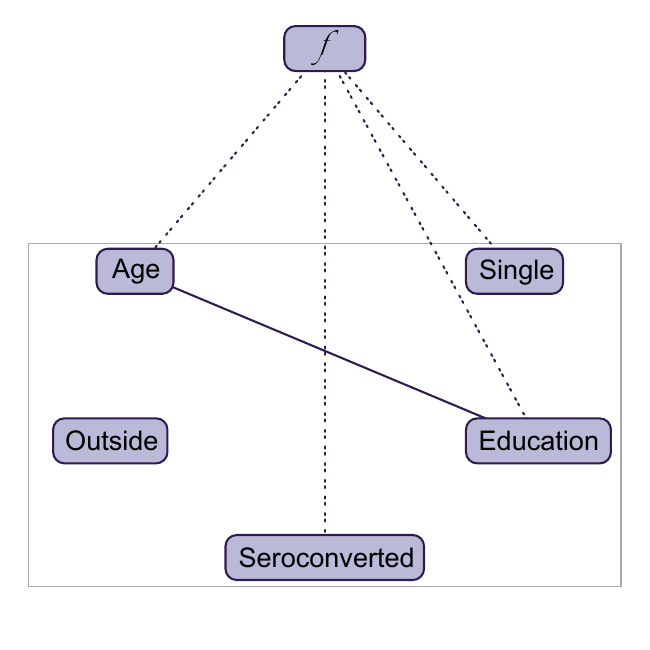} 
\includegraphics[scale=.6]{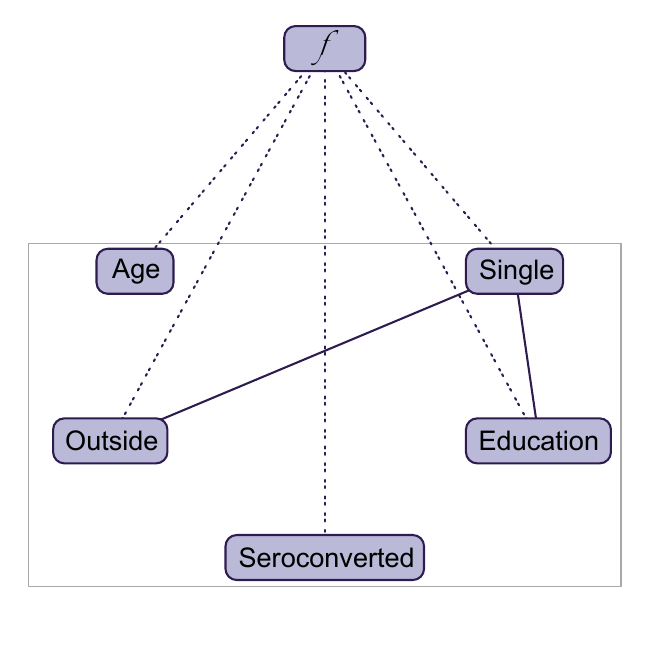} 
    \caption{Estimated factor loadings and residual graph in the HIV data for year 2006 for men (left panel)and women (right panel). The latent variable is denoted with $f$. Solid lines represent edges in the residual graph with a posterior inclusion probability greater than 0.5, and dotted lines represent factor loadings that have a Bayes factor $B_{1,0}$ above $3.2$.}
    \label{fig:hiv2006}
\end{figure}

\begin{figure}
\includegraphics[scale=.6]{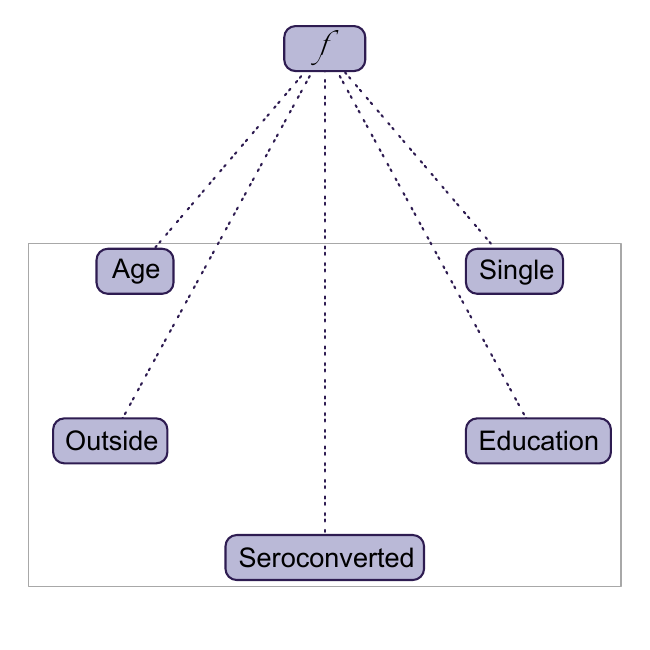} 
\includegraphics[scale=.6]{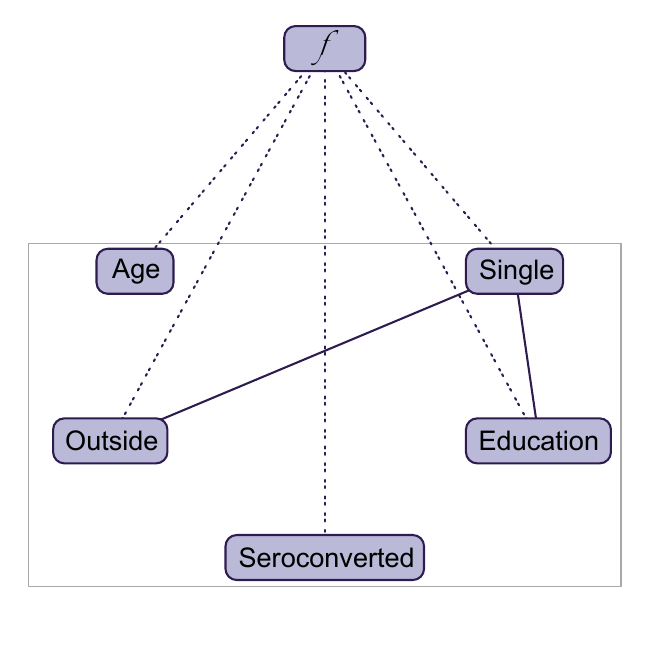} 
    \caption{Estimated factor loadings and residual graph in the HIV data for year 2007 for men (left panel)and women (right panel). The latent variable is denoted with $f$. Solid lines represent edges in the residual graph with a posterior inclusion probability greater than 0.5, and dotted lines represent factor loadings that have a Bayes factor $B_{1,0}$ above $3.2$.}
    \label{fig:hiv2007}
\end{figure}

\begin{figure}
\includegraphics[scale=.6]{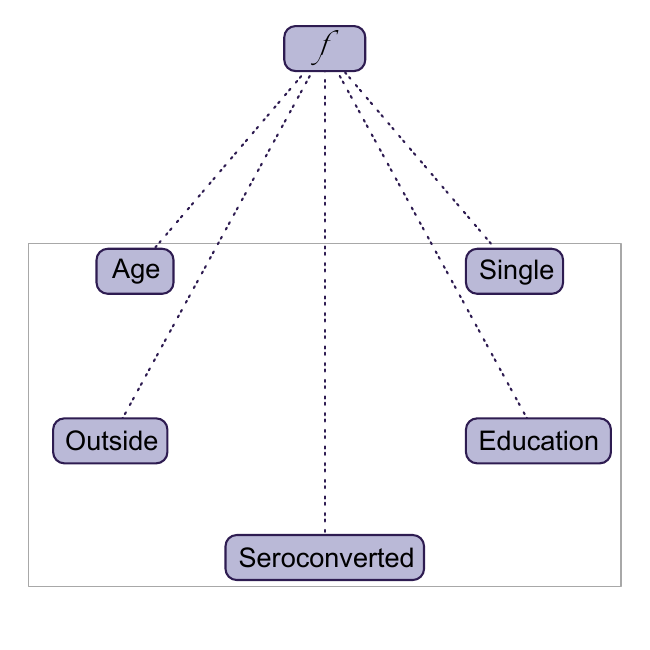} 
\includegraphics[scale=.6]{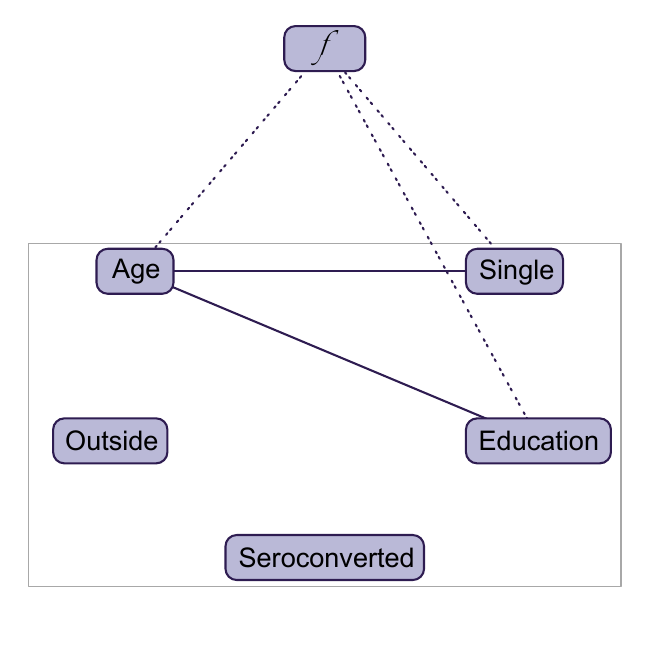} 
    \caption{Estimated factor loadings and residual graph in the HIV data for year 2008 for men (left panel)and women (right panel). The latent variable is denoted with $f$. Solid lines represent edges in the residual graph with a posterior inclusion probability greater than 0.5, and dotted lines represent factor loadings that have a Bayes factor $B_{1,0}$ above $3.2$.}
    \label{fig:hiv2008}
\end{figure}

\begin{figure}
\includegraphics[scale=.6]{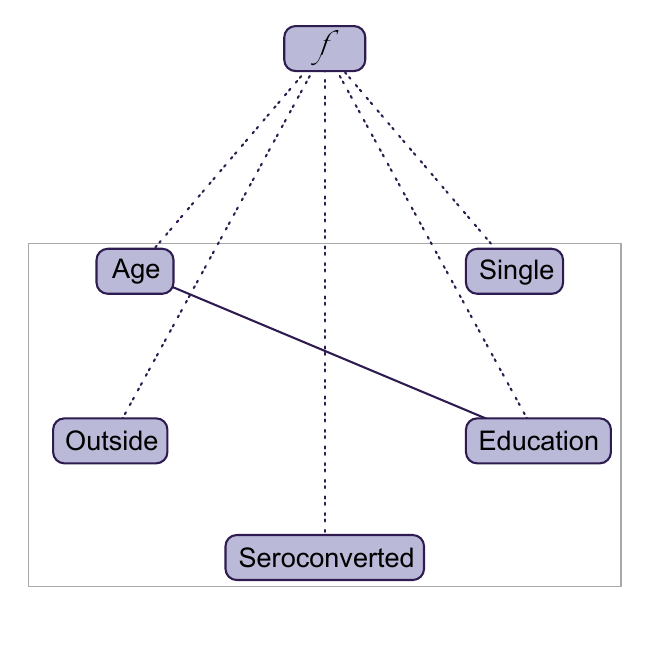} 
\includegraphics[scale=.6]{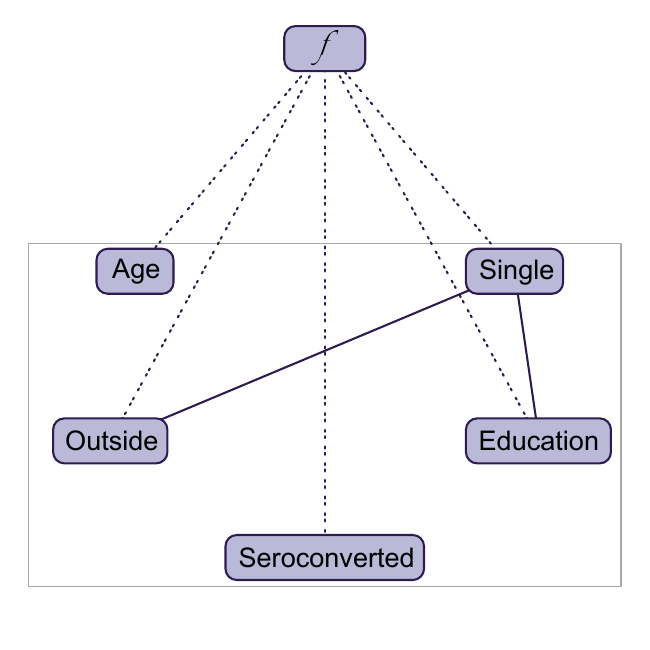} 
    \caption{Estimated factor loadings and residual graph in the HIV data for year 2009 for men (left panel)and women (right panel). The latent variable is denoted with $f$. Solid lines represent edges in the residual graph with a posterior inclusion probability greater than 0.5, and dotted lines represent factor loadings that have a Bayes factor $B_{1,0}$ above $3.2$.}
    \label{fig:hiv2009}
\end{figure}

\begin{figure}
\includegraphics[scale=.6]{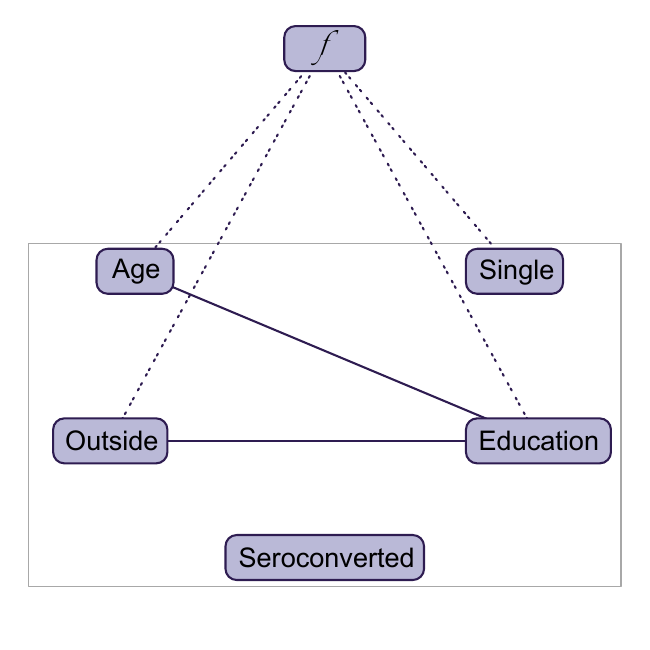} 
\includegraphics[scale=.6]{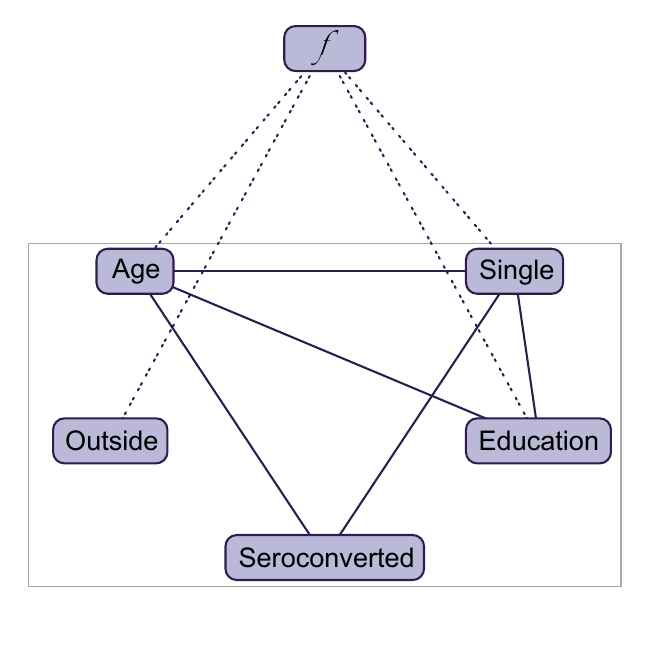} 
    \caption{Estimated factor loadings and residual graph in the HIV data for year 2010 for men (left panel)and women (right panel). The latent variable is denoted with $f$. Solid lines represent edges in the residual graph with a posterior inclusion probability greater than 0.5, and dotted lines represent factor loadings that have a Bayes factor $B_{1,0}$ above $3.2$.}
    \label{fig:hiv2010}
\end{figure}

\begin{figure}
\includegraphics[scale=.6]{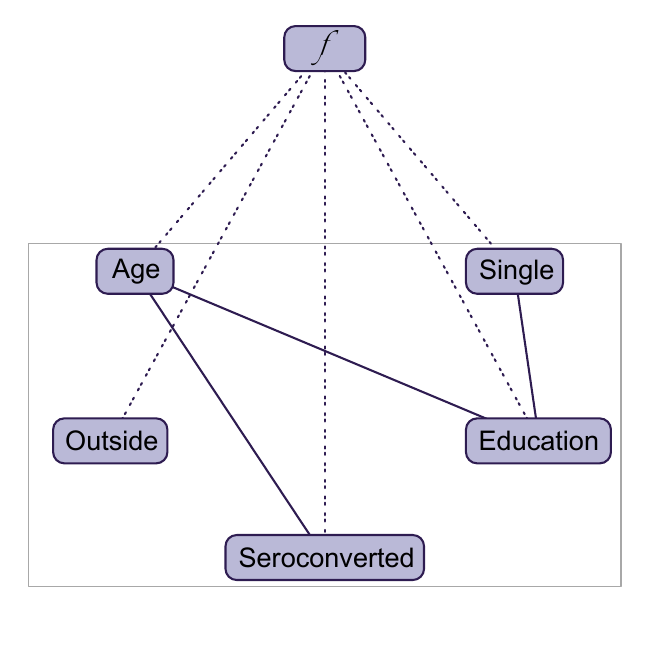} 
\includegraphics[scale=.6]{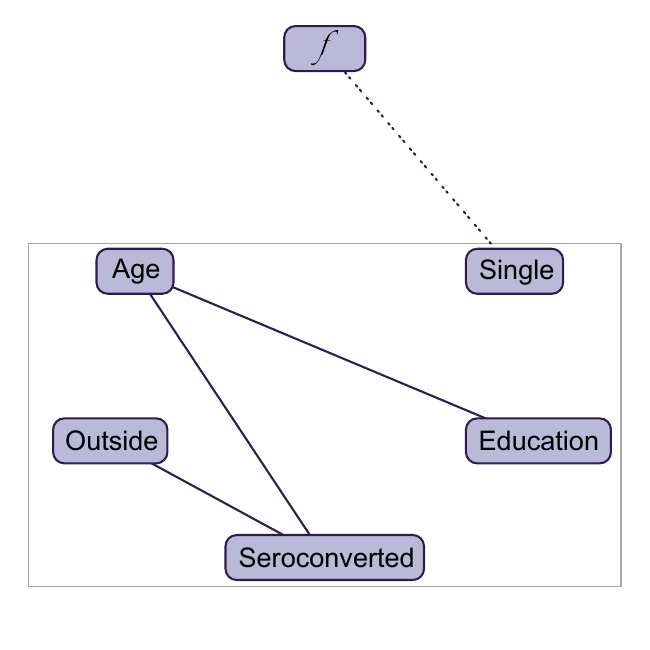} 
    \caption{Estimated factor loadings and residual graph in the HIV data for year 2011 for men (left panel)and women (right panel). The latent variable is denoted with $f$. Solid lines represent edges in the residual graph with a posterior inclusion probability greater than 0.5, and dotted lines represent factor loadings that have a Bayes factor $B_{1,0}$ above $3.2$.}
    \label{fig:hiv2011}
\end{figure}

\begin{figure}
\includegraphics[scale=.6]{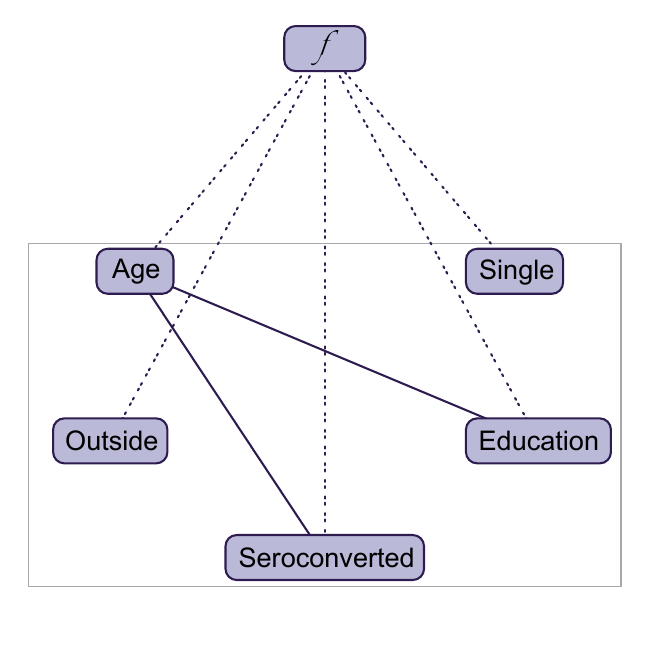} 
\includegraphics[scale=.6]{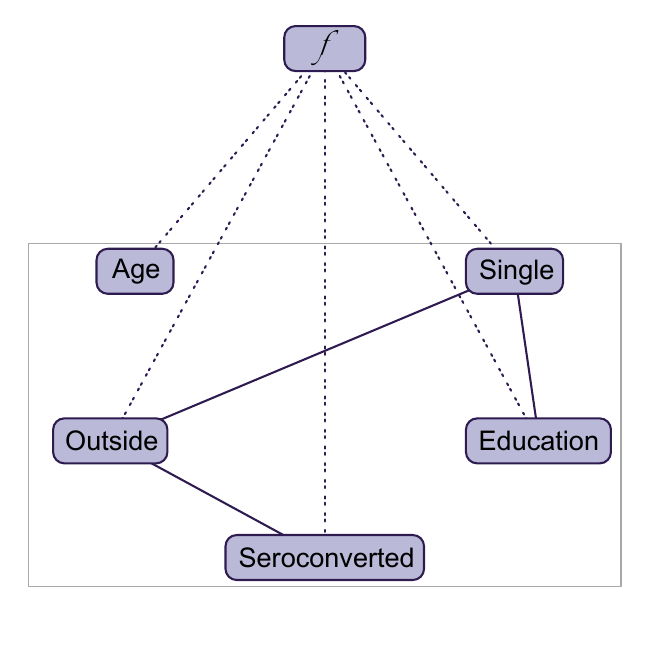} 
    \caption{Estimated factor loadings and residual graph in the HIV data for year 2012 for men (left panel)and women (right panel). The latent variable is denoted with $f$. Solid lines represent edges in the residual graph with a posterior inclusion probability greater than 0.5, and dotted lines represent factor loadings that have a Bayes factor $B_{1,0}$ above $3.2$.}
    \label{fig:hiv2012}
\end{figure}

\begin{figure}
\includegraphics[scale=.6]{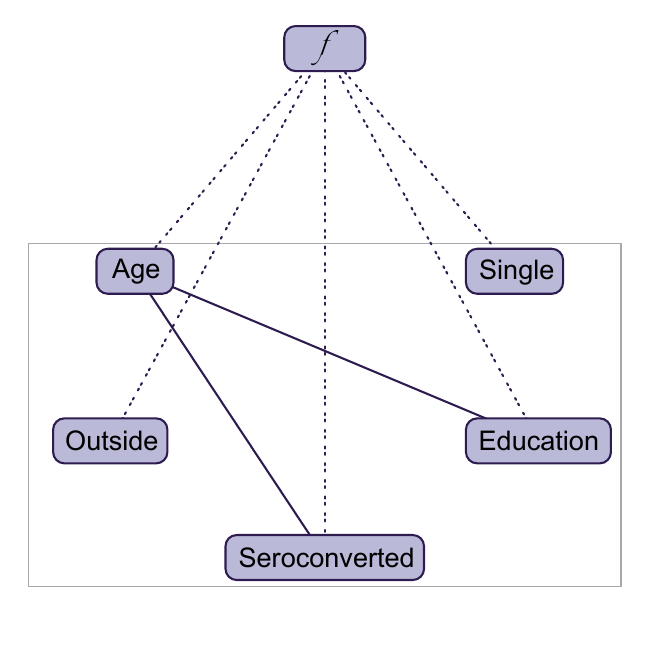} 
\includegraphics[scale=.6]{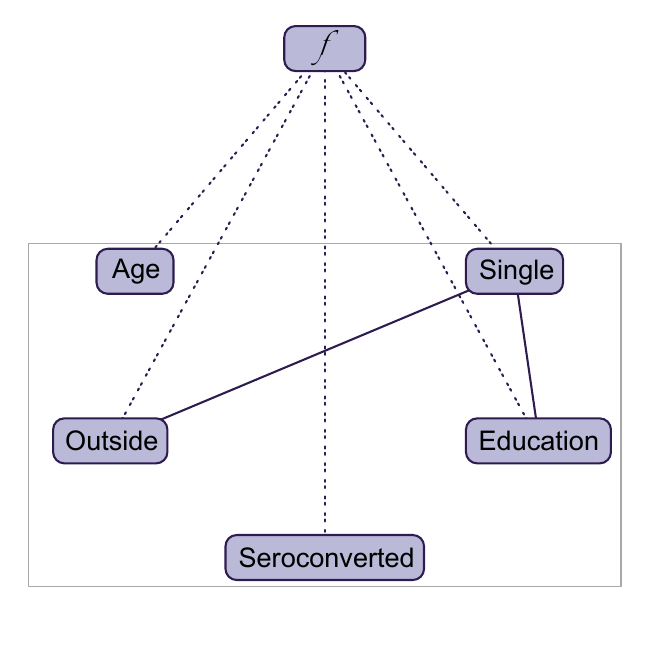} 
    \caption{Estimated factor loadings and residual graph in the HIV data for year 2014 for men (left panel)and women (right panel). The latent variable is denoted with $f$. Solid lines represent edges in the residual graph with a posterior inclusion probability greater than 0.5, and dotted lines represent factor loadings that have a Bayes factor $B_{1,0}$ above $3.2$.}
    \label{fig:hiv2014}
\end{figure}

\begin{figure}
\includegraphics[scale=.6]{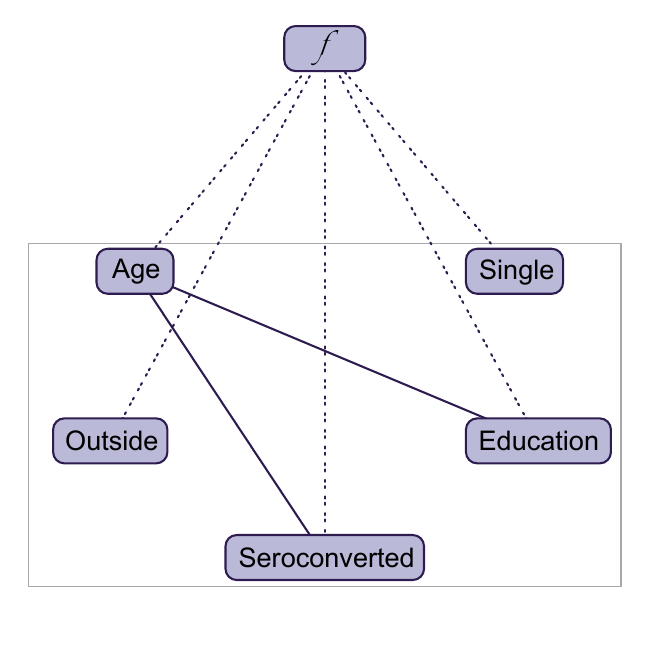} 
\includegraphics[scale=.6]{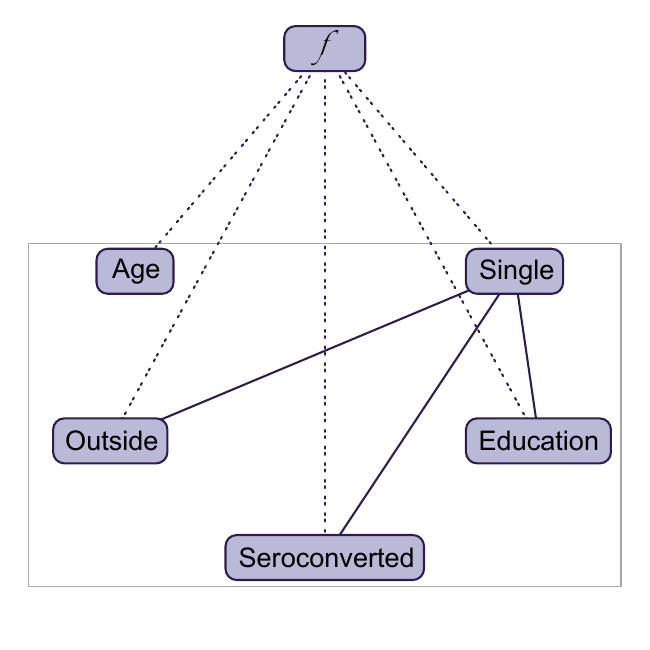} 
    \caption{Estimated factor loadings and residual graph in the HIV data for year 2015 for men (left panel)and women (right panel). The latent variable is denoted with $f$. Solid lines represent edges in the residual graph with a posterior inclusion probability greater than 0.5, and dotted lines represent factor loadings that have a Bayes factor $B_{1,0}$ above $3.2$.}
    \label{fig:hiv2015}
\end{figure}

\begin{figure}
\includegraphics[scale=.6]{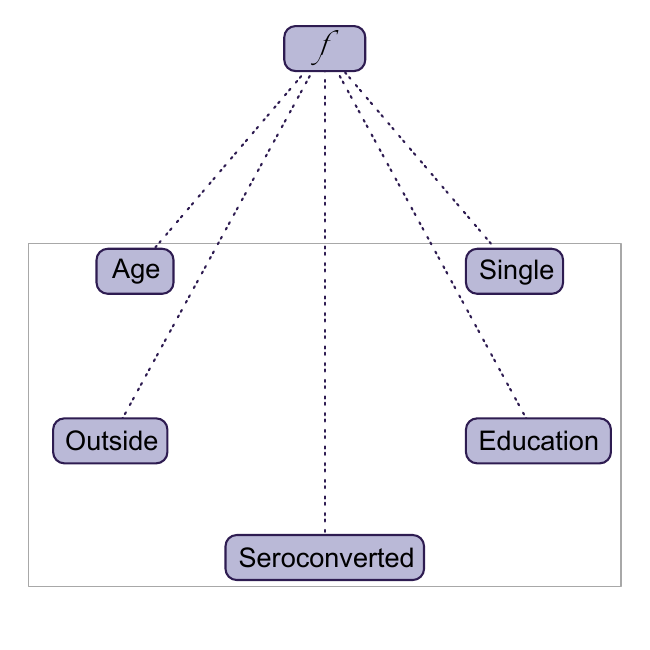} 
\includegraphics[scale=.6]{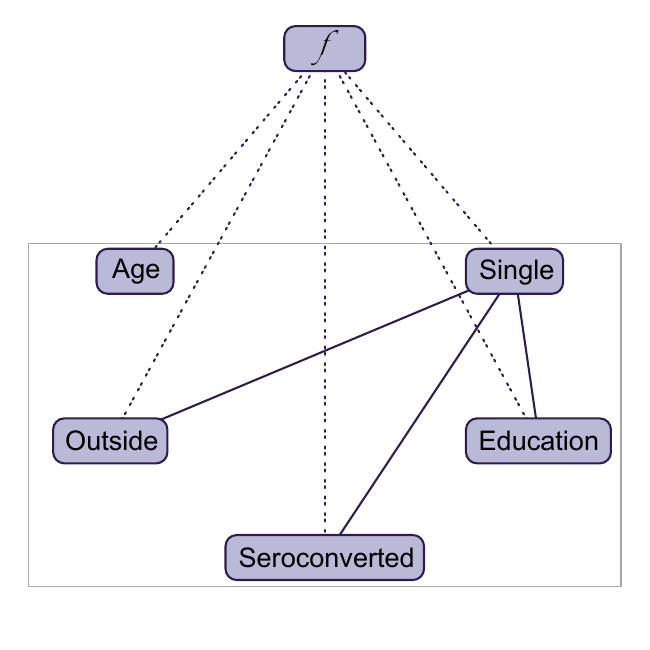} 
    \caption{Estimated factor loadings and residual graph in the HIV data for year 2016 for men (left panel)and women (right panel). The latent variable is denoted with $f$. Solid lines represent edges in the residual graph with a posterior inclusion probability greater than 0.5, and dotted lines represent factor loadings that have a Bayes factor $B_{1,0}$ above $3.2$.}
    \label{fig:hiv2016}
\end{figure}

\begin{figure}
\includegraphics[width=\textwidth]{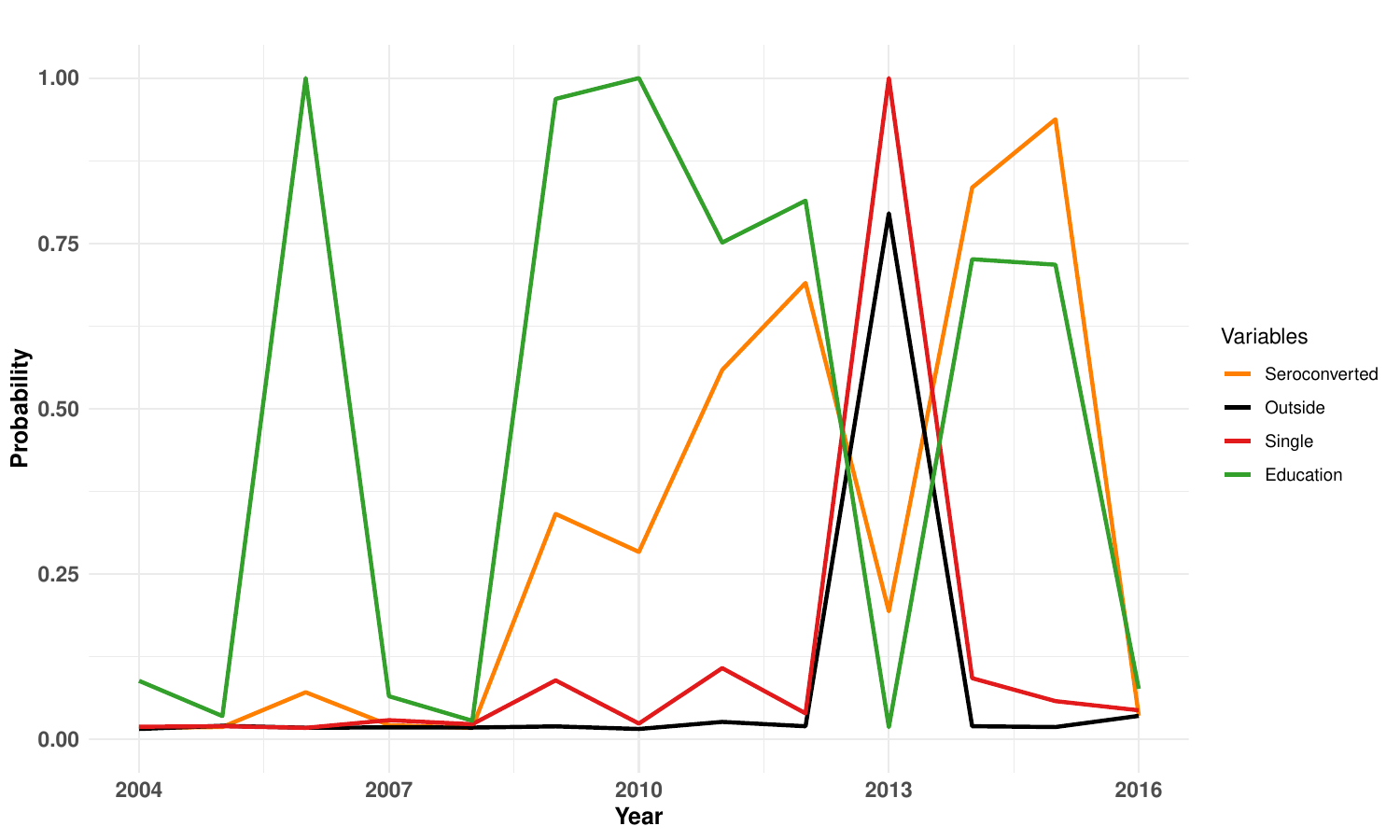} 
\includegraphics[width=\textwidth]{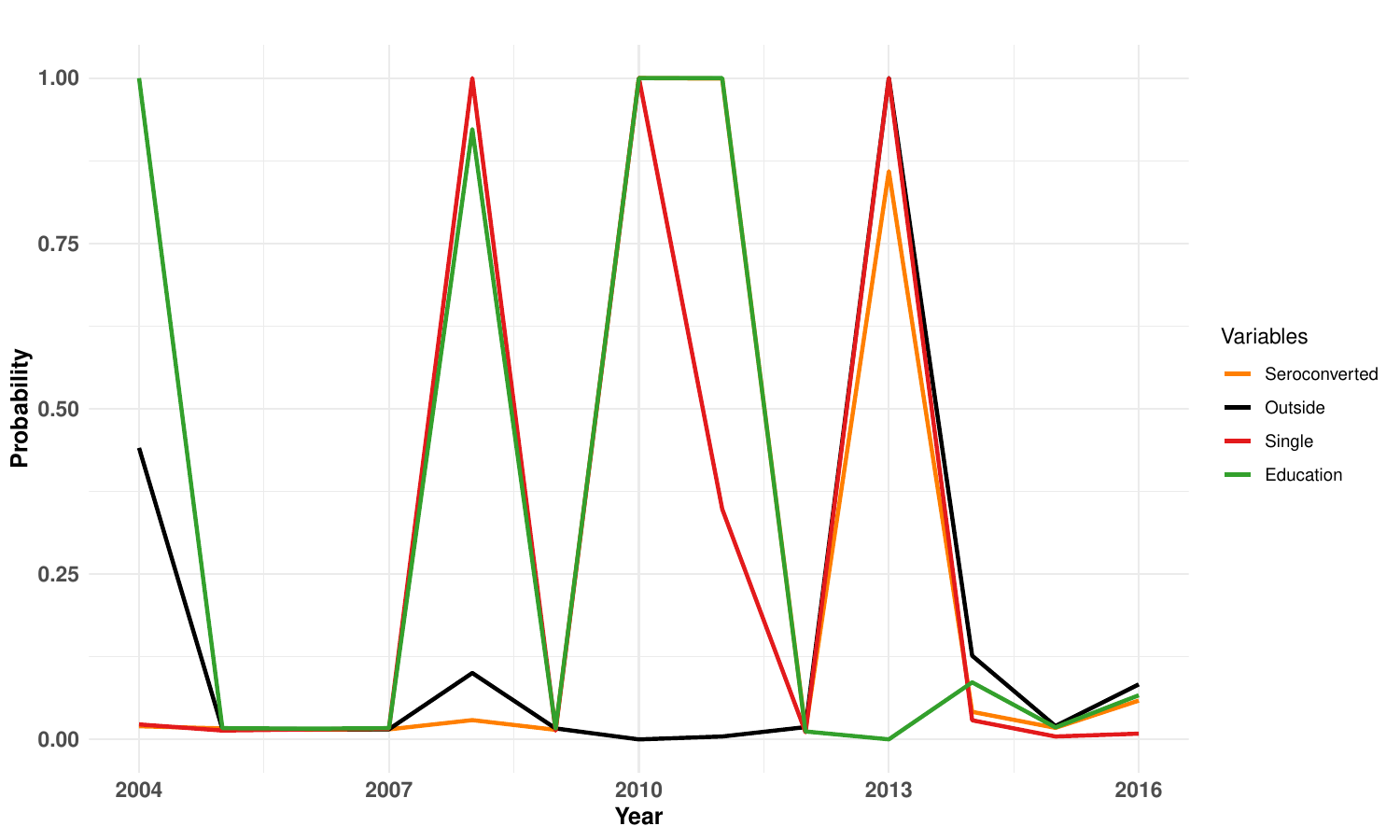} 
    \caption{Posterior inclusion probabilities for edges that involve variable Age in the residual graphs in the HIV data for men (top panel) and women (bottom panel).}
    \label{fig:HIVppiAge}
\end{figure}

\begin{figure}
\centering
\includegraphics[width=\textwidth]{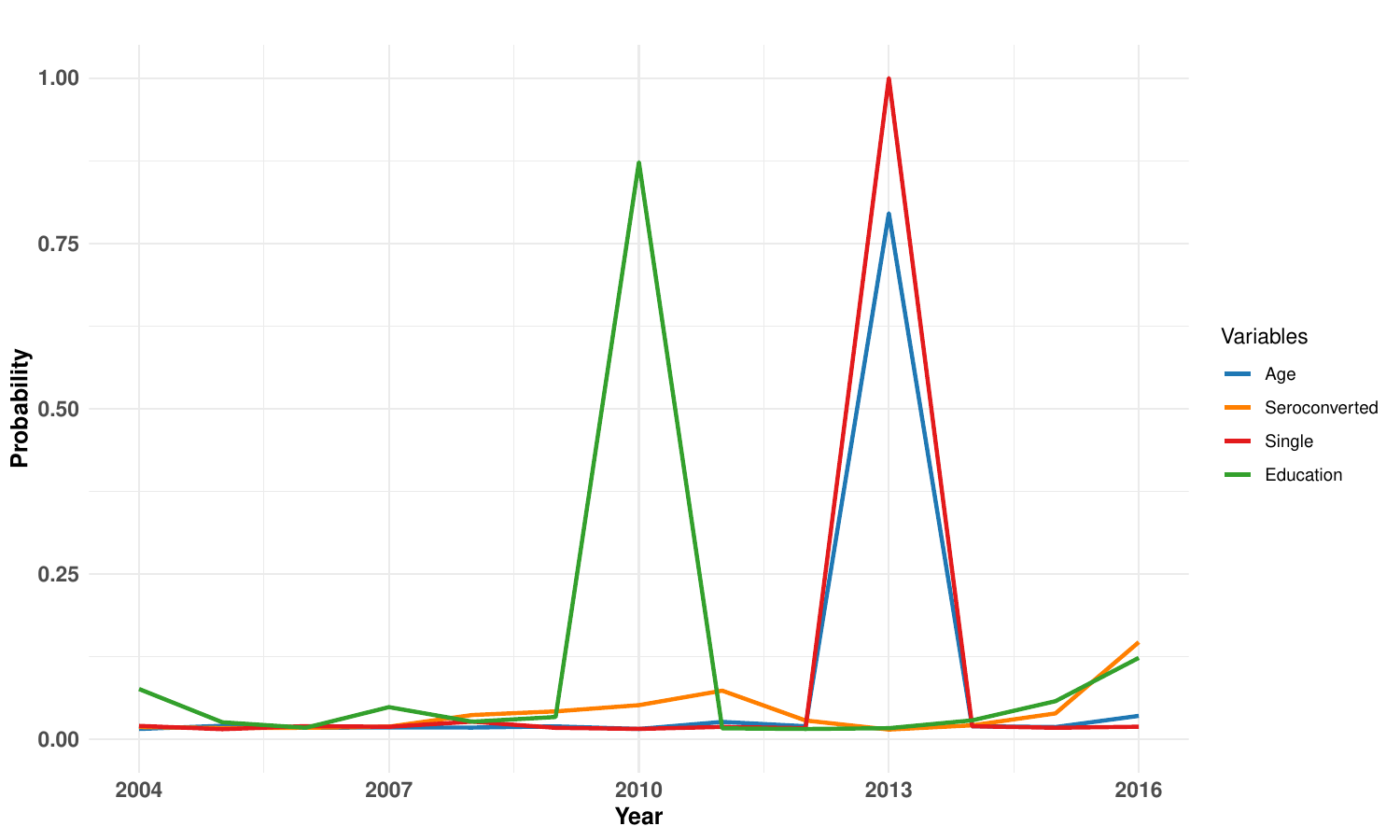} 
\includegraphics[width=\textwidth]{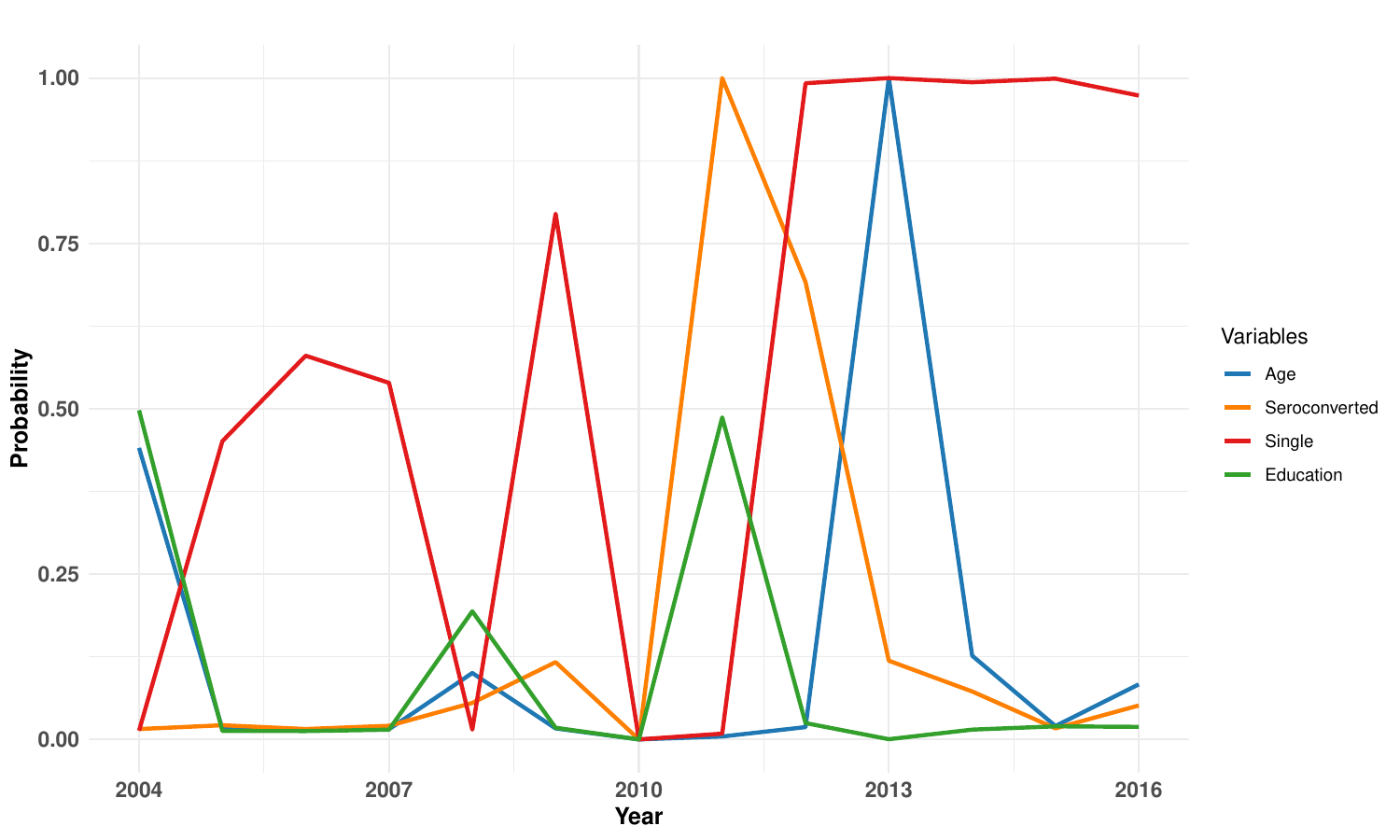} 
    \caption{Posterior inclusion probabilities for edges that involve variable Outside in the residual graphs in the HIV data for men (top panel) and women (bottom panel).}
    \label{fig:HIVppiOutside}
\end{figure}

\begin{figure}
\centering
\includegraphics[width=\textwidth]{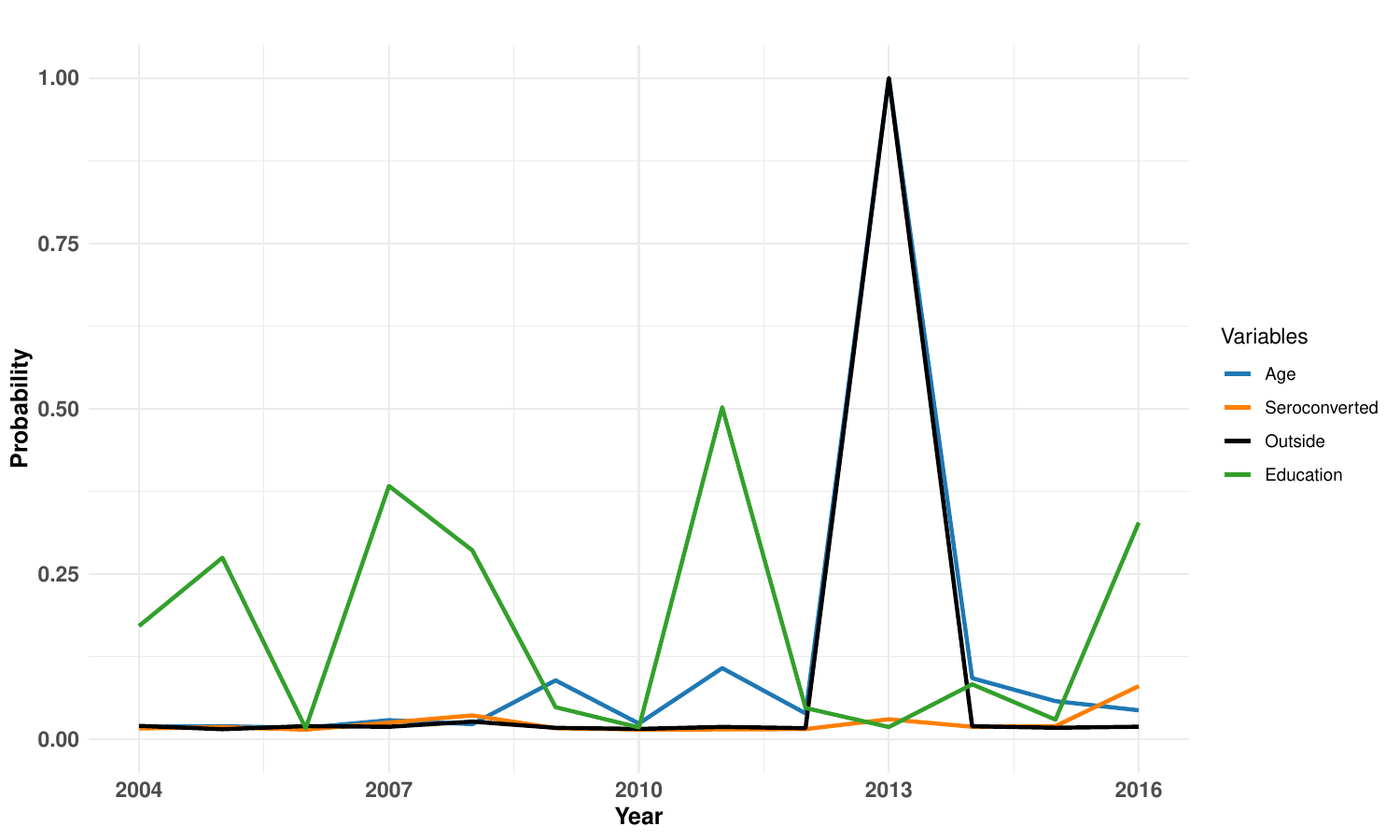} 
\includegraphics[width=\textwidth]{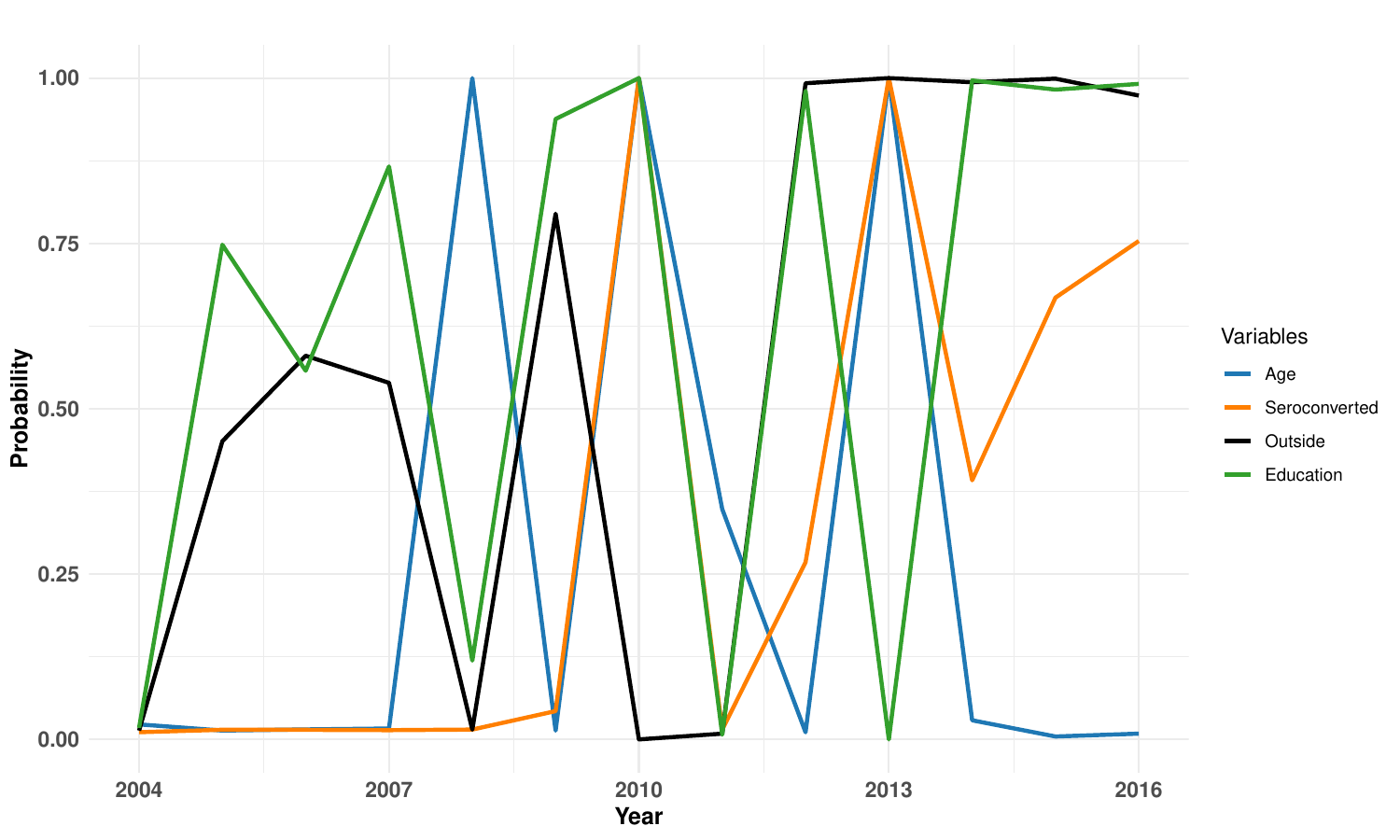} 
    \caption{Posterior inclusion probabilities for edges that involve variable Single in the residual graphs in the HIV data for men (top panel) and women (bottom panel).}
    \label{fig:HIVppiSingle}
\end{figure}

\begin{figure}
\centering
\includegraphics[width=\textwidth]{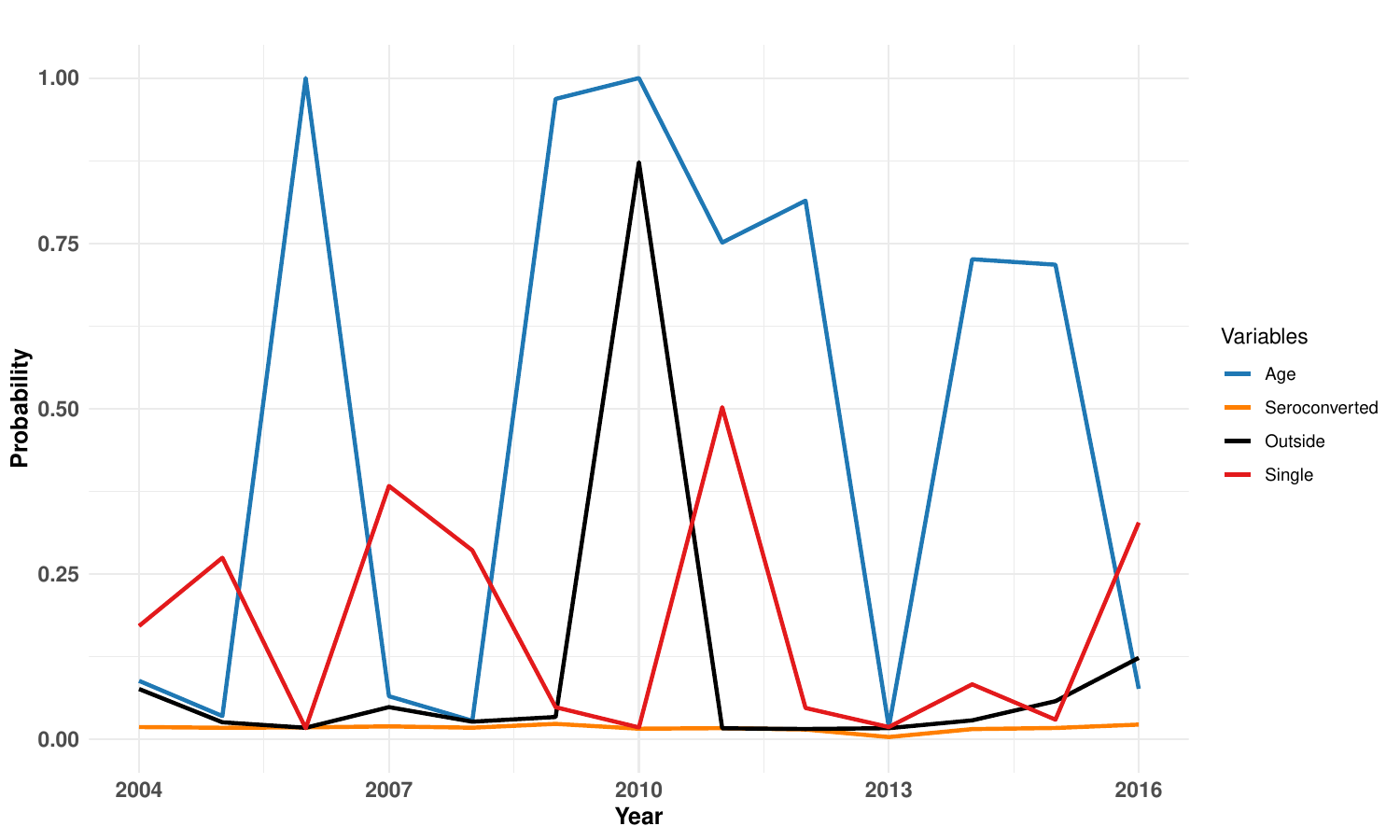} 
\includegraphics[width=\textwidth]{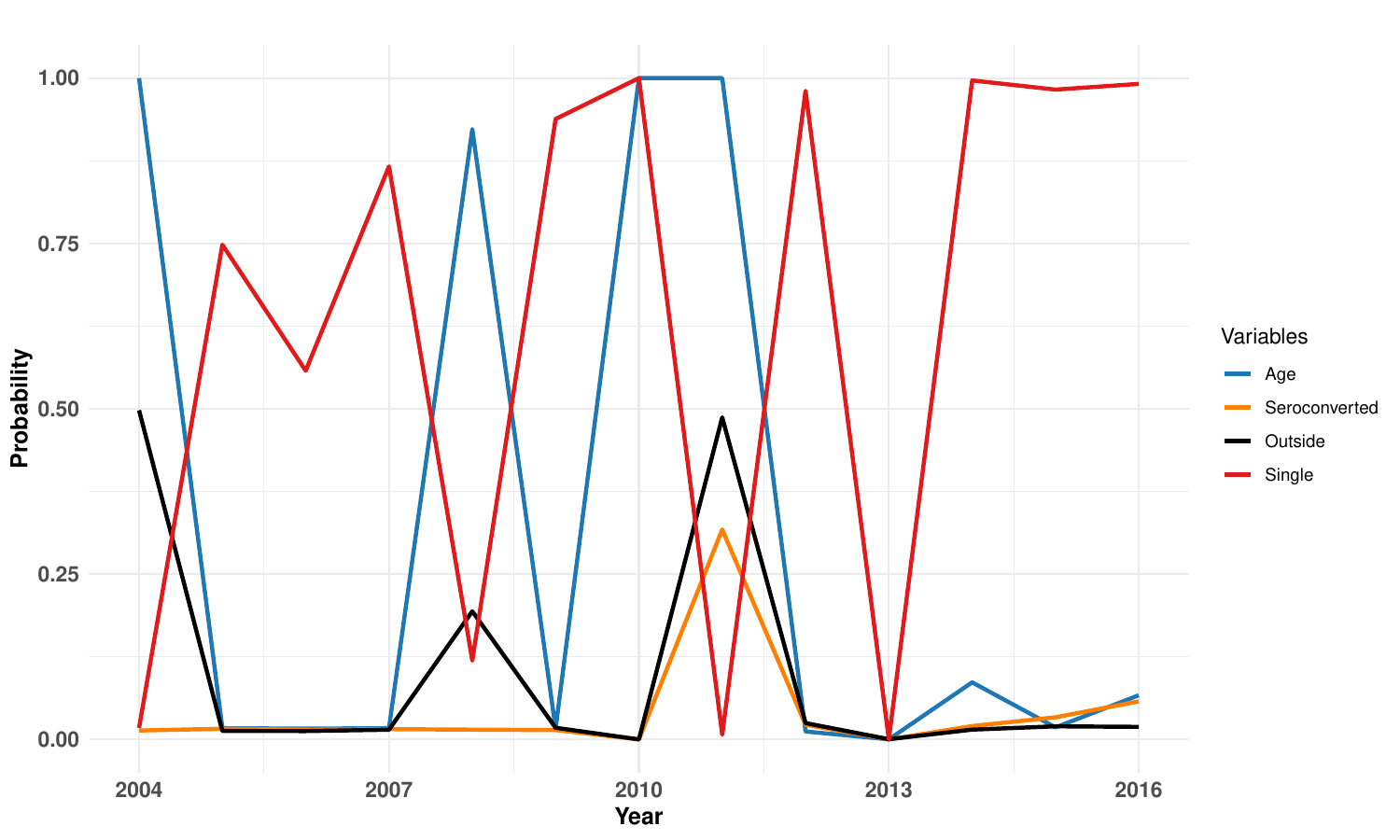} 
    \caption{Posterior inclusion probabilities for edges that involve variable Education in the residual graphs in the HIV data for men (top panel) and women (bottom panel).}
    \label{fig:HIVppiEducation}
\end{figure}

%\bibliographystyle{plainnat}
%\bibliographystyle{plain} % Style BST file
%\bibliography{SingleFactorGraphical}

\end{document}